\documentclass[a4paper,onecolumn,11pt,accepted=2024-02-28]{quantumarticle}
\pdfoutput=1
\usepackage[utf8]{inputenc}
\usepackage[english]{babel}
\usepackage[mathscr]{eucal}
\usepackage[dvipsnames]{xcolor}
\usepackage[T1]{fontenc}
\usepackage{amsmath,tikz,arydshln,array,caption,csquotes,lipsum,hyperref}
\usepackage[section]{placeins}
\usetikzlibrary{decorations.markings}
\usepackage[backend=biber, style=numeric, doi=true, url=true, hyperref=true]{biblatex}
\begin{document}
\title{Quantum circuits for toric code and X-cube fracton model}
\author{Penghua Chen}
\email{chen3014@purdue.edu}
\affiliation{Department of Physics and Astronomy, Purdue University, West Lafayette}
\author{Bowen Yan}
\email{yan312@purdue.edu}
\affiliation{Department of Physics and Astronomy, Purdue University, West Lafayette}
\thanks{The first two authors contributed equally to this work.}
\author{Shawn X. Cui}
\email{cui177@purdue.edu}
\affiliation{Department of Physics and Astronomy, Purdue University, West Lafayette}
\affiliation{Department of Mathematics, Purdue University, West Lafayette}
\thanks{Corresponding author.}
\maketitle

\begin{abstract}
We propose a systematic and efficient quantum circuit composed solely of Clifford gates for simulating the ground state of the surface code model. This approach yields the ground state of the toric code in $\lceil 2L+2+log_{2}(d)+\frac{L}{2d} \rceil$ time steps, where $L$ refers to the system size and $d$ represents the maximum distance to constrain the application of the CNOT gates. Our algorithm reformulates the problem into a purely geometric one, facilitating its extension to attain the ground state of certain 3D topological phases, such as the 3D toric model in $3L+8$ steps and the X-cube fracton model in $12L+11$ steps. Furthermore, we introduce a gluing method involving measurements, enabling our technique to attain the ground state of the 2D toric code on an arbitrary planar lattice and paving the way to more intricate 3D topological phases.
\end{abstract}

\section{Introduction}
The subject of topological phases of matter (TPMs) has been under extensive study for the past few decades. Topological phases are gapped spin liquids at low temperatures which are not described by the conventional Landau theory of spontaneous symmetry breaking and local order parameters; instead, they are characterized by a new order, \textit{topological order}. The ground states of a topological phase have stable degeneracy and robust long range entanglement. Topological phases in 2D also support quasi-particle excitations with anyonic exchange statistics which make them an appealing platform to fault-tolerantly store and process quantum information. Two peculiar features among others are that the ground state degeneracy is a topological invariant of the underlying system, and that the quasi-particles can freely move without costing energy. A large class of topological phases is realized by exactly solvable spin lattice models with bosonic degrees of freedom. A paradigmatic example in 2D is the toric code, and more generally Kitaev's quantum double model based on finite groups \cite{dennis2002topological, kitaev2003fault}, and yet even more generally the Levin-Wen string-net model based on fusion categories \cite{levin2005string}. Examples of 3D topological phases include 3D toric model and the Walker-Wang model based on premodular categories \cite{walker20123+}.

In recent years, more exotic phases in 3D,  called fracton phases, have been discovered \cite{haah2011local, vijay2015new, vijay2016fracton}. Fractons also possess stable ground state degeneracy and long range entanglement.  However, the ground state degeneracy of fractons depends on the system size, and hence is not a topological invariant. Moreover, the mobility of excitations is constrained. The excitations can only move in certain subsystems or cannot move at all. Well known examples of fractons include the Haah code \cite{haah2011local} and the X-cube model \cite{vijay2016fracton}. While regular topological phases are described by topological quantum field theories, it is still an open question what theories mathematically characterize fractons. Since fractons also satisfy the topological order conditions in the sense of \cite{bravyi2010topological}, we call the ground states of a fracton topologically ordered states, in the same way as those of regular topological phases.

Realizing topological phases in physical systems remains an extremely challenging task. On the other hand, there now exist quantum processors based on a number of platforms such as superconducting qubits \cite{satzinger2021realizing}, Rydberg atomic arrays \cite{ebadi2021quantum}, etc. These devices can host physical qubits at the scale of $10^2$, and this number is expected to increase significantly in the near future. Hence, it is both feasible and interesting to simulate topological phases in quantum processors. Thanks to the intrinsic robustness of topological phases, the simulation is relatively less sensitive to the noises in the current quantum processors. We may also gain more insight in topological phases by engineering them in processors.

The toric code ground states were realized in the superconducting-qubit-based systems \cite{satzinger2021realizing} and the Rydberg-atom systems \cite{verresen2021prediction}.  In \cite{satzinger2021realizing}, the authors gave a quantum circuit consisting of Clifford gates to realize the ground states of the planar toric code (a.k.a. surface code \cite{bravyi1998quantum}). Quantum circuits realizing non-Abelian topological orders such as Levin-Wen string-net model and Kitaev quantum double model have  also been studied. See for instance \cite{liu2021methods, tantivasadakarn2021long, verresen2021efficiently, bravyi2022adaptive, tantivasadakarn2022theshortest, tantivasadakarn2022ahierarchy}, though in these cases, the gates utilized are no longer in the Clifford group and measurements are required.

In this paper, we develop quantum circuits realizing the ground states for a number of topological phases. In \cite{satzinger2021realizing}, only planar toric code is considered where the lattice is defined on a planar surface. Here we generalize their method to apply to a large class of surfaces with or without boundary. The quantum circuit consists of only Clifford gates. In toric code, the Hamiltonian consists of two types of operators, the term $A_v$ for each vertex $v$ and the term $B_p$ for each plaquette $p$. See Figure \ref{fig:avandbp}. The key idea of constructing the ground state in \cite{satzinger2021realizing} is as follows.  Start with the product state $|\phi_0\rangle =|00...0\rangle$ (also written as $|0\rangle^{\otimes}$) which is the $+1$ eigenstate for all vertex terms. The ground state is then obtained by projecting $|\phi_0\rangle$ to the $+1$ eigenstate of all plaquette operators,
\begin{equation}
    |GS\rangle \sim \prod_p \frac{1 + B_p}{2} |\phi_0\rangle.
\end{equation}
The effect of $\frac{1 + B_p}{2}$ acting on certain states can be simulated by an appropriate combination of the Hadamard gate and the CNOT gate. For this method to work, the control qubit for CNOT has to be in the $|0\rangle$ state prior to applying the Hadamard and CNOT. Hence, it is critical to choose the right sequence for the plaquettes so that, immediately before simulating the term corresponding to each plaquette $p$, there is always at least one edge on the boundary of $p$ with the state $|0\rangle$. When the lattice is a simple planar lattice, the problem can be easily solved by dividing the lattice into several parts and applying the CNOT gates in a specific order. In this paper, since we consider lattices on arbitrary surfaces, this question is much subtler. 

Here we provide an explicit algorithm to determine the sequence in which the plaquette operators are simulated. We show that this is always possible for a large class of lattices with or without boundary. The result of the algorithm is a quantum circuit consisting of Clifford gates realizing the ground state of the toric code. Moreover, we also adapt this method to 3D phases including the 3D toric model and the X-cube fracton model. For the X-cube model, we again initialize the state to the product of $|0\rangle$ state and simulate the projectors corresponding to cube terms. A similar issue arises that we need to choose the correct sequence to simulate the cube terms. We note that the circuit we provide here realizes an {\it exact} ground state of the X-cube model. By comparison, using cluster states and measurements, the authors in \cite{verresen2021efficiently} gave an {\it approximate} realization of the model.  

In addition to the above method using only quantum gates, we also provide a different way of realizing the same states. The alternative way, called gluing method, combines Clifford gates and measurement of the Pauli $X$ gate. The resulting circuit has a shorter depth than the first one. Of course, for the toric code or X-cube model, it is possible to only use measurement to obtain the ground state. Considering that frequent measurements in near-term quantum processors are costly, our method is a trade-off between circuit depth and degree of measurements.

\section{Realizing ground state of 2D toric code}
\subsection{Toric code}

It is well known that for any Hamiltonian of the form
\begin{equation} \label{eqn:generalformH}
    H = - \sum_{i} P_i,
\end{equation}
where all elements in ${P_i}$ are projectors and mutually commuting, $|GS\rangle$ as defined below is a ground state as long as it is non-zero:
\begin{equation} \label{eqn:generalGS}
    |GS\rangle = \prod_i P_i |\phi\rangle,
\end{equation}
where $|\phi\rangle$ stands for an arbitrary state. Specifically, in a given connected lattice $\Gamma$, $V$ refers to the set of vertices, $P$ refers to the set of plaquettes and $E$ refers to the set of edges. We define $Bo(p) \subseteq E, p \in P$ to be  the set of boundary edges of the plaquette $p$, $\tau(e) \subseteq P, e \in E$ to be the set of plaquettes for which $e$ is a boundary edge, and $\sigma(e) \in V, e \in E$ to be the set of vertices attached with the edge $e$. When each edge is associated with a Hilbert space, we may abuse the notation and use $e$ to represent both an edge and the Hilbert space associated to the edge. As an example, if an edge $e$ appears as a subscript of an operator, it means the operator acts on the Hilbert space attached to the edge $e$.

\begin{table}[ht]
\centering
\begin{tabular}{m{5.6cm} m{5.6cm}}
\begin{center}
\begin{tikzpicture}[scale=0.8]
\foreach \i in {0, 1, 2}
\foreach \j in {1, 2, 3, 4}
        \filldraw[black] ({\i},{\j}) circle (0.8pt);
\foreach \i in {0, 1, 2}
        \draw[black] ({\i},1) -- ({\i},4);
\foreach \j in {1, 2, 3, 4}
        \draw[black] (0,{\j}) -- (3,{\j});
\draw[very thick, black] (0,3) -- (0,4);
\draw[->, >=stealth, black] (-0.7,2.8)--(-0.1,3.5);
\node[black] at (-0.7,2.6) {z-boundary};
\draw[very thick, black] (2,3) -- (3,3);
\draw[->, >=stealth, black] (3.4,3.4)--(2.5,3.1);
\node[black] at (3.4,3.6) {x-boundary};
\draw[very thick, red] (0,2) -- (3,2);
\draw[->, >=stealth, black] (3.4,1.6)--(2.5,1.9);
\node[black] at (3.4,1.4) {direct string};
\end{tikzpicture}
\end{center}
&
\begin{center}
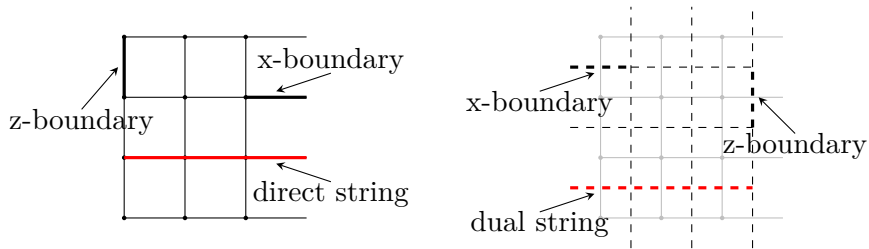

\begin{tikzpicture}[scale=0.8]
\foreach \i in {0, 1, 2}
\foreach \j in {1, 2, 3, 4}
        \filldraw[gray!50] ({\i},{\j}) circle (0.8pt);
\foreach \i in {0, 1, 2}
        \draw[gray!50] ({\i},1) -- ({\i},4);
\foreach \j in {1, 2, 3, 4}
        \draw[gray!50] (0,{\j}) -- (3,{\j});
\foreach \i in {0.5, 1.5}
        \draw[black, dashed] ({\i},0.5) -- ({\i},4.5);
\draw[black, dashed] (2.5,0.5) -- (2.5,2.5);
\draw[black, dashed] (2.5,3.5) -- (2.5,4.5);
\draw[black, dashed] (-0.5,2.5) -- (2.5,2.5);
\draw[black, dashed] (0.5,3.5) -- (2.5,3.5);
\draw[very thick, dashed, black] (-0.5,3.5) -- (0.5,3.5);
\draw[->, >=stealth, black] (-1,3.1)--(-0.1,3.4);
\node[black] at (-1,2.9) {x-boundary};

\draw[very thick, dashed, black] (2.5,2.5) -- (2.5,3.5);
\draw[->, >=stealth, black] (3.2,2.4)--(2.6,3);
\node[black] at (3.2,2.2) {z-boundary};

\draw[very thick, dashed, red] (-0.5,1.5) -- (2.5,1.5);
\draw[->, >=stealth, black] (-1,1.1)--(-0.1,1.4);
\node[black] at (-1,0.9) {dual string};
\end{tikzpicture}
\end{center}
\end{tabular}
\captionof{figure}[foo]{The black solid net on the left represents the lattice $\Gamma$ and the black dashed net on the right represents the dual of $\Gamma$ induced by the gray net.} 
\label{fig:lattice}
\end{table}

As shown in Figure \ref{fig:lattice}, an edge $e$ is a z-boundary when $\tau(e)$ contains only one element, and it is an x-boundary if $\sigma(e)$ contains only one element (see \cite{bravyi1998quantum} for details). On the lattice $\Gamma$, a direct string $S$ is a series of edges $e_i$, $i= 1 \cdots n$ such that $\tau(e_i) \bigcap \tau(e_{i+1}) \neq \oslash$ for $1 \leq i < n$. A direct string operator $F(S)$ is one operator applies $X$ on all edges along the string $S$, thereby generating two electric charges at both ends. Likewise, the concept of a dual string $S'$ can be introduced, representing a direct string in the dual lattice of $\Gamma$.  A dual ribbon operator $F(S')$ applies $Z$ to all edges intersected by the dual string $S'$, resulting in the creation of two magnetic charges at its endpoints.  It is worth noting that dual string operators that encounter a z-boundary at one end will generate (or annihilate) a magnetic charge at the other end.

\begin{table}[ht]
\centering
\begin{tabular}{m{4.5cm} m{4.5cm}} 
\begin{center}
\begin{tikzpicture}[scale=0.8]
\draw[thick, gray] (-2,0) -- (2,0);
\draw[thick, gray] (0,-2) -- (0,2);
\filldraw[gray!50] (0,1) circle (3pt) node[anchor=south west,black] {$Z$};
\filldraw[gray!50] (0,-1) circle (3pt) node[anchor=north west,black] {$Z$};
\filldraw[gray!50] (1,0) circle (3pt) node[anchor=north west,black] {$Z$};
\filldraw[gray!50] (-1,0) circle (3pt) node[anchor=north east,black] {$Z$};
\node[black] at (-0.5,0.5) {$A_{v}$};
\end{tikzpicture}
\end{center}
&
\begin{center}
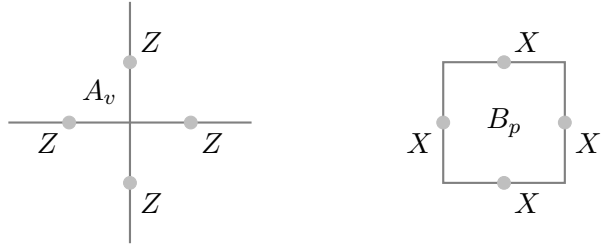

\begin{tikzpicture}[scale=0.8]
\draw[thick, gray] (-1,-1) rectangle (1,1);
\filldraw[gray!50] (0,1) circle (3pt) node[anchor=south west,black] {$X$};
\filldraw[gray!50] (0,-1) circle (3pt) node[anchor=north west,black] {$X$};
\filldraw[gray!50] (1,0) circle (3pt) node[anchor=north west,black] {$X$};
\filldraw[gray!50] (-1,0) circle (3pt) node[anchor=north east,black] {$X$};
\node[black] at (0,0) {$B_{p}$};
\end{tikzpicture}
\end{center}
\end{tabular}
\captionof{figure}[foo]{Definitions of $A_{v}$ and $B_{p}$ operators in toric code.} 
\label{fig:avandbp}
\end{table}

The toric code Hamiltonian $H$ is composed of operators defined in Figure \ref{fig:avandbp}:
\begin{equation} \label{eqn:ground state}
    H = - \sum_{v\in V} A_v-\sum_{p \in P} B_p. 
\end{equation}
The action of $A_v$ (vertex term) is to apply Pauli matrix $Z$ over edges $e$ if $v \in \sigma(e)$, and $B_p$ (plaquette term) acts to apply Pauli matrix $X$ over edges $e$ if $p \in Bo(e)$. Given that $A_v^2 = B_p^2 = 1$ and $[A_v,B_p] = 0$ for all $v \in V$ and $p\in P$, it can be easily verified that $\frac{1+A_v}{2}$ and $\frac{1+B_p}{2}$ function as projectors. By substituting $A_v$ with $\frac{1+A_v}{2}$ and $B_p$ with $\frac{1+B_p}{2}$ in the Hamiltonian, we achieve a equivalent form that matches the form of Equation \ref{eqn:generalformH}. This is due to a natural one-to-one correspondence between their spectra. So we get a ground state \footnote{The ground state degeneracy of the 2D toric code on a torus is four, and the state $|GS\rangle$ corresponds to $|00\rangle$. A comprehensive explanation can be found in Section \ref{sec:arbitrary}.} by Equation \ref{eqn:generalGS}: 
\begin{equation}
    |GS\rangle = \prod_{p \in P} \frac{1+B_p}{2} |\phi_0\rangle,
\end{equation}
where $|\phi_0\rangle=|00...0\rangle$ represents a product state where each qubit is in the $|0\rangle$ state, and we drop all $\frac{1+A_v}{2}$s due to its trivial action on $|\phi_0\rangle$. This state is non-zero due to the presence of positive coefficients of each components. 

\subsection{Single plaquette}

To systematically introduce our ground state simulation method, we initiate with the most elementary scenario: applying $\frac{1+B_p}{2}$ on a single plaquette, which is the \textit{basic structure} in 2D toric code. A Hadamard gate $H$ is naturally described by $\frac{X+Z}{\sqrt{2}}$, and CNOT gate $C_{i \to j}$ is defined as
\begin{equation}
    C_{i \to j} |ij\rangle = (\frac{1-Z_i}{2}X_j + \frac{1+Z_i}{2} )|ij\rangle,
\end{equation}
where $i$ is the control qubit and $j$ is the target qubit. 

\begin{figure}[ht]
\centering
\begin{tikzpicture}[scale=0.9]
\draw[thick, gray] (-4.5,0) rectangle (-2.5,2);
\draw[thick, gray] (-2,0) rectangle (0,2);
\draw[thick, gray] (0.5,0) rectangle (2.5,2);
\draw[thick, gray] (3,0) rectangle (5,2);
\filldraw[gray!50] (-3.5,2) circle (3pt);
\draw[thick, black] (-3.5,2) circle (3pt);
\node[black] at (-3.5,2.3) {$H_{1}$};
\filldraw[gray!50] (-4.5,1) circle (3pt);
\node[black] at (-3.5,1.7) {$1$};
\node[black] at (-4.2,1) {$2$};
\node[black] at (-3.5,0.3) {$3$};
\node[black] at (-2.8,1) {$4$};
\filldraw[gray!50] (-2.5,1) circle (3pt);
\filldraw[gray!50] (-3.5,0) circle (3pt);
\draw[double, ->, >=stealth, black] (-1,2)--(-1.9,1.1);
\node[black] at (-0.9,1.3) {$C_{1 \to 2}$};
\filldraw[black] (-1,2) circle (3pt);
\node[black] at (-0.7,2.3) {$\frac{|0\rangle+|1\rangle}{2}$};
\node[black] at (-1.7,0.7) {$|0\rangle$};
\filldraw[gray!50] (-2,1) circle (3pt);
\filldraw[gray!50] (0,1) circle (3pt);
\filldraw[gray!50] (-1,0) circle (3pt);
\draw[double, ->, >=stealth, black] (1.5,2)--(1.5,0.14);
\draw[double, ->, >=stealth, black] (1.5,2)--(2.4,1.1);
\filldraw[black] (0.5,1) circle (3pt);
\filldraw[black] (1.5,2) circle (3pt);
\filldraw[gray!50] (2.5,1) circle (3pt);
\filldraw[gray!50] (1.5,0) circle (3pt);
\node[black] at (1,1.5) {{$\frac{|00\rangle+|11\rangle}{2}$}};
\filldraw[black] (3,1) circle (3pt);
\filldraw[black] (4,2) circle (3pt);
\filldraw[black] (5,1) circle (3pt);
\filldraw[black] (4,0) circle (3pt);
\node[black] at (4,1) {{$\frac{|0000\rangle+|1111\rangle}{2}$}};
\end{tikzpicture}
\caption{Initially, a qubit in the state $|0\rangle$ is situated at each gray dot. As quantum gates are applied to these qubits, their color changes to black. A circle positioned on a dot signifies the application of a Hadamard gate to the corresponding qubit, while an arrow indicates a CNOT gate, with the arrowhead pointing from the control qubit to the target qubit.}
\label{fig:2d1plaquette}
\end{figure}
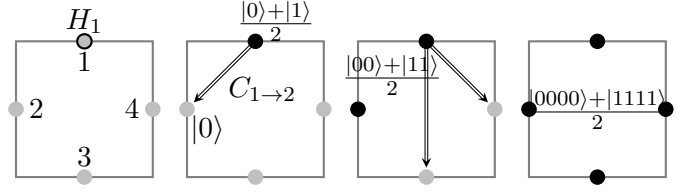

In the single plaquette shown in Figure \ref{fig:2d1plaquette}, four qubits labeled $1,2,3,$ and $4$ are initialized to the state $|0\rangle$. Subsequently, we will systematically implement Hadamard and CNOT gates in a specific sequence, as outlined in the figure. After the application of $H_1$ and $C_{1 \to 2}$, we have
\begin{equation} \label{eqn:the action of CNOT gate}
C_{1 \to 2} H_1 |0000\rangle = (\frac{1-Z_1}{2}X_2 + \frac{1+Z_1}{2}) \frac{X_1+Z_1}{\sqrt{2}} |0000\rangle = \frac{X_1 X_2 + 1}{\sqrt{2}} |0000\rangle.
\end{equation}
Explicitly, we can insert a $\frac{1+Z}{2}$ into the equation as $\frac{1+Z}{2}|0\rangle=|0\rangle$:
\begin{align}
     &(\frac{1-Z_1}{2}X_2 + \frac{1+Z_1}{2}) \frac{X_1+Z_1}{\sqrt{2}} \frac{1+Z_1}{2} \label{eqn:line 1}\\
    =&\frac{1}{\sqrt{2}} \{ (X_1 \frac{1+Z_1}{2}X_2 + X_1 \frac{1-Z_1}{2})+( \frac{1-Z_1}{2}X_2 + \frac{1+Z_1}{2}) \} \frac{1+Z_1}{2} \label{eqn:line 2}\\
    =& \frac{1}{\sqrt{2}}(X_1 X_2+1 ) \frac{1+Z_1}{2} \label{eqn:line 3}
\end{align}
Notice $\frac{1+Z_1}{2}$ survives within $\{ \; \}$ in Equation \ref{eqn:line 2}. After reverting to the original expression and substituting $\frac{1+Z_1}{2}$ with $1$, we verified the accuracy of Equation \ref{eqn:the action of CNOT gate}.  Importantly, this equation holds for any quantum state $|\phi\rangle$:
\begin{equation}
    C_{1 \to 2} H_1 |0\rangle \otimes |\phi\rangle =  \frac{X_1 X_2 + 1}{\sqrt{2}} |0\rangle \otimes |\phi\rangle,
\end{equation}
since the key step only requires that the initial state must be the eigenstate of $Z_1$ with an eigenvalue $+1$. Finally, applying the other CNOT gates results in
\begin{equation}
\prod_{i=2}^{4} C_{1 \to i} H_1 |0000\rangle =\frac{X_1 X_2 X_3 X_4 + 1}{\sqrt{2}} |0000\rangle = \frac{1+B_p}{\sqrt{2}} |0000\rangle,
\end{equation}
which is the desired ground state. It is important to observe that this procedure remains effective as long as a qubit from $B_o(p)$ is initially in the state $|0\rangle$. We term such qubits as \textit{free qubits}, and their presence is pivotal when considering scenarios involving multiple plaquettes.

\subsection{Developing to a surface with boundary}

Given a complicated lattice $\Gamma$ in the state $|\phi_0\rangle$, we need to find a path (termed permissible order in \cite{liu2021methods}) through all plaquettes ${p_i}$, such that $\bigcup_{i} p_i = P$, using a sequence of edges $e_i \in Bo(p_i)$ where $e_i \notin \bigcup_{j=1}^{i-1}p_j$. Each $e_i$ is then utilized as a free qubit to apply the introduced basic structure, resulting in the accumulation of $\prod_{i} \frac{1+B_{p_i}}{\sqrt{2}}$ over $|0 \cdots 0\rangle$, which represents the ground state of the toric code on lattice $\Gamma$. To illustrate the procedure, we take four plaquettes as an example depicted in Figure \ref{fig:2d4plaquettes}. A path featuring four free qubits $e_1$ to $e_4$ has been chosen, where $e_i$ starts in the state $|0\rangle$ at the onset of every step. Upon completing the path, the desired ground state is eventually obtained.

\begin{table}[ht]
\centering
\begin{tabular}{m{2.6cm} m{2.6cm} m{2.6cm} m{2.6cm}} 
\begin{center}
\begin{tikzpicture}[scale=0.6]
\draw[thick, gray] (-2,-2) rectangle (2,2);
\draw[thick, gray] (-2,0) -- (2,0);
\draw[thick, gray] (0,-2) -- (0,2);
\foreach \i in {-1, 1}
\foreach \j in {-2, 0, 2}
        \filldraw[gray!50] ({\i},{\j}) circle (3pt);
\foreach \i in {-2, 0, 2}
\foreach \j in {-1, 1}
        \filldraw[gray!50] ({\i},{\j}) circle (3pt);
\node[black] at (-1,2.3) {$e_{1}$};
\node[black] at (1,2.3) {$e_{2}$};
\node[black] at (1,-2.3) {$e_{3}$};
\node[black] at (-1,-2.3) {$e_{4}$};
\draw[thick, black] (-1,2) circle (3pt);
\draw[double, ->, >=stealth, black] (-1,2)--(-1.9,1.1);
\draw[double, ->, >=stealth, black] (-1,2)--(-0.1,1.1);
\draw[double, ->, >=stealth, black] (-1,2)--(-1,0.14);
\end{tikzpicture}
\end{center}
&
\begin{center}
\begin{tikzpicture}[scale=0.6]
\draw[thick, gray] (-2,-2) rectangle (2,2);
\draw[thick, gray] (-2,0) -- (2,0);
\draw[thick, gray] (0,-2) -- (0,2);
\foreach \i in {-1, 1}
\foreach \j in {-2, 0, 2}
        \filldraw[gray!50] ({\i},{\j}) circle (3pt);
\foreach \i in {-2, 0, 2}
\foreach \j in {-1, 1}
        \filldraw[gray!50] ({\i},{\j}) circle (3pt);
\node[black] at (-1,2.3) {$e_{1}$};
\node[black] at (1,2.3) {$e_{2}$};
\node[black] at (1,-2.3) {$e_{3}$};
\node[black] at (-1,-2.3) {$e_{4}$};
\filldraw[black] (-2,1) circle (3pt);
\filldraw[black] (-1,2) circle (3pt);
\filldraw[black] (0,1) circle (3pt);
\filldraw[black] (-1,0) circle (3pt);
\draw[thick, black] (1,2) circle (3pt);
\draw[double, ->, >=stealth, black] (1,2)--(1.9,1.1);
\draw[double, ->, >=stealth, black] (1,2)--(0.1,1.1);
\draw[double, ->, >=stealth, black] (1,2)--(1,0.14);
\end{tikzpicture}
\end{center}
&
\begin{center}
\begin{tikzpicture}[scale=0.6]
\draw[thick, gray] (-2,-2) rectangle (2,2);
\draw[thick, gray] (-2,0) -- (2,0);
\draw[thick, gray] (0,-2) -- (0,2);
\foreach \i in {-1, 1}
\foreach \j in {-2, 0, 2}
        \filldraw[gray!50] ({\i},{\j}) circle (3pt);
\foreach \i in {-2, 0, 2}
\foreach \j in {-1, 1}
        \filldraw[gray!50] ({\i},{\j}) circle (3pt);
\node[black] at (-1,2.3) {$e_{1}$};
\node[black] at (1,2.3) {$e_{2}$};
\node[black] at (1,-2.3) {$e_{3}$};
\node[black] at (-1,-2.3) {$e_{4}$};
\filldraw[black] (-2,1) circle (3pt);
\filldraw[black] (-1,2) circle (3pt);
\filldraw[black] (0,1) circle (3pt);
\filldraw[black] (-1,0) circle (3pt);
\filldraw[black] (2,1) circle (3pt);
\filldraw[black] (1,2) circle (3pt);
\filldraw[black] (1,0) circle (3pt);
\draw[thick, black] (1,-2) circle (3pt);
\draw[double, ->, >=stealth, black] (1,-2)--(1.9,-1.1);
\draw[double, ->, >=stealth, black] (1,-2)--(0.1,-1.1);
\draw[double, ->, >=stealth, black] (1,-2)--(1,-0.14);
\end{tikzpicture}
\end{center}
&
\begin{center}
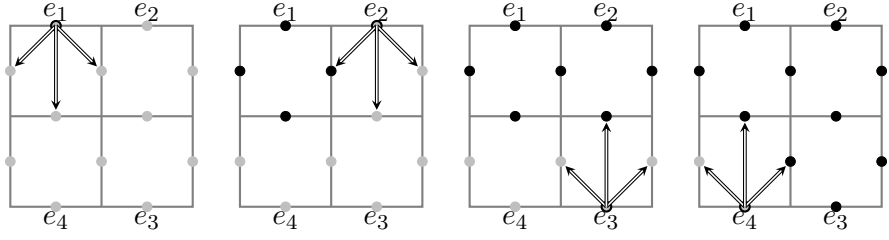

\begin{tikzpicture}[scale=0.6]
\draw[thick, gray] (-2,-2) rectangle (2,2);
\draw[thick, gray] (-2,0) -- (2,0);
\draw[thick, gray] (0,-2) -- (0,2);
\foreach \i in {-1, 1}
\foreach \j in {-2, 0, 2}
        \filldraw[gray!50] ({\i},{\j}) circle (3pt);
\foreach \i in {-2, 0, 2}
\foreach \j in {-1, 1}
        \filldraw[gray!50] ({\i},{\j}) circle (3pt);
\node[black] at (-1,2.3) {$e_{1}$};
\node[black] at (1,2.3) {$e_{2}$};
\node[black] at (1,-2.3) {$e_{3}$};
\node[black] at (-1,-2.3) {$e_{4}$};
\filldraw[black] (-2,1) circle (3pt);
\filldraw[black] (-1,2) circle (3pt);
\filldraw[black] (0,1) circle (3pt);
\filldraw[black] (-1,0) circle (3pt);
\filldraw[black] (2,1) circle (3pt);
\filldraw[black] (1,2) circle (3pt);
\filldraw[black] (1,0) circle (3pt);
\filldraw[black] (2,-1) circle (3pt);
\filldraw[black] (1,-2) circle (3pt);
\filldraw[black] (0,-1) circle (3pt);
\draw[thick, black] (-1,-2) circle (3pt);
\draw[double, ->, >=stealth, black] (-1,-2)--(-1.9,-1.1);
\draw[double, ->, >=stealth, black] (-1,-2)--(-0.1,-1.1);
\draw[double, ->, >=stealth, black] (-1,-2)--(-1,-0.14);
\end{tikzpicture}
\end{center}
\end{tabular}
\captionof{figure}[foo]{The procedure on the basics structure is applying Hadamard gate on any qubit at $|0\rangle$ first and CNOT gates to other qubits in any order.}
\label{fig:2d4plaquettes}
\end{table}

\subsection{Developing to a surface without boundary}\label{sec:2dtorus}

The scenario shifts when dealing with a surface without boundary. While the initial state remains $|\phi_0\rangle$, it becomes impossible to locate a path with sufficient free qubits to cover the entire lattice. Fortunately, as every edge sides two plaquettes, the equation holds: 
\begin{equation} \label{eqn:constraint for 2d toric code}
    \prod_{p \in P} B_p = 1,
\end{equation}
which implies that we can intentionally choose a specific $B_p$ as redundant. Consequently, we can select the final plaquette as the redundant one, effectively terminating the path. To illustrate, consider the lattices on a torus shown in Figure \ref{fig:2dtorus1}, there is no need to apply Hadamard and CNOT gates to the bottom left plaquette, as we have previously simulated the toric code's ground state.

\begin{table}[ht]
\centering
\begin{tabular}{m{3.3cm} m{3.3cm} m{3.3cm}} 
\begin{center}
\begin{tikzpicture}[scale=0.6]
\draw[thick, gray] (-2,-2) rectangle (2,2);
\draw[thick, red] (-2,-2) -- (2,-2);
\draw[thick, red] (-2,2) -- (2,2);
\draw[thick, blue] (-2,-2) -- (-2,2);
\draw[thick, blue] (2,-2) -- (2,2);
\draw[thick, gray] (-2,0) -- (2,0);
\draw[thick, gray] (0,-2) -- (0,2);
\foreach \i in {-1, 1}
\foreach \j in {-2, 0, 2}
        \filldraw[gray!50] ({\i},{\j}) circle (3pt);
\foreach \i in {-2, 0, 2}
\foreach \j in {-1, 1}
        \filldraw[gray!50] ({\i},{\j}) circle (3pt);
\node[black] at (-1,2.3) {$e_{1}$};
\node[black] at (1,2.3) {$e_{2}$};
\node[black] at (2.4,-1) {$e_{3}$};
\draw[thick, black] (-1,2) circle (3pt);
\draw[double, ->, >=stealth, black] (-1,2)--(-1.9,1.1);
\draw[double, ->, >=stealth, black] (-1,2)--(-0.1,1.1);
\draw[double, ->, >=stealth, black] (-1,2)--(-1,0.14);
\end{tikzpicture}
\end{center}
&
\begin{center}
\begin{tikzpicture}[scale=0.6]
\draw[thick, gray] (-2,-2) rectangle (2,2);
\draw[thick, red] (-2,-2) -- (2,-2);
\draw[thick, red] (-2,2) -- (2,2);
\draw[thick, blue] (-2,-2) -- (-2,2);
\draw[thick, blue] (2,-2) -- (2,2);
\draw[thick, gray] (-2,0) -- (2,0);
\draw[thick, gray] (0,-2) -- (0,2);
\foreach \i in {-1, 1}
\foreach \j in {-2, 0, 2}
        \filldraw[gray!50] ({\i},{\j}) circle (3pt);
\foreach \i in {-2, 0, 2}
\foreach \j in {-1, 1}
        \filldraw[gray!50] ({\i},{\j}) circle (3pt);
\node[black] at (-1,2.3) {$e_{1}$};
\node[black] at (1,2.3) {$e_{2}$};
\node[black] at (2.4,-1) {$e_{3}$};
\filldraw[black] (-2,1) circle (3pt);
\filldraw[black] (-1,2) circle (3pt);
\filldraw[black] (0,1) circle (3pt);
\filldraw[black] (-1,0) circle (3pt);
\filldraw[black] (-1,-2) circle (3pt);
\filldraw[black] (2,1) circle (3pt);
\draw[thick, black] (1,2) circle (3pt);
\draw[double, ->, >=stealth, black] (1,2)--(1.9,1.1);
\draw[double, ->, >=stealth, black] (1,2)--(0.1,1.1);
\draw[double, ->, >=stealth, black] (1,2)--(1,0.14);
\end{tikzpicture}
\end{center}
&
\begin{center}
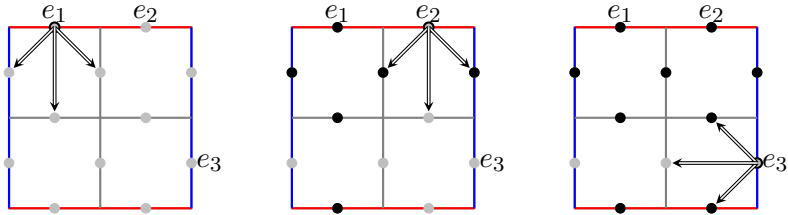

\begin{tikzpicture}[scale=0.6]
\draw[thick, gray] (-2,-2) rectangle (2,2);
\draw[thick, red] (-2,-2) -- (2,-2);
\draw[thick, red] (-2,2) -- (2,2);
\draw[thick, blue] (-2,-2) -- (-2,2);
\draw[thick, blue] (2,-2) -- (2,2);
\draw[thick, gray] (-2,0) -- (2,0);
\draw[thick, gray] (0,-2) -- (0,2);
\foreach \i in {-1, 1}
\foreach \j in {-2, 0, 2}
        \filldraw[gray!50] ({\i},{\j}) circle (3pt);
\foreach \i in {-2, 0, 2}
\foreach \j in {-1, 1}
        \filldraw[gray!50] ({\i},{\j}) circle (3pt);
\node[black] at (-1,2.3) {$e_{1}$};
\node[black] at (1,2.3) {$e_{2}$};
\node[black] at (2.4,-1) {$e_{3}$};
\filldraw[black] (-2,1) circle (3pt);
\filldraw[black] (-1,2) circle (3pt);
\filldraw[black] (0,1) circle (3pt);
\filldraw[black] (-1,0) circle (3pt);
\filldraw[black] (-1,-2) circle (3pt);
\filldraw[black] (2,1) circle (3pt);
\filldraw[black] (1,2) circle (3pt);
\filldraw[black] (1,0) circle (3pt);
\filldraw[black] (1,-2) circle (3pt);
\draw[thick, black] (2,-1) circle (3pt);
\draw[double, ->, >=stealth, black] (2,-1)--(1.1,-0.1);
\draw[double, ->, >=stealth, black] (2,-1)--(1.1,-1.9);
\draw[double, ->, >=stealth, black] (2,-1)--(0.14,-1);
\end{tikzpicture}
\end{center}
\end{tabular}
\captionof{figure}[foo]{Boundaries with the same color are identified to represent a torus.}
\label{fig:2dtorus1}
\end{table}

This method remains applicable to more intricate 2D surfaces with or without boundary, provided a suitable path can be identified. Additional examples are provided in Appendices \ref{sec:2dsphere} and \ref{sec:2dothers}. Furthermore, the gluing method described in Section \ref{sec:gluingmethod} empowers us to simulate ground states on arbitrary planar lattices.

\subsection{Simulate arbitrary ground state} \label{sec:arbitrary}
As stated in \cite{kitaev2003fault}, the degeneracy of ground states for 2D toric code on torus is four: $|00\rangle$, $|01\rangle$, $|10\rangle$ and $|11\rangle$. The ground state $|00\rangle$, presented in Section \ref{sec:2dtorus}, is simulated  from the initial state $\phi_0$. Due to the properties of logical operators, which can interchange ground states and commute with $B_{p}$, it is feasible to apply them to $\phi_0$ to obtain the remaining ground states.

\begin{table}[ht]
\centering
\begin{tabular}{m{2.6cm} m{2.6cm} m{2.6cm} m{2.6cm}} 
\begin{center}
\begin{tikzpicture}[scale=0.6]
\foreach \i in {-2, 2}
        \draw[thick, red] (-2,{\i}) -- (2,{\i});
\foreach \i in {-2, 2}
        \draw[thick, blue] ({\i},-2) -- ({\i},2);
\draw[thick, gray] (-2,0) -- (2,0);
\draw[thick, gray] (0,-2) -- (0,2);
\foreach \i in {-1, 1}
\foreach \j in {-2, 0, 2}
        \filldraw[gray!50] ({\i},{\j}) circle (3pt);
\foreach \i in {-2, 0, 2}
\foreach \j in {-1, 1}
        \filldraw[gray!50] ({\i},{\j}) circle (3pt);
\node[black] at (0,-2.5) {$\phi_0$};
\end{tikzpicture}
\end{center}
&
\begin{center}
\begin{tikzpicture}[scale=0.6]
\foreach \i in {-2, 2}
        \draw[thick, red] (-2,{\i}) -- (2,{\i});
\foreach \i in {-2, 2}
        \draw[thick, blue] ({\i},-2) -- ({\i},2);
\draw[thick, gray] (-2,0) -- (2,0);
\draw[thick, gray] (0,-2) -- (0,2);
\foreach \i in {-1, 1}
\foreach \j in {-2, 0, 2}
        \filldraw[gray!50] ({\i},{\j}) circle (3pt);
\foreach \i in {-2, 0, 2}
\foreach \j in {-1, 1}
        \filldraw[gray!50] ({\i},{\j}) circle (3pt);
\draw[thick, black] (0,-2) -- (0,2);
\filldraw[black] (0,1) circle (3pt);
\filldraw[black] (0,-1) circle (3pt);
\node[black] at (0,-2.5) {$\phi_{01}$};
\end{tikzpicture}
\end{center}
&
\begin{center}
\begin{tikzpicture}[scale=0.6]
\foreach \i in {-2, 2}
        \draw[thick, red] (-2,{\i}) -- (2,{\i});
\foreach \i in {-2, 2}
        \draw[thick, blue] ({\i},-2) -- ({\i},2);
\draw[thick, gray] (-2,0) -- (2,0);
\draw[thick, gray] (0,-2) -- (0,2);
\foreach \i in {-1, 1}
\foreach \j in {-2, 0, 2}
        \filldraw[gray!50] ({\i},{\j}) circle (3pt);
\foreach \i in {-2, 0, 2}
\foreach \j in {-1, 1}
        \filldraw[gray!50] ({\i},{\j}) circle (3pt);
\draw[thick, black] (-2,0) -- (2,0);
\filldraw[black] (1,0) circle (3pt);
\filldraw[black] (-1,0) circle (3pt);
\node[black] at (0,-2.5) {$\phi_{10}$};
\end{tikzpicture}
\end{center}
&
\begin{center}
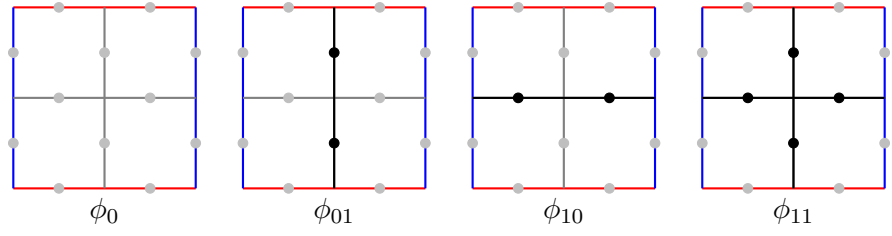

\begin{tikzpicture}[scale=0.6]
\foreach \i in {-2, 2}
        \draw[thick, red] (-2,{\i}) -- (2,{\i});
\foreach \i in {-2, 2}
        \draw[thick, blue] ({\i},-2) -- ({\i},2);
\draw[thick, gray] (-2,0) -- (2,0);
\draw[thick, gray] (0,-2) -- (0,2);
\foreach \i in {-1, 1}
\foreach \j in {-2, 0, 2}
        \filldraw[gray!50] ({\i},{\j}) circle (3pt);
\foreach \i in {-2, 0, 2}
\foreach \j in {-1, 1}
        \filldraw[gray!50] ({\i},{\j}) circle (3pt);
\draw[thick, black] (0,-2) -- (0,2);
\draw[thick, black] (-2,0) -- (2,0);
\filldraw[black] (0,1) circle (3pt);
\filldraw[black] (0,-1) circle (3pt);
\filldraw[black] (1,0) circle (3pt);
\filldraw[black] (-1,0) circle (3pt);
\node[black] at (0,-2.5) {$\phi_{11}$};
\end{tikzpicture}
\end{center}
\end{tabular}
\captionof{figure}[foo]{A qubit $|0\rangle$ is placed at each gray dot and the color changes to black when operator $X$ flips the qubit from $|0\rangle$ to $|1\rangle$.} 
\label{fig:others}
\end{table}

Illustrated in Figure \ref{fig:others}, a vertical loop and a horizontal loop of $X$ represent the two logical operators. They are capable of transforming $\phi_0$ into $\phi_{01}$, $\phi_{10}$ and $\phi_{11}$, which correspond to the initial states of $|01\rangle$, $|10\rangle$ and $|11\rangle$, respectively. Subsequently, we can repeat the same procedure detailed in Section \ref{sec:2dtorus}, but with a modification: utilize $X_{i}C_{i \to j}X_{i}$ instead of $C_{i \to j}$ when encountering a flipped qubit $e_i$.

One step further, in order to obtain an arbitrary ground state $\Phi=ae^{i\theta_{a}}|00\rangle+be^{i\theta_{b}}|01\rangle+ce^{i\theta_{c}}|10\rangle+de^{i\theta_{d}}|11\rangle$, we can implement the unitary operator $U$ outlined in Equation \ref{eqn:unitarymatrix} on an adjacent pair of vertical and horizontal edges of $\phi_0$ and subsequently utilize CNOT gates to transmit vertically and horizontally to get $\phi$. From there, we can proceed by repeating the aforementioned method and we must avoid qubits that have already been utilized, opting instead for free qubits.

\begin{align}
U_1= & 
\left(\begin{smallmatrix}
\frac{a}{\sqrt{a^2+b^2}} & \frac{-b}{\sqrt{a^2+b^2}} & 0 & 0\\
\frac{b}{\sqrt{a^2+b^2}} & \frac{a}{\sqrt{a^2+b^2}} & 0 & 0\\
0 & 0 & \frac{c}{\sqrt{c^2+d^2}} & \frac{-d}{\sqrt{c^2+d^2}}\\
0 & 0 & \frac{d}{\sqrt{c^2+d^2}} & \frac{c}{\sqrt{c^2+d^2}}
\end{smallmatrix}\right)
\left(\begin{smallmatrix}
\sqrt{a^2+b^2} & 0 & -\sqrt{c^2+d^2} & 0\\
0 & \sqrt{a^2+b^2} & 0 & -\sqrt{c^2+d^2}\\
\sqrt{c^2+d^2} & 0 & \sqrt{a^2+b^2} & 0\\
0 & \sqrt{c^2+d^2} & 0 & \sqrt{a^2+b^2}
\end{smallmatrix}\right), \nonumber\\
& \; \; \; \; \; \; \; \; \; \; \; \; U_2=
\left(\begin{smallmatrix}
e^{i\theta_{a}} & 0 & 0 & 0\\
0 & e^{i\theta_{b}} & 0 & 0\\
0 & 0 & e^{i\theta_{c}} & 0\\
0 & 0 & 0 & e^{i\theta_{d}}
\end{smallmatrix}\right)
\; \; \; \; \; \; and \; \; \; \; \; \; U=U_2U_1.
\label{eqn:unitarymatrix}
\end{align}

\subsection{Quantum circuit depth }
To simulate a toric code with length $L$, using local unitary gates requires at least linear size $O(L)$ depth circuits \cite{bravyi2006lieb}, and constant depth is achievable if measurement operations are allowed \cite{bravyi2022adaptive}. A recent work provided a systematic method to simulate an unknown toric code in $3L+2$ depth \cite{higgott2021optimal,aguado2008entanglement}. In comparison, on a $L \times L$ square lattice over a torus, we can simulate a known toric code state like $|00\rangle$ in $2L+2$ depth and an unknown toric code $\Phi$ in $\lceil 2L+2+log_{2}(d)+\frac{L}{2d} \rceil$ depth. Here, the quantum circuit is local and the $CNOT$ gate is restricted to be applied on two qubits with a distance $d$.

\begin{table}[ht]
\centering
\begin{tabular}{m{5.5cm} m{5.5cm}} 
\begin{center}
\begin{tikzpicture}[scale=0.6]
\foreach \i in {0, 2, 4, 6}
        \draw[thick, gray] (0,{\i}) -- (6,{\i});
\foreach \i in {0, 2, 4, 6}
        \draw[thick, gray] ({\i},0) -- ({\i},6);
\foreach \i in {1, 3 ,5}
\foreach \j in {0, 2, 4, 6}
        \filldraw[gray!50] ({\i},{\j}) circle (3pt);
\foreach \i in {0, 2, 4, 6}
\foreach \j in {1, 3 ,5}
        \filldraw[gray!50] ({\i},{\j}) circle (3pt);
\node[black] at (3,-0.5) {$|00\rangle$};
\foreach \j in {1, 3, 5}
        \draw[thick, black] (2,{\j}) circle (3pt);
\foreach \j in {1, 3, 5}
        \draw[double, ->, >=stealth, OliveGreen] (2,{\j})--(1.1,{\j+0.9});
\foreach \j in {1, 3, 5}
        \draw[double, ->, >=stealth, blue] (2,{\j})--(1.1,{\j-0.9});
\foreach \j in {1, 3, 5}
        \draw[double, ->, >=stealth, red] (2,{\j})--(0.14,{\j});
\foreach \j in {1, 3, 5}
        \node[black] at (1,{\j+0.2}) {$1$};
\foreach \j in {1, 3, 5}
        \draw[thick, black] (4,{\j}) circle (3pt);
\foreach \j in {1, 3, 5}
        \draw[double, ->, >=stealth, OliveGreen] (4,{\j})--(3.1,{\j+0.9});
\foreach \j in {1, 3, 5}
        \draw[double, ->, >=stealth, blue] (4,{\j})--(3.1,{\j-0.9});
\foreach \j in {1, 3, 5}
        \draw[double, ->, >=stealth, red] (4,{\j})--(2.14,{\j});
\foreach \j in {1, 3, 5}
        \node[black] at (3,{\j+0.2}) {$2$};
\draw[thick, black] (5,4) circle (3pt);
\draw[double, ->, >=stealth, OliveGreen] (5,4)--(5.9,4.9);
\draw[double, ->, >=stealth, red] (5,4)--(4.1,4.9);
\node[black] at (4.55,4.65) {$3$};
\draw[double, ->, >=stealth, red] (5,4)--(5,5.86);
\node[black] at (5.2,5) {$4$};
\draw[thick, black] (5,2) circle (3pt);
\draw[double, ->, >=stealth, OliveGreen] (5,2)--(5.9,2.9);
\draw[double, ->, >=stealth, red] (5,2)--(4.1,2.9);
\node[black] at (4.55,2.65) {$3$};
\draw[double, ->, >=stealth, red] (5,2)--(5,3.86);
\node[black] at (5.2,3) {$5$};
\end{tikzpicture}
\end{center}
&
\begin{center}
\begin{tikzpicture}[scale=0.6]
\foreach \i in {0, 4, 6}
        \draw[thick, gray] (0,{\i}) -- (6,{\i});
\foreach \i in {0, 2, 6}
        \draw[thick, gray] ({\i},0) -- ({\i},6);
\draw[thick, orange] (4,0) -- (4,6);
\draw[thick, orange] (0,2) -- (6,2);
\foreach \i in {1, 3 ,5}
\foreach \j in {0, 4, 6}
        \filldraw[gray!50] ({\i},{\j}) circle (3pt);
\foreach \i in {0, 2, 6}
\foreach \j in {1, 3 ,5}
        \filldraw[gray!50] ({\i},{\j}) circle (3pt);
\foreach \i in {1, 3 ,5}
        \filldraw[orange] ({\i},2) circle (3pt);
\foreach \j in {1, 3 ,5}
        \filldraw[orange] (4,{\j}) circle (3pt);
\node[black] at (3,-0.5) {$\Phi$};
\foreach \j in {1, 3, 5}
        \draw[thick, black] (2,{\j}) circle (3pt);
\foreach \j in {1, 3, 5}
        \draw[double, ->, >=stealth, OliveGreen] (2,{\j})--(2.9,{\j-0.9});
\foreach \j in {1, 3, 5}
        \draw[double, ->, >=stealth, blue] (2,{\j})--(2.9,{\j+0.9});
\foreach \j in {1, 3, 5}
        \draw[double, ->, >=stealth, red] (2,{\j})--(3.86,{\j});
\foreach \j in {1, 3, 5}
        \node[black] at (1,{\j+0.2}) {$2$};
\foreach \j in {1, 3, 5}
        \draw[thick, black] (0,{\j}) circle (3pt);
\foreach \j in {1, 3, 5}
        \draw[double, ->, >=stealth, OliveGreen] (0,{\j})--(0.9,{\j-0.9});
\foreach \j in {1, 3, 5}
        \draw[double, ->, >=stealth, blue] (0,{\j})--(0.9,{\j+0.9});
\foreach \j in {1, 3, 5}
        \draw[double, ->, >=stealth, red] (0,{\j})--(1.86,{\j});
\foreach \j in {1, 3, 5}
        \node[black] at (3,{\j+0.2}) {$1$};
\draw[thick, black] (5,4) circle (3pt);
\draw[double, ->, >=stealth, red] (5,4)--(4.1,3.1);
\node[black] at (4.55,3.35) {$3$};
\draw[double, ->, >=stealth, blue] (5,4)--(5.9,3.1);
\draw[double, ->, >=stealth, red] (5,4)--(5,2.14);
\node[black] at (5.2,3) {$4$};
\draw[thick, black] (5,6) circle (3pt);
\draw[double, ->, >=stealth, red] (5,6)--(4.1,5.1);
\node[black] at (4.55,5.35) {$3$};
\draw[double, ->, >=stealth, blue] (5,6)--(5.9,5.1);
\draw[double, ->, >=stealth, red] (5,6)--(5,4.14);
\node[black] at (5.2,5) {$5$};
\end{tikzpicture}
\end{center}
\end{tabular}

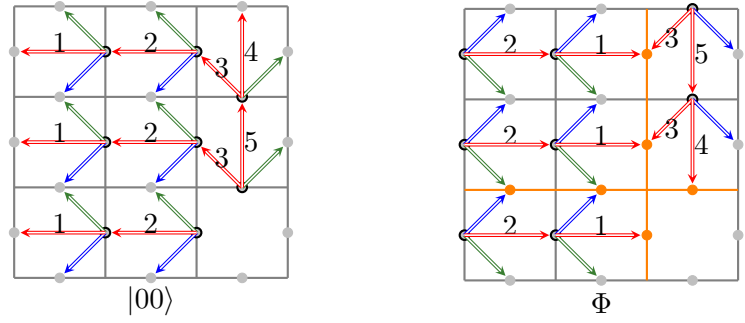
\captionof{figure}[foo]{Opposite boundaries are identified and the case of L=3 is provided as an example. In both figures, the prescribed order for gate application is as follows: 1 Apply Hadamard gates to the qubits encircled by circles; 2 Execute CNOT gates, indicated by green arrows, followed by those with blue arrows; 3 Implement CNOT gates denoted by red arrows, following their numerical order. In the right figure, orange dots signify qubits that hold the information of $\Phi$.} 
\label{fig:depth}
\end{table}

To simulate the state $|00\rangle$, we initiate the process with $\phi_0$ and designate the plaquette at the bottom right corner as redundant. Subsequently, we proceed the quantum gates step by step, following the instructions outlined on the left side of Figure \ref{fig:depth}.  On the other hand, as elaborated upon in Section \ref{sec:arbitrary}, an unknown toric code state $\Phi$ can be attained by substituting $\phi_0$ with $\phi$, which is obtained from two logical qubits through a sequence of CNOT gates. This procedure demands $\lceil log_{2}(d)+\frac{L}{2d} \rceil$ steps \footnote{For detailed discussion on the local CNOT gate, see Section \ref{sec:clocalCNOT}.}, where $d$ represents the maximum distance between the two qubits that CNOT gate could apply without breaking locality. Additionally, a slight variation in the order, as demonstrated on the right side of Figure \ref{fig:depth}, is essential to initiate with $\phi$.

\section{Gluing method for 2D toric code} \label{sec:gluingmethod}
\subsection{Gluing method for two single plaquettes}
The method introduced in the preceding sections is efficient; however, it hinges on the selection of a suitable path. This choice could prove challenging for intricate surfaces. To address this concern, we propose a gluing method designed to overcome this complexity.  To exemplify the essence of the gluing approach, we will commence with a straightforward example.  To simulate the ground state of toric code on the two plaquettes in Figure \ref{fig:glue2plaquettes}, we can employ an ancilla qubit to partition it into two independent plaquettes $p_1$ and $p_2$. The edges within $Bo(p_1)$ are denoted by ${1,2,3,4}$, while those within $Bo(p_2)$ are denoted by ${5,6,7,8}$. We initiate the process with $\phi_0$ and ignore the overall normalization constant to simplify subsequent calculations.

\begin{table}[ht]
\centering
\begin{tabular}{m{4cm} m{7cm}} 
\begin{center}
\begin{tikzpicture}[scale=0.9]
\draw[thick, gray] (-2,-1) rectangle (0,1);
\draw[thick, gray] (0,-1) rectangle (2,1);
\filldraw[gray!50] (0,0) circle (3pt);
\filldraw[gray!50] (2,0) circle (3pt);
\filldraw[gray!50] (-2,0) circle (3pt);
\filldraw[gray!50] (1,1) circle (3pt);
\filldraw[gray!50] (1,-1) circle (3pt);
\filldraw[gray!50] (-1,1) circle (3pt);
\filldraw[gray!50] (-1,-1) circle (3pt);
\node[black] at (-1,0) {$p_1$};
\node[black] at (1,0) {$p_2$};
\end{tikzpicture}
\end{center}
&
\begin{center}
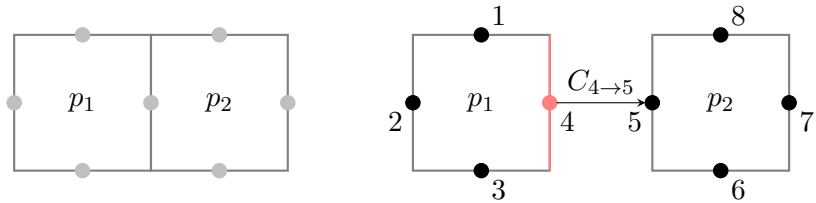

\begin{tikzpicture}[scale=0.9]
\draw[thick, gray] (-2,-1) rectangle (0,1);
\draw[thick, red!50] (0,1) -- (0,-1);
\draw[thick, gray] (1.5,-1) rectangle (3.5,1);
\filldraw[red!50] (0,0) circle (3pt) node[anchor=north west,black] {$4$};
\filldraw[black] (1.5,0) circle (3pt) node[anchor=north east,black] {$5$};
\filldraw[black] (-2,0) circle (3pt) node[anchor=north east,black] {$2$};
\filldraw[black] (-1,1) circle (3pt) node[anchor=south west,black] {$1$};
\filldraw[black] (-1,-1) circle (3pt) node[anchor=north west,black] {$3$};
\filldraw[black] (3.5,0) circle (3pt) node[anchor=north west,black] {$7$};
\filldraw[black] (2.5,1) circle (3pt) node[anchor=south west,black] {$8$};
\filldraw[black] (2.5,-1) circle (3pt) node[anchor=north west,black] {$6$};
\node[black] at (-1,0) {$p_1$};
\node[black] at (2.5,0) {$p_2$};
\draw[->, >=stealth, black] (0.1,0)--(1.4,0);
\node[black] at (0.75,0.3) {$C_{4 \to 5}$};
\end{tikzpicture}
\end{center}
\end{tabular}
\captionof{figure}[foo]{The lattice of two plaquettes is divided into two independent plaquettes by introducing the ancilla qubit in red.} 
\label{fig:glue2plaquettes}
\end{table}

First apply $1 + B_{p_1}$ and $1 + B_{p_2}$ independently to get
\begin{equation}
    (1 + X_1 X_2 X_3 X_4)(1 + X_5 X_6 X_7 X_8) |00...0\rangle.
\end{equation}
Then apply $C_{4 \to 5}$ and notice this operator commutes with $1 + B_{p_2}$: 
\begin{align}
    &(\frac{1-Z_4}{2}X_5 + \frac{1+Z_4}{2})(1 + X_1 X_2 X_3 X_4)(1 + X_5 X_6 X_7 X_8)|00...0\rangle \nonumber \\
    &= (1 + X_1 X_2 X_3 X_4 X_5)(1 + X_5 X_6 X_7 X_8)|00...0\rangle.
\end{align}
Finally, make a measurement over the ancilla qubit with basis $|+\rangle = \frac{|0\rangle + |1\rangle}{\sqrt{2}}$ and $|-\rangle = \frac{|0\rangle - |1\rangle}{\sqrt{2}}$. If we get +1, it is equivalent to applying $\frac{1 + X_4}{2}$ and thus
\begin{align} \label{eqn:+1measurementoutcome}
    &\frac{1 + X_4}{2} (1 + X_1 X_2 X_3 X_4 X_5)(1 + X_5 X_6 X_7 X_8)|00...0\rangle \nonumber \\
    &= (\frac{1 + X_4}{2}) (1 + X_1 X_2 X_3 X_5)(1 + X_5 X_6 X_7 X_8)|00...0\rangle.
\end{align}
The ancilla qubit is disentangled, leaving us with the ground state of the two plaquettes. Observing that, when two boundaries $e_i$ and $e_j$ are glued together, all plaquettes terms commute with each other and $C_{i \to j}$ commutes with all plaquette terms except $1 + B_{p_k}$ where $e_i \in Bo(p_k)$. This observation underscores that the resultant combination remains a ground state even when multiple plaquettes are fused together concurrently.

On the other hand, if we get -1, it is equivalent to applying $\frac{1 - X_4}{2}$ and thus 
\begin{align} \label{eqn:-1measurementoutcome}
    &\frac{1 - X_4}{2} (1 + X_1 X_2 X_3 X_4 X_5)(1 + X_5 X_6 X_7 X_8)|00...0\rangle \nonumber \\
    &= (\frac{1 - X_4}{2}) (1 - X_1 X_2 X_3 X_5)(1 + X_5 X_6 X_7 X_8)|00...0\rangle,
\end{align}
which is not the expected ground state.  It is worth noting that the resulting state is an excited state if a magnetic charge exists at $p_1$.  Fortunately, we can correct it by applying $Z_1$, $Z_2$ or $Z_3$, each of which is a short dual ribbon operator. In the subsequent section, we will establish a proof demonstrating that a correcting operator invariably exists for any planar lattice.

Our method, can be naturally extended to more general scenarios where projectors only involve tensor products of Pauli $X$ (given that tensor products of Pauli $Z$ operators are automatically satisfied by the state $|00...0\rangle$), such as 3D toric model or X cube model to be addressed below.  For instance, let us consider two edges from different lattices, labeled as $m$ and $n$ (note that we abuse the notation referring to both edges and lattices).  The state of these two lattices can be expressed as $\prod \frac{1+H_m}{2} |0\rangle_{m} \otimes |0\rangle_{res_m}$ or $\prod \frac{1+H_n}{2} |0\rangle_{n} \otimes |0\rangle_{res_n}$. Here, $res_{m(n)}$ signifies the remaining system of lattice $m(n)$, and $H_{m(n)}$ denotes the projector onto lattice $m(n)$. Given that $C_{m\rightarrow n}$ only relies on $|m\rangle$, expanding the product of $H_m$ yields
\begin{equation}
    C_{m\rightarrow n} \sum_{i} ( X \otimes 1 \otimes A_{i} + 1 \otimes 1 \otimes B_{i} ) \prod \frac{1+H_n}{2} |0\rangle_m |0\rangle_n |0\rangle_{res_m} |0\rangle_{res_n},
\end{equation}
where $A_i$ and $B_i$ acts only on $res_m$, and we have not expanded $H_n$ since it has a trivial impact on $m$. Upon applying $C_{m\rightarrow n}$, we obtain
\begin{equation}
    \sum_{i} ( X \otimes X \otimes A_{i} + 1 \otimes 1 \otimes B_{i} ) \prod \frac{1+H_n}{2} |0\rangle_m |0\rangle_n |0\rangle_{res_m} |0\rangle_{res_n}.
\end{equation}
Essentially, this signifies that the CNOT gate transfers all actions from $m$ to $n$ after disentangling $m$. Akin to Equation \ref{eqn:+1measurementoutcome} and \ref{eqn:-1measurementoutcome}, we count the excitations and employ correction operators to obtain the ground state. Consequently, we can attain the expected ground state of the glued lattice by gluing the edges correspondingly , as long as the projectors consist of tensor products of Pauli $X$ operators .

\subsection{Gluing method for an arbitrary lattice}

When transitioning from the gluing method's application on two single plaquettes to the broader context of numerous arbitrary plaquettes, our focus should not be on the edges measured +1, but rather on establishing a systematic method to correct address edges measured -1. 

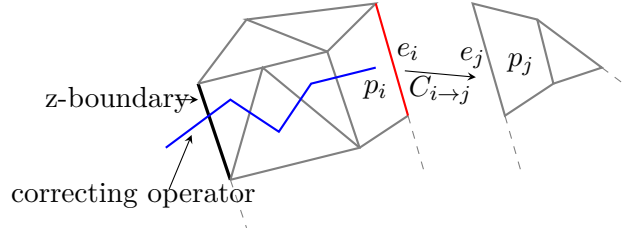
\begin{figure}[ht]
\centering
\begin{tikzpicture}[scale=0.85]
\coordinate (O) at (0,0);
\coordinate (A) at (2,0.5);
\coordinate (B) at (2.5,-1);
\coordinate (C) at (0.5,-1.5);
\coordinate (D) at (1,0.25);
\coordinate (E) at (0.75,1);
\coordinate (F) at (2.75,1.25);
\coordinate (G) at (3.25,-0.5);
\coordinate (H) at (4.25,1.25);
\coordinate (I) at (4.75,-0.5);
\coordinate (J) at (5.5,0);
\coordinate (K) at (5.25,1);
\coordinate (L) at (6.25,0.25);
\coordinate (M) at (0.75,-2.25);
\coordinate (N) at (3.5,-1.375);
\coordinate (P) at (5,-1.375);
\coordinate (Q) at (6.75,-0.125);
\draw[very thick, black] (O) -- (C);
\draw[thick, gray] (O) -- (A) -- (B) -- (C);
\draw[thick, gray] (B) -- (D) -- (C);
\draw[thick, gray] (O) -- (E);
\draw[thick, gray] (E) -- (F) -- (A) -- cycle;
\draw[thick, gray] (B) -- (G);
\draw[thick, red] (G) -- (F);
\draw[thick, gray] (H) -- (I) -- (J) -- (K) -- cycle;
\draw[thick, gray] (J) -- (L) -- (K);
\draw[dashed, gray] (C) -- (M);
\draw[dashed, gray] (G) -- (N);
\draw[dashed, gray] (I) -- (P);
\draw[dashed, gray] (L) -- (Q);
\draw[thick, blue] (-0.5,-1) -- (0.5,-0.25) -- (1.25,-0.75) -- (1.75,0) -- (2.75,0.25);
\node[black] at (-1.25,-0.25) {z-boundary};
\draw[->, >=stealth, black] (-0.35,-0.25)--(0,-0.25);
\node[black] at (-1,-1.75) {correcting operator};
\draw[->, >=stealth, black] (-0.45,-1.65)--(-0.1,-0.8);
\node[black] at (2.75,-0.1) {$p_{i}$};
\node[black] at (5,0.3) {$p_{j}$};
\node[black] at (3.25,0.5) {$e_i$};
\node[black] at (4.25,0.4) {$e_j$};
\draw[->, >=stealth, black] (3.2,0.2)--(4.3,0.1);
\node[black] at (3.75,-0.1) {$C_{i \to j}$};
\end{tikzpicture}
\caption{Correcting procedure after gluing two arbitrary plaquettes.}
\label{fig:Z-boundary}
\end{figure}

In the instance presented in Figure \ref{fig:Z-boundary}, if we apply $C_{i \to j}$ to glue two boundaries and get -1 after measuring qubit $e_i$, the correcting operator must anti-commute with $B_{p_i}$ and exhibit commutativity with everything else \footnote{It is worth noting that this correcting operator effortlessly commutes with all vertex terms.}. One intuitive approach is to apply a dual string operator starting at $p_{i}$ and ending crossing a z-boundary.

Expanding upon this notion, let us consider a situation involving any connected planar lattice $\gamma=\bigcup_{i=1}^{n} p_i$ with z-boundary $e_0$. For a series $T = {p_i}$, $Bo(p_i) \bigcap \bigcup_{j=1}^{i-1} Bo(p_j)\neq \oslash$ for any $i \in [2,n]$, we can insert ancilla qubits to separate $T$ into multiple plaquettes and subsequently glue them back. To illustrate this idea, let us delve into an example consist of four plaquettes, as depicted in Figure \ref{fig:glue4plaquettes1}.

\begin{table}[ht]
\centering
\begin{tabular}{m{5cm} m{5cm}} 
\begin{center}
\begin{tikzpicture}[scale=0.7]
\draw[thick, gray] (0,0) rectangle (2,2);
\draw[thick, gray] (0,0) rectangle (2,-2);
\draw[thick, gray] (0,0) rectangle (-2,2);
\draw[thick, gray] (0,0) rectangle (-2,-2);
\draw[very thick, black] (-2,0) rectangle (-2,2);
\node[black] at (-1.7,1.3) {$e_0$};
\node[black] at (-0.3,1.3) {$e_1$};
\node[black] at (1.3,0.3) {$e_2$};
\node[black] at (0.3,-1.3) {$e_3$};
\node[black] at (-1.3,-0.3) {$e_{4}$};
\node[black] at (-1,1) {$p_1$};
\node[black] at (1,1) {$p_2$};
\node[black] at (1,-1) {$p_3$};
\node[black] at (-1,-1) {$p_4$};
\foreach \i in {-1, 1}
\foreach \j in {-2, 0, 2}
        \filldraw[gray!50] ({\i},{\j}) circle (3pt);
\foreach \i in {-2, 0, 2}
\foreach \j in {-1, 1}
        \filldraw[gray!50] ({\i},{\j}) circle (3pt);
\end{tikzpicture}
\end{center}
&
\begin{center}
\begin{tikzpicture}[scale=0.7]
\draw[thick, gray] (0.5,0.5) rectangle (2.5,2.5);
\draw[thick, gray] (0.5,-0.5) rectangle (2.5,-2.5);
\draw[thick, gray] (-0.5,0.5) rectangle (-2.5,2.5);
\draw[thick, gray] (-0.5,-0.5) rectangle (-2.5,-2.5);
\draw[very thick, black] (-2.5,0.5) rectangle (-2.5,2.5);
\node[black] at (-1.5,1.5) {$p_1$};
\node[black] at (1.5,1.5) {$p_2$};
\node[black] at (1.5,-1.5) {$p_3$};
\node[black] at (-1.5,-1.5) {$p_4$};
\foreach \i in {-1.5, 1.5}
\foreach \j in {-2.5, -0.5, 0.5, 2.5}
        \filldraw[gray!50] ({\i},{\j}) circle (3pt);
\foreach \i in {-2.5, -0.5, 0.5, 2.5}
\foreach \j in {-1.5, 1.5}
        \filldraw[gray!50] ({\i},{\j}) circle (3pt);
\draw[thick, red!50] (-0.5,0.5) -- (-0.5,2.5);
\filldraw[red!50] (-0.5,1.5) circle (3pt);
\draw[thick, red!50] (0.5,0.5) -- (2.5,0.5);
\filldraw[red!50] (1.5,0.5) circle (3pt);
\draw[thick, red!50] (0.5,-0.5) -- (0.5,-2.5);
\filldraw[red!50] (0.5,-1.5) circle (3pt);
\draw[thick, red!50] (-0.5,0.5) -- (-2.5,0.5);
\filldraw[red!50] (-1.5,0.5) circle (3pt);
\node[black] at (-2.2,1.8) {$e_0$};
\node[black] at (0.8,1.8) {$e_{1}$};
\node[black] at (-0.8,1.8) {$e'_{1}$};
\node[black] at (0.8,-1.8) {$e'_{3}$};
\node[black] at (-0.8,-1.8) {$e_{3}$};
\node[black] at (1.8,-0.8) {$e_{2}$};
\node[black] at (1.8,0.8) {$e'_{2}$};
\node[black] at (-1.8,-0.8) {$e_{4}$};
\node[black] at (-1.8,0.8) {$e'_{4}$};
\node[black] at (0,1) {$C_{e'_{1} \to e_{1}}$};
\draw[->, >=stealth, black] (-0.4,1.5)--(0.4,1.5);
\end{tikzpicture}
\end{center}\\

\begin{center}
\begin{tikzpicture}[scale=0.7]
\draw[thick, gray] (-0.5,0.5) rectangle (1.5,2.5);
\draw[thick, gray] (0.5,-0.5) rectangle (2.5,-2.5);
\draw[thick, gray] (-0.5,0.5) rectangle (-2.5,2.5);
\draw[thick, gray] (-0.5,-0.5) rectangle (-2.5,-2.5);
\draw[very thick, black] (-2.5,0.5) rectangle (-2.5,2.5);
\node[black] at (-1.5,1.5) {$p_1$};
\node[black] at (0.5,1.5) {$p_2$};
\node[black] at (1.5,-1.5) {$p_3$};
\node[black] at (-1.5,-1.5) {$p_4$};
\foreach \j in {-2.5, -0.5, 0.5, 2.5}
        \filldraw[gray!50] (-1.5,{\j}) circle (3pt);
\foreach \i in {-2.5, -0.5}
\foreach \j in {-1.5, 1.5}
        \filldraw[gray!50] ({\i},{\j}) circle (3pt);
\filldraw[gray!50] (0.5,2.5) circle (3pt);
\filldraw[gray!50] (1.5,1.5) circle (3pt);
\filldraw[gray!50] (2.5,-1.5) circle (3pt);
\filldraw[gray!50] (1.5,-0.5) circle (3pt);
\filldraw[gray!50] (1.5,-2.5) circle (3pt);
\draw[thick, red!50] (0.5,-0.5) -- (0.5,-2.5);
\filldraw[red!50] (0.5,-1.5) circle (3pt);
\draw[thick, red!50] (-0.5,0.5) -- (1.5,0.5);
\filldraw[red!50] (0.5,0.5) circle (3pt);
\draw[thick, red!50] (-0.5,0.5) -- (-2.5,0.5);
\filldraw[red!50] (-1.5,0.5) circle (3pt);
\node[black] at (-2.2,1.8) {$e_0$};
\node[black] at (1.8,-0.8) {$e_{2}$};
\node[black] at (0.8,0.8) {$e'_{2}$};
\node[black] at (0.8,-1.8) {$e'_{3}$};
\node[black] at (-0.8,-1.8) {$e_{3}$};
\node[black] at (-1.8,-0.8) {$e_{4}$};
\node[black] at (-1.8,0.8) {$e'_{4}$};
\node[black] at (1.7,0.1) {$C_{e'_{2} \to e_{2}}$};
\draw[->, >=stealth, black] (0.6,0.4)--(1.4,-0.4);
\end{tikzpicture}
\end{center}
&
\begin{center}
\begin{tikzpicture}[scale=0.7]
\draw[thick, gray] (-0.5,0.5) rectangle (1.5,2.5);
\draw[thick, gray] (-0.5,0.5) rectangle (1.5,-1.5);
\draw[thick, gray] (-0.5,0.5) rectangle (-2.5,2.5);
\draw[thick, gray] (-1.5,-0.5) rectangle (-3.5,-2.5);
\draw[very thick, black] (-2.5,0.5) rectangle (-2.5,2.5);
\node[black] at (-1.5,1.5) {$p_1$};
\node[black] at (0.5,1.5) {$p_2$};
\node[black] at (0.5,-0.5) {$p_3$};
\node[black] at (-2.5,-1.5) {$p_4$};
\foreach \j in {-1.5, 0.5, 2.5}
        \filldraw[gray!50] (0.5,{\j}) circle (3pt);
\foreach \i in {-0.5, 1.5}
\foreach \j in {-0.5, 1.5}
        \filldraw[gray!50] ({\i},{\j}) circle (3pt);
\filldraw[gray!50] (-2.5,-0.5) circle (3pt);
\filldraw[gray!50] (-2.5,1.5) circle (3pt);
\filldraw[gray!50] (-1.5,2.5) circle (3pt);
\filldraw[gray!50] (-1.5,-1.5) circle (3pt);
\filldraw[gray!50] (-2.5,-2.5) circle (3pt);
\filldraw[gray!50] (-3.5,-1.5) circle (3pt);
\draw[thick, red!50] (-0.5,0.5) -- (-0.5,-1.5);
\filldraw[red!50] (-0.5,-0.5) circle (3pt);
\draw[thick, red!50] (-2.5,0.5) -- (-0.5,0.5);
\filldraw[red!50] (-1.5,0.5) circle (3pt);
\node[black] at (-2.2,1.8) {$e_0$};
\node[black] at (-0.2,-0.8) {$e'_{3}$};
\node[black] at (-1.8,-1.8) {$e_{3}$};
\node[black] at (-2.8,-0.8) {$e_{4}$};
\node[black] at (-1.8,0.8) {$e'_{4}$};
\node[black] at (-0.5,-1.7) {$C_{e'_{3} \to e_{3}}$};
\draw[->, >=stealth, black] (-0.6,-0.6)--(-1.4,-1.4);
\node[black] at (-2.8,0.1) {$C_{e'_{4} \to e_{4}}$};
\draw[->, >=stealth, black] (-1.6,0.4)--(-2.4,-0.4);
\end{tikzpicture}
\end{center}
\end{tabular}

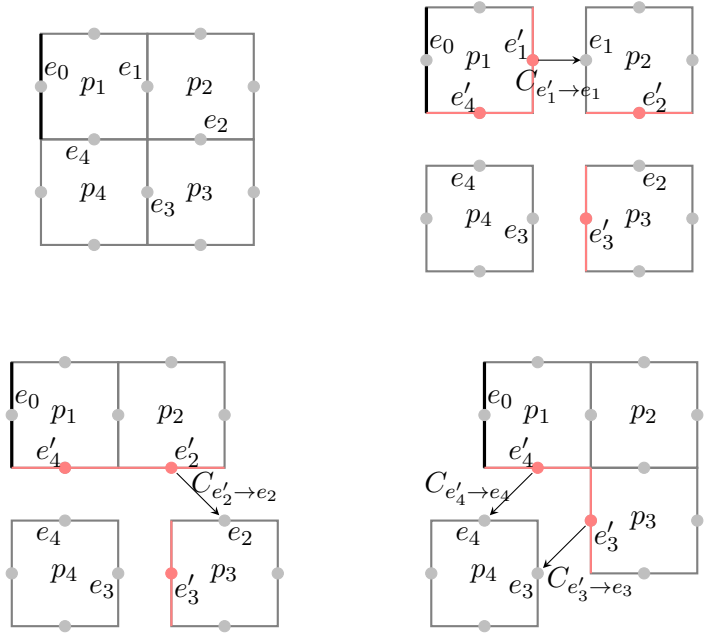
\captionof{figure}[foo]{$e_0$ is a z-boundary and $e'$ in red represents an ancilla qubit.} 
\label{fig:glue4plaquettes1}
\end{table}

First, we initiate by utilizing ancilla qubits to fragment the lattice into single plaquettes. For $\tau(e_k)$ containing $p_i$ and $p_j$, where $1 \le i < j \le n$ according to the series $T$, insert an ancilla edge $e'_k$ into $p_i$ while retaining $e_k$ in $p_j$. Then we apply $1 + B_{p}$ to every single plaquette $p \in P$. Subsequently, glue them together piece by piece. For $p_i, 1 < i \le n$, it becomes necessary to apply $C_{e' \to e}$ to all pairs of $e' \in \bigcup_{j=1}^{i-1} Bo(p_j)$ and $e \in Bo(p_i)$. Finally, we measure and disentangle $e'$. If we get -1, apply a dual string operator connecting $p_i$ and the z-boundary of $p_1$ to correct it. It is noteworthy that all plaquettes can be glued simultaneously, allowing a dual string operator to annihilate two magnetic charges by connecting them. In this example, if we get -1 for $e'_1$ and $e'_4$ concurrently, the correcting operator will effectively nullify their impact.

In the case of a lattice without boundary, we can choose a specific plaquette $p$ to be redundant, effectively transforming $Bo(p)$ into z-boundaries. Subsequently,this situation mirrors the scenario depicted in the lattice with boundaries, and further details are left for readers to explore. If the lattice solely contains x-boundaries, a viable solution is to consider the dual lattice of it. The process remains unchanged, except for the inversion of plaquette and vertex operators. Thus we can confidently assert that our method is universally capable of simulating the ground state of a toric code on any planar lattice configuration.

In the case of the 2D toric code, the gluing method might initially resemble a simple measurement process, especially when we break down the lattice into pieces, attach ancilla qubits, and then fuse them to obtain the ground state for the entire lattice.  However, its capabilities extend significantly when we consider scenarios like 3D models, as discussed thoroughly in Section \ref{sec:gluing3Dmodel}, or when we have two lattices in their ground states to be joined. In such cases, a stabilizer measurement can not glue two lattice and get the ground state of the glued lattice.

\FloatBarrier
\section{Simulate ground state for 3D models}
\subsection{3D toric model} \label{sec:3d toric}
The 3D toric model bears strong resemblance to the 2D toric code and is established on  an arbitrary 3D lattice. To enhance clarity, a cubic lattice is adopted, as depicted in Figure \ref{fig:3dtoricdefinition}. Within this lattice, $V$ represents the set of all vertices, while $P$ corresponds to the set of all plaquettes; each edge accommodates a single qubit. Moreover,  for the sake of convenience, we have affixed labels to each edge, denoting them as $x$, $y$, or $z$ based on their alignment with the respective axis (i.e., parallel to the $x$, $y$, or $z$ axis). Notice this labeling maintains consistency even when applied to a 3-dimensional torus. The Hamiltonian is defined as
\begin{equation}
    H=-\sum_{v \in V} A_v-\sum_{p \in P} B_p,
\end{equation}
where $A_v$ pertains to the application of the Pauli operator $X$ over six edges connected to the vertex $v$, and $B_p$ pertains to the application of the Pauli operator $Z$ over the four edges encompassing the plaquette $p$. It is straightforward to see these new-defined $A_v$ and $B_p$ operators also satisfy $A_v^2 = B_p^2 = 1$ and $[A_v,B_p] = 0$. So this 3D toric Hamiltonian is equivalent to the equation expressed as a summation of local projectors. We get the ground state
\begin{equation}
    |GS\rangle=\prod_v \frac{1+A_v}{2} |\phi_0\rangle,
\end{equation}
where $|\phi_0\rangle=|00...0\rangle$, and we drop $\frac{1+B_p}{2}$s since its action on $|\phi_0\rangle$ is $+1$. It is important to highlight that the constancy of ground state degeneracy endures with fluctuations in system size, a pivotal characteristic of topological phases of matter. Additionally, Figure \ref{fig:3dtoricdefinition} presents a comprehensive depiction of a pair of conjugate logical operators. Notice that, the definition of logical operators only depends on the nontrivial loop or non-cotractable planes. Consequently, we have three pairs of conjugate logical operators, each acting on edges labeled by $x$, $y$, or $z$ respectively.

\begin{table}[ht]
\centering
\begin{tabular}{m{5cm} m{5cm}} 
\begin{center}
\begin{tikzpicture}[scale=0.6]
\foreach \i in {0, 2}
\foreach \j in {0, 2}
        \draw[gray] ({\i},{\j},-2) -- ({\i},{\j},0);
\foreach \i in {0, 2}
\foreach \k in {-2, 0}
        \draw[gray] ({\i},0,{\k}) -- ({\i},2,{\k});
\foreach \j in {0, 2}
\foreach \k in {-2, 0}
        \draw[gray] (0,{\j},{\k}) -- (2,{\j},{\k});
\draw[gray] (0,0,0) -- (-2,0,0);
\draw[gray] (0,0,0) -- (0,-2,0);
\draw[gray] (0,0,0) -- (0,0,2);
\foreach \j in {0, 2}
\foreach \k in {-2, 0}
        \filldraw[gray!50] (1,{\j},{\k}) circle (3pt);
\foreach \i in {0, 2}
\foreach \k in {-2, 0}
        \filldraw[gray!50] ({\i},1,{\k}) circle (3pt);
\foreach \i in {0, 2}
\foreach \j in {0, 2}
        \filldraw[gray!50] ({\i},{\j},-1) circle (3pt);
\filldraw[gray!50] (-1,0,0) circle (3pt);
\filldraw[gray!50] (0,-1,0) circle (3pt);
\filldraw[gray!50] (0,0,1) circle (3pt);
\draw[thick, blue] (0,1,0) -- (1,0,0) -- (2,1,0) -- (1,2,0) -- cycle;
\node[black] at (1,1,0) {$B_p$};
\draw[gray,->] (-2,-2,0) -- (-1,-2,0) node[anchor=south]{$x$};
\draw[gray,->] (-2,-2,0) -- (-2,-1,0) node[anchor=west]{$y$};
\draw[gray,->] (-2,-2,0) -- (-2,-2,1) node[anchor=west]{$z$};
\draw[thick, red] (-1,0,0) -- (1,0,0);
\draw[thick, red] (0,-1,0) -- (0,1,0);
\draw[thick, red] (0,0,-1) -- (0,0,1);
\node[black] at (0.5,-0.4,0) {$A_v$};
\end{tikzpicture}
\end{center}
&
\begin{center}
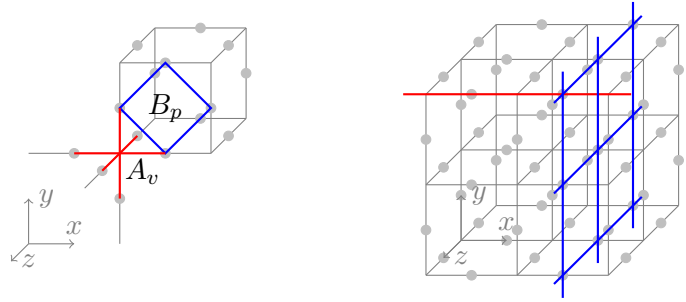

\begin{tikzpicture}[scale=0.6]
\foreach \i in {-2, 0, 2}
\foreach \j in {-2, 0, 2}
        \draw[gray] ({\i},{\j},-2) -- ({\i},{\j},2);
\foreach \i in {-2, 0, 2}
\foreach \k in {-2, 0, 2}
        \draw[gray] ({\i},-2,{\k}) -- ({\i},2,{\k});
\foreach \j in {-2, 0, 2}
\foreach \k in {-2, 0, 2}
        \draw[gray] (-2,{\j},{\k}) -- (2,{\j},{\k});
\foreach \i in {-1, 1}
\foreach \j in {-2, 0, 2}
\foreach \k in {-2, 0, 2}
        \filldraw[gray!50] ({\i},{\j},{\k}) circle (3pt);
\foreach \i in {-2, 0, 2}
\foreach \j in {-1, 1}
\foreach \k in {-2, 0, 2}
        \filldraw[gray!50] ({\i},{\j},{\k}) circle (3pt);
\foreach \i in {-2, 0, 2}
\foreach \j in {-2, 0, 2}
\foreach \k in {-1, 1}
        \filldraw[gray!50] ({\i},{\j},{\k}) circle (3pt);
\draw[gray,->] (-2,-2,0) -- (-1,-2,0) node[anchor=south]{$x$};
\draw[gray,->] (-2,-2,0) -- (-2,-1,0) node[anchor=west]{$y$};
\draw[gray,->] (-2,-2,0) -- (-2,-2,1) node[anchor=west]{$z$};
\draw[thick, red] (-2.5,2,2) -- (2.5,2,2);
\foreach \j in {-2, 0, 2}
        \draw[thick, blue] (1,{\j},-2.5) -- (1,{\j},2.5);
\foreach \k in {-2, 0, 2}
        \draw[thick, blue] (1,-2.5,{\k}) -- (1,2.5,{\k});
\end{tikzpicture}
\end{center}
\end{tabular}
\captionof{figure}[foo]{The left sub-figure illustrates the definitions of $A_v$ and $B_p$ operators. Meanwhile, the right sub-figure displays a pair of conjugate logical operators composed of edges labeled by $x$. The red string is a nontrivial circle parallel to $x$ axis, and a logical $Z$ operator is to apply Pauli $Z$ over edges along the string. Conversely, the blue plane is a non-contractable plane perpendicular to $x$ axis, and a logical $X$ operator is to apply Pauli $X$ over edges within the plane.} 
\label{fig:3dtoricdefinition}
\end{table}

We can extend the method of 2D toric code to 3D toric model with boundary directly utilizing a plaquette as the basic structure. It is complicated yet straightforward, so its details are outlined in Section \ref{sec:3dtoricwith}. However, applying this approach to the 3D toric model without boundaries presents challenges, as the absence of free edges in the final step poses an issue. To circumvent this challenge, we must adopt a basic structure, illustrated in Figure \ref{fig:basicstructures}. We still initiate the process with $|\phi_0\rangle$. Then we execute the quantum circuits as illustrated in the figure to achieve the action of $\frac{1+A_{v}}{2}$.

\begin{table}[ht]
\centering
\begin{tabular}{m{5cm} m{5cm}} 
\begin{center}
\begin{tikzpicture}[scale=0.8]
\coordinate (OO) at (0,0,0);
\coordinate (A) at (0,2,0);
\coordinate (B) at (0,2,2);
\coordinate (C) at (0,0,2);
\coordinate (D) at (2,0,0);
\coordinate (E) at (2,2,0);
\coordinate (F) at (2,2,2);
\coordinate (G) at (2,0,2);
\draw[thick, gray] (OO) -- (C) -- (G) -- (D) -- cycle;
\draw[thick, gray] (OO) -- (A) -- (E) -- (D) -- cycle;
\draw[thick, gray] (OO) -- (A) -- (B) -- (C) -- cycle;
\draw[thick, gray] (D) -- (E) -- (F) -- (G) -- cycle;
\draw[thick, gray] (C) -- (B) -- (F) -- (G) -- cycle;
\draw[thick, gray] (A) -- (B) -- (F) -- (E) -- cycle;
\filldraw[gray!50] (0,1,0) circle (3pt); 
\filldraw[gray!50] (1,0,0) circle (3pt);
\filldraw[gray!50] (2,1,0) circle (3pt);
\filldraw[gray!50] (1,2,0) circle (3pt);
\filldraw[gray!50] (0,1,2) circle (3pt);
\filldraw[gray!50] (1,0,2) circle (3pt);
\filldraw[gray!50] (2,1,2) circle (3pt);
\filldraw[gray!50] (1,2,2) circle (3pt);
\filldraw[gray!50] (0,0,1) circle (3pt);
\filldraw[gray!50] (2,0,1) circle (3pt);
\filldraw[gray!50] (0,2,1) circle (3pt);
\filldraw[gray!50] (2,2,1) circle (3pt);
\draw[thick, black] (2,2,1) circle (3pt); 
\draw[double, ->, >=stealth, black] (2,2,1)--(2,0.14,1); 
\draw[double, ->, >=stealth, black] (2,2,1)--(2,1.1,0.1);
\draw[double, ->, >=stealth, black] (2,2,1)--(2,1.1,1.9);
\node[black] at (2.5,0.6,0) {$\frac{1+B_{p}}{2}$};
\end{tikzpicture}
\end{center}
&
\begin{center}
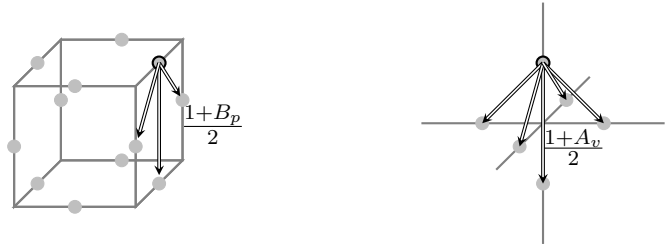

\begin{tikzpicture}[scale=0.8]
\coordinate (O) at (0,0,0);
\coordinate (A) at (0,2,0);
\coordinate (B) at (0,-2,0);
\coordinate (C) at (2,0,0);
\coordinate (D) at (-2,0,0);
\coordinate (E) at (0,0,2);
\coordinate (F) at (0,0,-2);
\coordinate (G) at (0,1,0);
\coordinate (H) at (0,-1,0);
\coordinate (I) at (1,0,0);
\coordinate (J) at (-1,0,0);
\coordinate (K) at (0,0,1);
\coordinate (L) at (0,0,-1);
\draw[thick, gray] (O) -- (A);
\draw[thick, gray] (O) -- (B);
\draw[thick, gray] (O) -- (C);
\draw[thick, gray] (O) -- (D);
\draw[thick, gray] (O) -- (E);
\draw[thick, gray] (O) -- (F);
\filldraw[gray!50] (G) circle (3pt);
\filldraw[gray!50] (H) circle (3pt);
\filldraw[gray!50] (I) circle (3pt);
\filldraw[gray!50] (J) circle (3pt);
\filldraw[gray!50] (K) circle (3pt);
\filldraw[gray!50] (L) circle (3pt);
\draw[thick, black] (G) circle (3pt); 
\draw[double, ->, >=stealth, black] (G)--(H); 
\draw[double, ->, >=stealth, black] (G)--(I);
\draw[double, ->, >=stealth, black] (G)--(J);
\draw[double, ->, >=stealth, black] (G)--(K);
\draw[double, ->, >=stealth, black] (G)--(L);
\node[black] at (0.5,-0.4,0) {$\frac{1+A_{v}}{2}$};
\end{tikzpicture}
\end{center}
\end{tabular}
\captionof{figure}[foo]{Comparison of two different basic structures: An example consisting of eight cubes with boundary is shown in Section \ref{sec:3dtoricwith} and a similar example without boundary is shown in Section \ref{sec:3dtoricwithout}.} 
\label{fig:basicstructures}
\end{table}

Using this basic structure to develop the lattice vertex by vertex, we will end with a redundant vertex as
\begin{equation}
    \prod_{v \in V} A_v = 1.
\end{equation}
In Section \ref{sec:3dtoricwithout}, we present a straightforward example comprising eight cubes to illustrate the method. To address the general case, we delineate the procedures required for constructing a quantum circuit for the 3D toric model on an $L \times L \times L$ lattice over a 3-dimensional torus in Figure \ref{fig:3dtoricsteps}. The process consists of several carefully orchestrated steps to efficiently realize the circuit, amounting to a total of $3L+8$ steps. The quantum circuit is purely local since all applied quantum gates  are either acting on a single qubit or on nearest two qubits. Certain non-interacting gates offer the potential for further parallelization, but this would only result in a constant difference in circuit depth. 

In a manner akin to the procedure detailed in Section \ref{sec:arbitrary}, we employ certain qubits to generate a particular initial state that encodes information about the logical qubits, as depicted in Figure \ref{fig:3dtoricdegeneracy}. However, it is important to note that these selected qubits are unnecessary when we opt for free qubits during the ground state preparation. In conclusion, our method can prepare an arbitrary ground state of the 3D toric model with linear depth.

\begin{table}[ht]
\centering
\begin{tabular}{m{5.6cm} m{5.4cm}}
\begin{center}
\begin{tikzpicture}[scale=0.72]

\draw[gray,->] (0,0,0) -- (5.3,0,0) node[anchor=north east]{$x$};
\draw[gray,->] (0,0,0) -- (0,5.3,0) node[anchor=north west]{$y$};
\draw[gray,->] (0,0,0) -- (0,0,5.3) node[anchor=south]{$z$};

\draw[gray, fill=gray!20, opacity=0.3] (0,0,0) -- (0,5,0) -- (4,5,0) -- (4,0,0) -- cycle;
\draw[gray, fill=gray!20, opacity=0.3] (0,0,5) -- (0,0,0) -- (4,0,0) -- (4,0,5) -- cycle;
\draw[gray, fill=gray!20, opacity=0.3] (0,0,5) -- (0,5,5) -- (0,5,0) -- (0,0,0) -- cycle;
\draw[black, fill=blue!10, opacity=0.5] (4,0,5) -- (4,5,5) -- (4,5,0) -- (4,0,0) -- cycle;
\draw[black, fill=blue!10, opacity=0.5] (4,5,0) -- (5,5,0) -- (5,0,0) -- (4,0,0) -- cycle;
\draw[black, fill=blue!10, opacity=0.5] (4,0,5) -- (5,0,5) -- (5,0,0) -- (4,0,0) -- cycle;

\draw[gray] (0,0,3) -- (0,2,3) -- (2,2,3) -- (2,0,3) -- cycle;
\draw[gray] (1,0,3) -- (1,2,3);
\draw[gray] (2,1,3) -- (0,1,3);
\foreach \i in {0, 1, 2}
\foreach \j in {0, 1, 2}
        \draw[gray] ({\i},{\j},3) -- ({\i},{\j},4);

\filldraw[red!50] (0.5,1,3) circle (3pt);
\filldraw[red!50] (1.5,1,3) circle (3pt);
\filldraw[red!50] (0.5,2,3) circle (3pt);
\filldraw[red!50] (1.5,2,3) circle (3pt);

\draw[black] (0,0,4) -- (0,2,4) -- (2,2,4) -- (2,0,4) -- cycle;
\draw[black] (1,0,4) -- (1,2,4);
\draw[black] (2,1,4) -- (0,1,4);
\foreach \i in {0, 1, 2}
\foreach \j in {0, 1, 2}
        \draw[black] ({\i},{\j},4) -- ({\i},{\j},5);
\filldraw[red!50] (0.5,1,3) circle (3pt);
\filldraw[red!50] (1.5,1,3) circle (3pt);
\filldraw[red!50] (0.5,2,3) circle (3pt);
\filldraw[red!50] (1.5,2,3) circle (3pt);

\filldraw[red] (0.5,1,4) circle (3pt);
\filldraw[red] (1.5,1,4) circle (3pt);
\filldraw[red] (0.5,2,4) circle (3pt);
\filldraw[red] (1.5,2,4) circle (3pt);

\draw[gray] (4,4,3) -- (4,4,4);
\draw[gray] (4,4,3) -- (4,5,3);
\draw[gray] (4,4,3) -- (5,4,3);

\foreach \j in {0, 1, 2}
        \draw[gray] (4,{\j},3) -- (5,{\j},3);
\draw[gray] (4,0,3) -- (4,2,3);
\foreach \j in {0, 1, 2}
        \draw[gray] (4,{\j},3) -- (4,{\j},4);

\filldraw[OliveGreen!50] (4,0.5,3) circle (3pt);
\filldraw[OliveGreen!50] (4,1.5,3) circle (3pt);
\filldraw[blue!50] (4,4,3.5) circle (3pt);

\foreach \j in {0, 1, 2}
        \draw[black] (4,{\j},4) -- (5,{\j},4);
\draw[black] (4,0,4) -- (4,2,4);
\foreach \j in {0, 1, 2}
        \draw[black] (4,{\j},4) -- (4,{\j},5);

\draw[black] (4,4,4) -- (4,4,5);
\draw[black] (4,4,4) -- (4,5,4);
\draw[black] (4,4,4) -- (5,4,4);

\filldraw[OliveGreen] (4,0.5,4) circle (3pt);
\filldraw[OliveGreen] (4,1.5,4) circle (3pt);
\draw[double, ->, >=stealth, blue] (4,4,4.5)--(4.5,4,5);
\draw[double, ->, >=stealth, blue] (4,4,4.5)--(3.5,4,5);
\filldraw[blue] (4,4,4.5) circle (3pt);

\draw[black, fill=gray!20, opacity=0.3] (0,0,5) -- (0,5,5) -- (4,5,5) -- (4,0,5) -- cycle;
\draw[black, fill=gray!20, opacity=0.3] (0,5,5) -- (0,5,0) -- (4,5,0) -- (4,5,5) -- cycle;
\draw[black, fill=blue!10, opacity=0.5] (5,0,5) -- (5,5,5) -- (5,5,0) -- (5,0,0) -- cycle;
\draw[black, fill=blue!10, opacity=0.5] (4,5,5) -- (5,5,5) -- (5,0,5) -- (4,0,5) -- cycle;
\draw[black, fill=blue!10, opacity=0.5] (4,5,5) -- (5,5,5) -- (5,5,0) -- (4,5,0) -- cycle;

\foreach \i in {0, 1, 2}
        \draw[black] ({\i},0,5) -- ({\i},2,5);
\foreach \j in {0, 1, 2}
        \draw[black] (0,{\j},5) -- (2,{\j},5);

\foreach \j in {0, 1, 2, 4, 5}
        \draw[black] (5,{\j},3) -- (5,{\j},5);
\foreach \k in {3, 4, 5}
        \draw[black] (5,0,{\k}) -- (5,2,{\k});
\foreach \k in {3, 4, 5}
        \draw[black] (4,5,{\k}) -- (5,5,{\k});
\foreach \k in {3, 4, 5}
        \draw[black] (5,4,{\k}) -- (5,5,{\k});
\foreach \j in {0, 1, 2, 4, 5}
        \draw[black] (4,{\j},5) -- (5,{\j},5);
\draw[black] (4,0,5) -- (4,2,5);
\draw[black] (4,4,5) -- (4,5,5);
\draw[black] (4,5,5) -- (4,5,3);

\draw[double, ->, >=stealth, red] (0.5,1,5)--(0,1.5,5);
\draw[double, ->, >=stealth, red] (0.5,1,5)--(0,0.5,5);
\draw[double, ->, >=stealth, red] (0.5,1,5)--(0,1,4.5);
\draw[double, ->, >=stealth, red] (0.5,1,5)--(0,1,6);
\draw[double, ->, >=stealth, red] (0.5,1,5)--(-0.5,1,5);

\draw[double, ->, >=stealth, OliveGreen] (4,0.5,5)--(3.5,0,5);
\draw[double, ->, >=stealth, OliveGreen] (4,0.5,5)--(4.5,0,5);

\draw[black, fill=yellow!50, opacity=0.4] (5,4,1)--(5,4,0)--(4,4,0)--(4,4,1)--cycle;
\draw[black, fill=yellow!50, opacity=0.4] (5,5,1)--(4,5,1)--(4,4,1)--(5,4,1)--cycle;
\draw[black, fill=yellow!50, opacity=0.4] (5,5,1)--(4,5,1)--(4,5,0)--(5,5,0)--cycle;
\draw[black, fill=yellow!50, opacity=0.4] (5,5,1)--(5,5,0)--(5,4,0)--(5,4,1)--cycle;

\filldraw[black] (2.8,1,4) circle (1pt);
\filldraw[black] (3.0,1,4) circle (1pt);
\filldraw[black] (3.2,1,4) circle (1pt);

\filldraw[black] (1,2.8,4) circle (1pt);
\filldraw[black] (1,3.0,4) circle (1pt);
\filldraw[black] (1,3.2,4) circle (1pt);

\filldraw[black] (1,1,1.8) circle (1pt);
\filldraw[black] (1,1,2.0) circle (1pt);
\filldraw[black] (1,1,2.2) circle (1pt);

\filldraw[black] (4.5,2.8,4) circle (1pt);
\filldraw[black] (4.5,3.0,4) circle (1pt);
\filldraw[black] (4.5,3.2,4) circle (1pt);

\filldraw[black] (4.5,1,1.8) circle (1pt);
\filldraw[black] (4.5,1,2.0) circle (1pt);
\filldraw[black] (4.5,1,2.2) circle (1pt);

\filldraw[black] (4.5,4.5,1.8) circle (1pt);
\filldraw[black] (4.5,4.5,2.0) circle (1pt);
\filldraw[black] (4.5,4.5,2.2) circle (1pt);

\filldraw[red] (0.5,1,5) circle (3pt);
\filldraw[red] (1.5,1,5) circle (3pt);
\filldraw[OliveGreen] (4,0.5,5) circle (3pt);
\filldraw[red] (0.5,2,5) circle (3pt);
\filldraw[red] (1.5,2,5) circle (3pt);
\filldraw[OliveGreen] (4,1.5,5) circle (3pt);

\draw[double, ->, >=stealth, red] (0.5,0,6)--(2.5,0,6);
\end{tikzpicture}
\end{center}
&
\begin{center}
\begin{tikzpicture}[scale=0.67]

\foreach \i in {3, 4}
\foreach \j in {0, 1, 2, 4}
        \draw[black] ({\i},{\j},0) -- ({\i},{\j},1);
\foreach \i in {3, 4}
        \draw[black] ({\i},0,0) -- ({\i},2,0);
\foreach \i in {3, 4}
        \draw[black] ({\i},4,0) -- ({\i},5,0);
\foreach \j in {1, 2, 4}
        \draw[black] (3,{\j},0) -- (5,{\j},0);
\draw[black, fill=blue!10, opacity=0.5] (0,0,0) -- (0,0,1) -- (0,5,1) -- (0,5,0) -- cycle;
\draw[black, fill=blue!10, opacity=0.5] (0,0,0) -- (0,0,1) -- (5,0,1) -- (5,0,0) -- cycle;
\draw[black, fill=blue!10, opacity=0.5] (0,0,0) -- (5,0,0) -- (5,5,0) -- (0,5,0) -- cycle;
\draw[black, fill=blue!10, opacity=0.5] (5,5,0) -- (5,5,1) -- (0,5,1) -- (0,5,0) -- cycle;
\draw[black, fill=blue!10, opacity=0.5] (5,5,0) -- (5,5,1) -- (5,0,1) -- (5,0,0) -- cycle;
\draw[black, fill=blue!10, opacity=0.5] (0,0,1) -- (5,0,1) -- (5,5,1) -- (0,5,1) -- cycle;
\foreach \i in {3, 4, 5}
        \draw[black] ({\i},5,0) -- ({\i},5,1);
\foreach \j in {0, 1, 2, 4, 5}
        \draw[black] (5,{\j},0) -- (5,{\j},1);
\foreach \i in {3, 4}
        \draw[black] ({\i},0,1) -- ({\i},2,1);
\foreach \i in {3, 4}
        \draw[black] ({\i},4,1) -- ({\i},5,1);
\foreach \j in {1, 2, 4}
        \draw[black] (3,{\j},1) -- (5,{\j},1);

\draw[double, ->, >=stealth, OliveGreen] (5,0.5,1)--(4.5,0,1);
\draw[double, ->, >=stealth, OliveGreen] (5,0.5,1)--(5.5,0,1);
\draw[double, ->, >=stealth, OliveGreen] (5,0.5,1)--(5,-0.5,1);

\draw[double, ->, >=stealth, blue] (4.5,4,1)--(5.5,4,1);
\draw[double, ->, >=stealth, blue] (4.5,4,1)--(5,4.5,1);
\draw[double, ->, >=stealth, blue] (4.5,4,1)--(5,3.5,1);

\foreach \i in {3, 4, 5}
\foreach \j in {0.5, 1.5}
        \filldraw[OliveGreen] ({\i},{\j},1) circle (3pt);
\foreach \i in {3.5, 4.5}
        \filldraw[blue] ({\i},4,1) circle (3pt);

\draw[black, fill=yellow!50, opacity=0.4] (1,4,1)--(1,4,0)--(0,4,0)--(0,4,1)--cycle;
\draw[black, fill=yellow!50, opacity=0.4] (1,5,1)--(0,5,1)--(0,4,1)--(1,4,1)--cycle;
\draw[black, fill=yellow!50, opacity=0.4] (1,5,1)--(0,5,1)--(0,5,0)--(1,5,0)--cycle;
\draw[black, fill=yellow!50, opacity=0.4] (1,5,1)--(1,5,0)--(1,4,0)--(1,4,1)--cycle;

\draw[double, ->, >=stealth, OliveGreen] (5.5,0.5,0)--(5.5,2.5,0);
\draw[double, ->, >=stealth, blue] (4.5,5.5,0)--(2.5,5.5,0);

\draw[gray,->] (2,0,5) -- (2,0,4) node[anchor=west]{$x$};
\draw[gray,->] (2,0,5) -- (2,1,5) node[anchor=west]{$y$};
\draw[gray,->] (2,0,5) -- (3,0,5) node[anchor=west]{$z$};

\filldraw[black] (1.8,1,0.5) circle (1pt);
\filldraw[black] (2.0,1,0.5) circle (1pt);
\filldraw[black] (2.2,1,0.5) circle (1pt);

\filldraw[black] (4,3.2,0.5) circle (1pt);
\filldraw[black] (4,3.0,0.5) circle (1pt);
\filldraw[black] (4,2.8,0.5) circle (1pt);

\filldraw[black] (1.8,4.5,1) circle (1pt);
\filldraw[black] (2.0,4.5,1) circle (1pt);
\filldraw[black] (2.2,4.5,1) circle (1pt);

\end{tikzpicture}
\end{center}
\end{tabular}

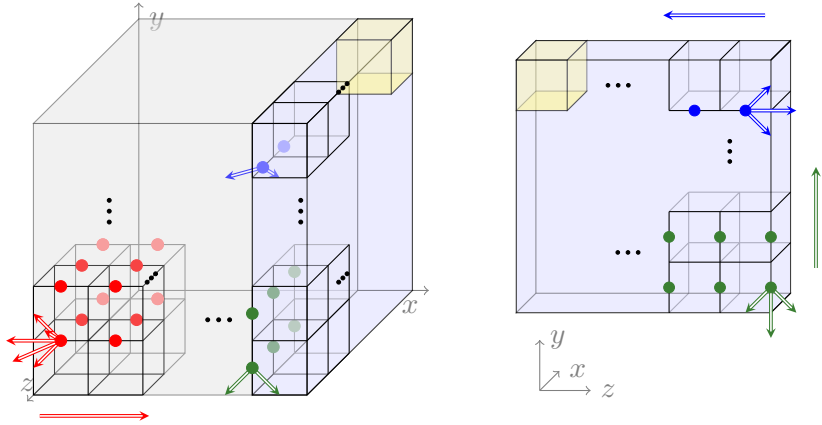
\captionof{figure}[foo]{We initiate with slicing the 3D torus into layers along the x-direction and applying H gates to all the colored dots. Subsequently, we apply all CNOT gates from red dots to non-red dots simultaneously and between adjacent red dots layer by layer, which requires $L+3$ steps.  The last layer needs special treatment, which simplifies into a 2D problem after applying 2 CNOT gates from all green and blue dots. Further progression involves applying CNOT gates concurrently from green dots to non-green dots and between adjacent green dots row by row, necessitating $L+1$ steps. Similarly, another $L+1$ steps applying CNOT gates from blue dots completes the procedure and leaves a redundant vertex in the yellow cube.} 
\label{fig:3dtoricsteps}
\end{table}

\begin{table}[ht]
\centering
\begin{tabular}{m{5.5cm} m{5.5cm}}
\begin{center}
\begin{tikzpicture}[scale=0.67]

\draw[gray,->] (0,0,0) -- (5.3,0,0) node[anchor=north east]{$x$};
\draw[gray,->] (0,0,0) -- (0,5.3,0) node[anchor=north west]{$y$};
\draw[gray,->] (0,0,0) -- (0,0,5.3) node[anchor=south]{$z$};

\draw[gray, fill=gray!20, opacity=0.3] (0,0,0) -- (0,5,0) -- (4,5,0) -- (4,0,0) -- cycle;
\draw[gray, fill=gray!20, opacity=0.3] (0,0,5) -- (0,0,0) -- (4,0,0) -- (4,0,5) -- cycle;
\draw[gray, fill=gray!20, opacity=0.3] (0,0,5) -- (0,5,5) -- (0,5,0) -- (0,0,0) -- cycle;
\draw[black, fill=blue!10, opacity=0.5] (4,0,5) -- (4,5,5) -- (4,5,0) -- (4,0,0) -- cycle;
\draw[black, fill=blue!10, opacity=0.5] (4,5,0) -- (5,5,0) -- (5,0,0) -- (4,0,0) -- cycle;
\draw[black, fill=blue!10, opacity=0.5] (4,0,5) -- (5,0,5) -- (5,0,0) -- (4,0,0) -- cycle;

\draw[gray] (0,0,3) -- (0,2,3) -- (2,2,3) -- (2,0,3) -- cycle;
\draw[gray] (1,0,3) -- (1,2,3);
\draw[gray] (2,1,3) -- (0,1,3);
\foreach \i in {0, 1, 2}
\foreach \j in {0, 1, 2}
        \draw[gray] ({\i},{\j},3) -- ({\i},{\j},4);

\filldraw[red!50] (0.5,1,3) circle (3pt);
\filldraw[red!50] (1.5,1,3) circle (3pt);
\filldraw[red!50] (0.5,2,3) circle (3pt);
\filldraw[red!50] (1.5,2,3) circle (3pt);

\draw[black] (0,0,4) -- (0,2,4) -- (2,2,4) -- (2,0,4) -- cycle;
\draw[black] (1,0,4) -- (1,2,4);
\draw[black] (2,1,4) -- (0,1,4);
\foreach \i in {0, 1, 2}
\foreach \j in {0, 1, 2}
        \draw[black] ({\i},{\j},4) -- ({\i},{\j},5);
\filldraw[red!50] (0.5,1,3) circle (3pt);
\filldraw[red!50] (1.5,1,3) circle (3pt);
\filldraw[red!50] (0.5,2,3) circle (3pt);
\filldraw[red!50] (1.5,2,3) circle (3pt);

\filldraw[red] (0.5,1,4) circle (3pt);
\filldraw[red] (1.5,1,4) circle (3pt);
\filldraw[red] (0.5,2,4) circle (3pt);
\filldraw[red] (1.5,2,4) circle (3pt);

\draw[gray] (4,4,3) -- (4,4,4);
\draw[thick, Purple] (4,4,3) -- (4,5,3);
\draw[thick, Purple] (4,4,3) -- (5,4,3);

\foreach \j in {0, 1, 2}
        \draw[thick, Purple] (4,{\j},3) -- (5,{\j},3);
\draw[thick, Purple] (4,0,0) -- (5,0,0);
\draw[gray] (4,0,3) -- (4,2,3);
\foreach \j in {0, 1, 2}
        \draw[gray] (4,{\j},3) -- (4,{\j},4);

\filldraw[OliveGreen!50] (4,0.5,3) circle (3pt);
\filldraw[OliveGreen!50] (4,1.5,3) circle (3pt);
\filldraw[blue!50] (4,4,3.5) circle (3pt);

\foreach \j in {0, 1, 2}
        \draw[thick, Purple] (4,{\j},4) -- (5,{\j},4);
\draw[black] (4,0,4) -- (4,2,4);
\foreach \j in {0, 1, 2}
        \draw[black] (4,{\j},4) -- (4,{\j},5);

\draw[black] (4,4,4) -- (4,4,5);
\draw[thick, Purple] (4,4,4) -- (4,5,4);
\draw[thick, Purple] (4,4,4) -- (5,4,4);

\filldraw[OliveGreen] (4,0.5,4) circle (3pt);
\filldraw[OliveGreen] (4,1.5,4) circle (3pt);
\filldraw[blue] (4,4,4.5) circle (3pt);

\draw[black, fill=gray!20, opacity=0.3] (0,0,5) -- (0,5,5) -- (4,5,5) -- (4,0,5) -- cycle;
\draw[black, fill=gray!20, opacity=0.3] (0,5,5) -- (0,5,0) -- (4,5,0) -- (4,5,5) -- cycle;
\draw[black, fill=blue!10, opacity=0.5] (5,0,5) -- (5,5,5) -- (5,5,0) -- (5,0,0) -- cycle;
\draw[black, fill=blue!10, opacity=0.5] (4,5,5) -- (5,5,5) -- (5,0,5) -- (4,0,5) -- cycle;
\draw[black, fill=blue!10, opacity=0.5] (4,5,5) -- (5,5,5) -- (5,5,0) -- (4,5,0) -- cycle;

\foreach \i in {0, 1, 2}
        \draw[black] ({\i},0,5) -- ({\i},2,5);
\foreach \j in {0, 1, 2}
        \draw[black] (0,{\j},5) -- (2,{\j},5);

\foreach \j in {0, 1, 2, 4, 5}
        \draw[black] (5,{\j},3) -- (5,{\j},5);
\foreach \k in {3, 4, 5}
        \draw[black] (5,0,{\k}) -- (5,2,{\k});
\foreach \k in {3, 4, 5}
        \draw[black] (4,5,{\k}) -- (5,5,{\k});
\foreach \k in {3, 4, 5}
        \draw[black] (5,4,{\k}) -- (5,5,{\k});
\foreach \j in {0, 1, 2, 4, 5}
        \draw[thick, Purple] (4,{\j},5) -- (5,{\j},5);
\draw[black] (4,0,5) -- (4,2,5);
\draw[black] (4,4,5) -- (4,5,5);
\draw[black] (4,5,5) -- (4,5,3);

\draw[thick, Purple] (4,4,0) -- (4,5,0);
\draw[thick, Purple] (4,4,0) -- (4,4,1);
\draw[thick, Purple] (4,4,1) -- (4,5,1);
\draw[thick, Purple] (4,4,0) -- (5,4,0);
\draw[thick, Purple] (4,4,1) -- (5,4,1);
\draw[black, fill=yellow!50, opacity=0.4] (5,4,1)--(5,4,0)--(4,4,0)--(4,4,1)--cycle;
\draw[black, fill=yellow!50, opacity=0.4] (5,5,1)--(4,5,1)--(4,4,1)--(5,4,1)--cycle;
\draw[black, fill=yellow!50, opacity=0.4] (5,5,1)--(4,5,1)--(4,5,0)--(5,5,0)--cycle;
\draw[black, fill=yellow!50, opacity=0.4] (5,5,1)--(5,5,0)--(5,4,0)--(5,4,1)--cycle;

\filldraw[black] (2.8,1,4) circle (1pt);
\filldraw[black] (3.0,1,4) circle (1pt);
\filldraw[black] (3.2,1,4) circle (1pt);

\filldraw[black] (1,2.8,4) circle (1pt);
\filldraw[black] (1,3.0,4) circle (1pt);
\filldraw[black] (1,3.2,4) circle (1pt);

\filldraw[black] (1,1,1.8) circle (1pt);
\filldraw[black] (1,1,2.0) circle (1pt);
\filldraw[black] (1,1,2.2) circle (1pt);

\filldraw[black] (4.5,2.8,4) circle (1pt);
\filldraw[black] (4.5,3.0,4) circle (1pt);
\filldraw[black] (4.5,3.2,4) circle (1pt);

\filldraw[black] (4.5,1,1.8) circle (1pt);
\filldraw[black] (4.5,1,2.0) circle (1pt);
\filldraw[black] (4.5,1,2.2) circle (1pt);

\filldraw[black] (4.5,4.5,1.8) circle (1pt);
\filldraw[black] (4.5,4.5,2.0) circle (1pt);
\filldraw[black] (4.5,4.5,2.2) circle (1pt);

\filldraw[red] (0.5,1,5) circle (3pt);
\filldraw[red] (1.5,1,5) circle (3pt);
\filldraw[OliveGreen] (4,0.5,5) circle (3pt);
\filldraw[red] (0.5,2,5) circle (3pt);
\filldraw[red] (1.5,2,5) circle (3pt);
\filldraw[OliveGreen] (4,1.5,5) circle (3pt);

\draw[thick, Purple] (4,4,5) -- (4,5,5);

\foreach \j in {4, 5}
        \draw[thick, Purple] (5,{\j},0) -- (5,{\j},1);
\filldraw[Purple] (5,2.8,0.5) circle (1pt);
\filldraw[Purple] (5,3.0,0.5) circle (1pt);
\filldraw[Purple] (5,3.2,0.5) circle (1pt);

\draw[thick, Purple] (4,5,0) -- (4,5,1);
\filldraw[Purple] (2.8,5,0.5) circle (1pt);
\filldraw[Purple] (3.0,5,0.5) circle (1pt);
\filldraw[Purple] (3.2,5,0.5) circle (1pt);

\foreach \k in {0, 1, 3, 4}
        \draw[thick, Purple] (4,5,{\k}) -- (5,5,{\k});
\filldraw[Purple] (5,4.5,1.8) circle (1pt);
\filldraw[Purple] (5,4.5,2.0) circle (1pt);
\filldraw[Purple] (5,4.5,2.2) circle (1pt);

\foreach \k in {0, 1, 3, 4, 5}
        \draw[thick, Purple] (5,4,{\k}) -- (5,5,{\k});
\filldraw[Purple] (4.5,5,1.8) circle (1pt);
\filldraw[Purple] (4.5,5,2.0) circle (1pt);
\filldraw[Purple] (4.5,5,2.2) circle (1pt);

\end{tikzpicture}
\end{center}
&
\begin{center}
\begin{tikzpicture}[scale=0.67]

\draw[gray,->] (0,0,0) -- (5.3,0,0) node[anchor=north east]{$x$};
\draw[gray,->] (0,0,0) -- (0,5.3,0) node[anchor=north west]{$y$};
\draw[gray,->] (0,0,0) -- (0,0,5.3) node[anchor=south]{$z$};

\draw[gray, fill=gray!20, opacity=0.2] (0,0,1) -- (0,4,1) -- (4,4,1) -- (4,0,1) -- cycle;
\draw[gray, fill=gray!20, opacity=0.2] (0,0,5) -- (0,0,1) -- (4,0,1) -- (4,0,5) -- cycle;
\draw[gray, fill=gray!20, opacity=0.2] (0,0,5) -- (0,4,5) -- (0,4,1) -- (0,0,1) -- cycle;

\foreach \i in {0, 1, 2, 3, 4, 5}
\foreach \j in {0, 1, 2, 3, 4, 5}
        \draw[thick, Purple] ({\i},{\j},0) -- ({\i},{\j},1);

\foreach \i in {0, 1, 2, 3, 4, 5}
\foreach \k in {0, 1, 2, 3, 4, 5}
        \draw[thick, Purple] ({\i},4,{\k}) -- ({\i},5,{\k});

\foreach \j in {0, 1, 2, 3, 4, 5}
\foreach \k in {0, 1, 2, 3, 4, 5}
        \draw[thick, Purple] (4,{\j},{\k}) -- (5,{\j},{\k});

\draw[black, fill=gray!20, opacity=0.3] (4,4,5) -- (4,4,1) -- (0,4,1) -- (0,4,5) -- cycle;
\draw[black, fill=gray!20, opacity=0.3] (4,4,5) -- (4,4,1) -- (4,0,1) -- (4,0,5) -- cycle;
\draw[black, fill=gray!20, opacity=0.3] (0,0,5) -- (0,4,5) -- (4,4,5) -- (4,0,5) -- cycle;

\draw[gray] (0,4,0) -- (5,4,0);
\draw[gray] (0,5,1) -- (5,5,1);
\draw[gray] (5,0,1) -- (5,5,1);
\draw[gray] (5,4,1) -- (5,4,5);
\draw[gray] (4,5,1) -- (4,5,5);

\draw[black, fill=Purple!10, opacity=0.3] (0,5,0) -- (5,5,0) -- (5,5,5) -- (0,5,5) -- cycle;
\draw[black, fill=Purple!10, opacity=0.3] (5,0,5) -- (5,5,5) -- (5,5,0) -- (5,0,0) -- cycle;
\draw[black, fill=Purple!10, opacity=0.3] (4,5,5) -- (5,5,5) -- (5,0,5) -- (4,0,5) -- cycle;
\draw[black, fill=Purple!10, opacity=0.3] (0,4,5) -- (4,4,5) -- (4,5,5) -- (0,5,5) -- cycle;

\end{tikzpicture}
\end{center}
\end{tabular}

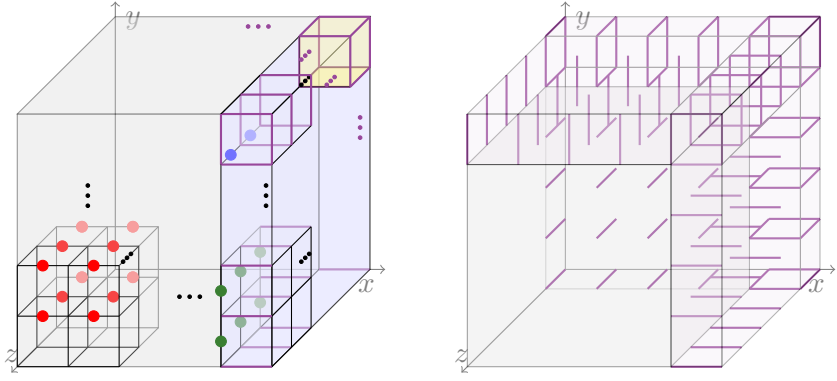
\captionof{figure}[foo]{The left sub-figure highlights the employed qubits (indicated by purple edges) that are utilized for the specific initial state within the ground state preparation process. Meanwhile, the right sub-figure sketches all employed edges within a cubic lattice of size $L=5$. All edges labeled by $z$ in the back layer, $x$ in the right layer, and $y$ in the top layer are employed to encode arbitrary ground states. Importantly, the procedure we introduced earlier remains uninterrupted since we do not designate any of these edges as free qubits.} 
\label{fig:3dtoricdegeneracy}
\end{table}

\FloatBarrier
\subsection{X-cube model}
The X-cube model is a fracton model defined on a 3D cubic lattice, as visually depicted in Figure \ref{fig:Xcubedefinition}. Within this lattice, $V$ represents the set of all vertices, while $C$ corresponds to the set of all cubes; each edge accommodates a single qubit. For the sake of convenience, we also affixed labels to each edge, denoting them as $x$, $y$, or $z$ based on their alignment with the respective axis. The Halmitonian is defined as
\begin{equation}
    H=-\sum_{v \in V} {A_v^{x}+A_v^{y}+A_v^{z}}-\sum_{c \in C} B_c,
\end{equation}
where $A_v^{i}, i= x,y,z$ is defined to implement Pauli operator $Z$ across the four edges oriented vertically to the $i$ axis and attached to vertex $v$, and $B_c$ is designated to effectuate Pauli operator $X$ across the twelve edges associated with cube $c$. Again these $A_v^i$s and $B_c$s operators satisfy $(A_v^i)^2 = B_c^2 = 1$ and $[A_v^i,B_c] = 0$. We get the ground state
\begin{equation}
    |GS\rangle=\prod_c \frac{1+B_c}{2} |\phi_0\rangle, 
\end{equation}
where $|\phi_0\rangle=|00...0\rangle$, and we drop $\frac{1+A_v^i}{2}$ since its action on $|\phi_0\rangle$ is $+1$.  It is important to emphasize, however, that the ground state degeneracy experiences exponential growth alongside the system size. Additionally, Figure \ref{fig:Xcubedefinition} presents a comprehensive depiction of a pair of logical operators of type $W$ and $T$\footnote{The complete set of logical operators are given in \cite{slagle2017quantum}, but we only use two types of them, which are not conjugate to each other.}. Similarly, we have three types of logical operator pairs, each acting on edges labeled by $x$, $y$, or $z$ respectively. Notably, distinct non-trivial loops exhibit identical homotopy while differ in terms of logical operators. This distinction is a crucial hallmark distinguishing the fracton model from conventional topological orders.

\begin{table}[ht]
\centering
\begin{tabular}{m{5cm} m{5cm}} 
\begin{center}
\begin{tikzpicture}[scale=0.6]
\foreach \i in {0, 2}
\foreach \j in {0, 2}
        \draw[thick, blue] ({\i},{\j},-2) -- ({\i},{\j},0);
\foreach \i in {0, 2}
\foreach \k in {-2, 0}
        \draw[thick, blue] ({\i},0,{\k}) -- ({\i},2,{\k});
\foreach \j in {0, 2}
\foreach \k in {-2, 0}
        \draw[thick, blue] (0,{\j},{\k}) -- (2,{\j},{\k});
\draw[gray] (0,0,0) -- (-2,0,0);
\draw[gray] (0,0,0) -- (0,-2,0);
\draw[gray] (0,0,0) -- (0,0,2);
\foreach \j in {0, 2}
\foreach \k in {-2, 0}
        \filldraw[gray!50] (1,{\j},{\k}) circle (3pt);
\foreach \i in {0, 2}
\foreach \k in {-2, 0}
        \filldraw[gray!50] ({\i},1,{\k}) circle (3pt);
\foreach \i in {0, 2}
\foreach \j in {0, 2}
        \filldraw[gray!50] ({\i},{\j},-1) circle (3pt);
\filldraw[gray!50] (-1,0,0) circle (3pt);
\filldraw[gray!50] (0,-1,0) circle (3pt);
\filldraw[gray!50] (0,0,1) circle (3pt);
\draw[thick, red] (0,1,0) -- (1,0,0) -- (0,-1,0) -- (-1,0,0) -- cycle;
\draw[gray,->] (-2,-2,0) -- (-1,-2,0) node[anchor=south]{$x$};
\draw[gray,->] (-2,-2,0) -- (-2,-1,0) node[anchor=west]{$y$};
\draw[gray,->] (-2,-2,0) -- (-2,-2,1) node[anchor=west]{$z$};
\node[black] at (1,-0.8,0) {$A_v^{z}$};
\node[black] at (1,1,-1) {$B_c$};
\end{tikzpicture}
\end{center}
&
\begin{center}
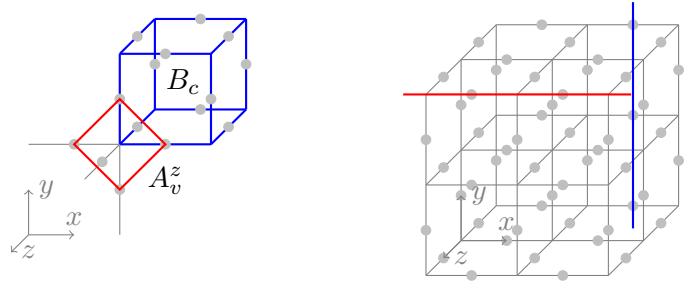

\begin{tikzpicture}[scale=0.6]
\foreach \i in {-2, 0, 2}
\foreach \j in {-2, 0, 2}
        \draw[gray] ({\i},{\j},-2) -- ({\i},{\j},2);
\foreach \i in {-2, 0, 2}
\foreach \k in {-2, 0, 2}
        \draw[gray] ({\i},-2,{\k}) -- ({\i},2,{\k});
\foreach \j in {-2, 0, 2}
\foreach \k in {-2, 0, 2}
        \draw[gray] (-2,{\j},{\k}) -- (2,{\j},{\k});
\foreach \i in {-1, 1}
\foreach \j in {-2, 0, 2}
\foreach \k in {-2, 0, 2}
        \filldraw[gray!50] ({\i},{\j},{\k}) circle (3pt);
\foreach \i in {-2, 0, 2}
\foreach \j in {-1, 1}
\foreach \k in {-2, 0, 2}
        \filldraw[gray!50] ({\i},{\j},{\k}) circle (3pt);
\foreach \i in {-2, 0, 2}
\foreach \j in {-2, 0, 2}
\foreach \k in {-1, 1}
        \filldraw[gray!50] ({\i},{\j},{\k}) circle (3pt);
\draw[thick, red] (-2.5,2,2) -- (2.5,2,2);
\draw[thick, blue] (1,-2.5,-2) -- (1,2.5,-2);
\draw[gray,->] (-2,-2,0) -- (-1,-2,0) node[anchor=south]{$x$};
\draw[gray,->] (-2,-2,0) -- (-2,-1,0) node[anchor=west]{$y$};
\draw[gray,->] (-2,-2,0) -- (-2,-2,1) node[anchor=west]{$z$};
\end{tikzpicture}
\end{center}
\end{tabular}
\captionof{figure}[foo]{The left sub-figure illustrates the definitions of $A_v$ and $B_p$ operators. Meanwhile, the right sub-figure displays a pair of conjugate logical operators composed of edges labeled by $x$. The red string is of type $W$, a nontrivial circle parallel to $x$ axis, and we apply Pauli $X$ over edges along the string. Conversely, the blue string is of type $T$, a nontrivial circle perpendicular to $x$ axis, and we apply Pauli $Z$ over edges along the string.} 
\label{fig:Xcubedefinition}
\end{table}

To simulate the ground state for the X-cube model, we outline\footnote{We also present a straightforward example comprising eight cubes to illustrate the method, as elaborated in Section \ref{sec:Xcubesimple}.} the procedures required to construct a quantum circuit for the X-cube model on an $L \times L \times L$ lattice, over a 3-dimensional torus, in Figure \ref{fig:xcubesteps}. The initial state is $|\phi_0\rangle$, and our target is to find quantum circuit to implement $\prod_c \frac{1+B_c}{2}$. We identify the redundancy by specifically selecting certain cubes, namely the three edges of the cubes in yellow, resulting from the requirement $\prod B_c=1$ of the involved layer of cubes. Given that layers can be independently sliced in three distinct directions, this selective arrangement of yellow-colored structures \footnote{While there may exist more redundant cubes, our focus is solely on the chosen ones.} is achieved. To start, we strategically partition the cube into distinct components: a central $(L-1) \times (L-1) \times (L-1)$ cube (colored gray), three $(L-1) \times (L-1) \times 1$ layers of cubes (colored blue), and three rows of redundant cubes (colored yellow). Then we further slice the central cube into $(L-1) \times (L-1) \times 1$ layers. Notice each gray and blue layer has the same boundaries up to rotation. This occurs because those yellow cubes are redundant, and the blue cubes are intended for the application of projectors in other layers. Neither of them interferes with the preparation of the layer structure. Consequently, we treat each layer of cubes as having the same structure, and their corresponding quantum circuits are outlined in Figure \ref{fig:xcubesteps}.

After the initial 9 steps, we apply CNOT from $(i,j)$ to $(i-1,j)$ in the $(3k+10)$-th step, apply CNOT from $(i,j)$ to $(i,j-1)$ in the $(3k+11)$-th step, and apply CNOT from $(i,j)$ to $(i-1,j-1)$ in the $(3k+12)$-th step, where $i+j=k+2$. This allows us to complete the layer structure in a total of $6L+6$ steps. These carefully orchestrated steps efficiently realize the circuit, requiring a total of $12L+11$ steps and the quantum circuit is purely local, similarly to the 3-dimensional toric code case. Certain non-interacting gates offer the potential for further parallelization, reducing the circuit depth by $2 \lfloor \frac{2L-3}{2} \rfloor$. It is worth noting that this method can be readily extended to the X-cube model on a lattice of dimensions $L_1 \times L_2 \times L_3$.

The ground state degeneracy can be resolved by the complete set of logical operators \cite{slagle2017quantum}. We can readily attain all bases of the ground space of the X-cube model by replacing the initial product state, as demonstrated in Section \ref{sec:arbitrary} and \ref{sec:3d toric}. However, it is not straightforward to see whether our method can be applied to prepare arbitrary ground states of the X-cube model. The comprehensive encoding of arbitrary ground states still remains an open question and is left as a topic for future research directions.

\begin{table}[ht]
\centering
\begin{tabular}{m{5.5cm} m{5.5cm}}
\begin{center}
\begin{tikzpicture}[scale=0.72]

\draw[black, fill=blue!10, opacity=0.5] (0,0,0) -- (4,0,0) -- (4,0,4) -- (0,0,4) -- cycle;
\draw[black, fill=blue!10, opacity=0.5] (0,0,0) -- (4,0,0) -- (4,1,0) -- (0,1,0) -- cycle;
\draw[black, fill=blue!10, opacity=0.5] (0,0,0) -- (0,0,4) -- (0,1,4) -- (0,1,0) -- cycle;
\draw[black, fill=blue!10, opacity=0.5] (0,1,0) -- (4,1,0) -- (4,1,4) -- (0,1,4) -- cycle;
\draw[gray, fill=gray!20, opacity=0.3] (0,1,0) -- (0,5,0) -- (4,5,0) -- (4,1,0) -- cycle;
\draw[gray, fill=gray!20, opacity=0.3] (0,1,0) -- (0,5,0) -- (0,5,4) -- (0,1,4) -- cycle;
\draw[black, fill=blue!10, opacity=0.5] (4,1,4) -- (4,5,4) -- (4,5,0) -- (4,1,0) -- cycle;
\draw[black, fill=blue!10, opacity=0.5] (4,5,0) -- (5,5,0) -- (5,1,0) -- (4,1,0) -- cycle;
\draw[black, fill=blue!10, opacity=0.5] (4,1,4) -- (5,1,4) -- (5,1,0) -- (4,1,0) -- cycle;
\draw[black, fill=blue!10, opacity=0.5] (0,1,4) -- (0,1,5) -- (0,5,5) -- (0,5,4) -- cycle;
\draw[black, fill=blue!10, opacity=0.5] (0,1,4) -- (0,1,5) -- (4,1,5) -- (4,1,4) -- cycle;
\draw[black, fill=yellow!20, opacity=0.4] (0,1,4) -- (0,1,5) -- (0,0,5) -- (0,0,4) -- cycle;
\draw[black, fill=yellow!20, opacity=0.4] (4,1,0) -- (5,1,0) -- (5,0,0) -- (4,0,0) -- cycle;
\draw[black, fill=yellow!20, opacity=0.4] (0,0,4) -- (0,0,5) -- (5,0,5) -- (5,0,0) -- (4,0,0) -- (4,0,4) -- (0,0,4) -- cycle;
\draw[black] (0,5,0) -- (4,5,0);

\draw[black, fill=yellow!20, opacity=0.4] (0,0,4) -- (4,0,4) -- (4,1,4) -- (0,1,4) -- cycle;
\draw[black, fill=yellow!20, opacity=0.4] (4,0,0) -- (4,0,4) -- (4,1,4) -- (4,1,0) -- cycle;
\draw[gray, fill=gray!20, opacity=0.3] (0,5,4) -- (0,5,0) -- (4,5,0) -- (4,5,4) -- cycle;
\draw[black, fill=blue!10, opacity=0.5] (0,1,4) -- (0,5,4) -- (4,5,4) -- (4,1,4) -- cycle;
\draw[black, fill=blue!10, opacity=0.5] (5,5,4) -- (4,5,4) -- (4,5,0) -- (5,5,0) -- cycle;
\draw[black, fill=yellow!20, opacity=0.4] (5,5,4) -- (4,5,4) -- (4,1,4) -- (5,1,4) -- cycle;
\draw[black, fill=blue!10, opacity=0.5] (5,5,4) -- (5,1,4) -- (5,1,0) -- (5,5,0) -- cycle;
\draw[black, fill=yellow!20, opacity=0.4] (4,1,4) -- (4,1,5) -- (4,5,5) -- (4,5,4) -- cycle;
\draw[black, fill=blue!10, opacity=0.5] (0,5,4) -- (0,5,5) -- (4,5,5) -- (4,5,4) -- cycle;
\draw[black, fill=blue!10, opacity=0.5] (0,5,5) -- (4,5,5) -- (4,1,5) -- (0,1,5) -- cycle;
\draw[black, fill=yellow!50, opacity=0.4] (5,5,5) -- (4,5,5) -- (4,5,4) -- (5,5,4) -- cycle;
\draw[black, fill=yellow!50, opacity=0.4] (5,5,5) -- (5,0,5) -- (5,0,0) -- (5,1,0) -- (5,1,4) -- (5,5,4) -- (5,5,5) -- cycle;
\draw[black, fill=yellow!50, opacity=0.4] (5,5,5) -- (5,0,5) -- (0,0,5) -- (0,1,5) -- (4,1,5) -- (4,5,5) -- (5,5,5) -- cycle;

\draw[gray,->] (0,5,5) -- (5.3,5,5) node[anchor=north]{$x$};
\draw[gray,->] (0,5,5) -- (0,-0.3,5) node[anchor=west]{$y$};
\draw[gray,->] (0,5,5) -- (0,5,-0.3) node[anchor=west]{$z$};
\end{tikzpicture}
\end{center}
&
\begin{center}
\begin{tikzpicture}[scale=0.67]

\draw[gray, fill=gray!20, opacity=0.3] (2.5,6,0) -- (4.5,6,0) -- (4.5,8,0) -- (2.5,8,0) -- cycle;
\draw[gray, fill=gray!20, opacity=0.3] (2.5,6,0) -- (4.5,6,0) -- (4.5,6,2) -- (2.5,6,2) -- cycle;
\draw[gray, fill=gray!20, opacity=0.3] (2.5,6,0) -- (2.5,8,0) -- (2.5,8,2) -- (2.5,6,2) -- cycle;
\draw[gray, fill=gray!20, opacity=0.3] (2.5,6,2) -- (4.5,6,2) -- (4.5,8,2) -- (2.5,8,2) -- cycle;
\draw[gray, fill=gray!20, opacity=0.3] (2.5,8,0) -- (4.5,8,0) -- (4.5,8,2) -- (2.5,8,2) -- cycle;
\draw[gray, fill=gray!20, opacity=0.3] (4.5,6,0) -- (4.5,8,0) -- (4.5,8,2) -- (4.5,6,2) -- cycle;

\draw[orange, thick] (4.5,6,0) -- (4.5,6,2);
\draw[orange, thick] (4.5,8,0) -- (4.5,8,2);
\draw[orange, thick] (2.5,8,0) -- (2.5,8,2);
\draw[orange, thick] (2.5,6,0) -- (2.5,6,2);

\filldraw[gray!50] (2.5,6,1) circle (3pt);
\filldraw[gray!50] (2.5,8,1) circle (3pt);

\foreach \i in {-1, 1}
\foreach \k in {0, 2}
        \filldraw[gray!50] ({\i+3.5},7,{\k}) circle (3pt);
\foreach \j in {-1, 1}
\foreach \k in {0, 2}
        \filldraw[gray!50] (3.5,{\j+7},{\k}) circle (3pt);
\foreach \i in {-1, 1}
\foreach \k in {0, 2}
        \draw[double, ->, >=stealth, black] (4.5,6,1)--({\i+3.5},7,{\k});
\foreach \j in {-1, 1}
\foreach \k in {0, 2}
        \draw[double, ->, >=stealth, black] (4.5,6,1)--(3.5,{\j+7},{\k});

\filldraw[gray!50] (4.5,6,1) circle (3pt);
\draw[thick, black] (4.5,6,1) circle (3pt);
\filldraw[gray!50] (4.5,8,1) circle (3pt);

\foreach \i in {0, 1, 2}
\foreach \j in {3, 4, 5}
        \draw[black] ({\i},{\j},0) -- ({\i},{\j},1);
\foreach \i in {1, 2}
        \draw[black] ({\i},3,0) -- ({\i},5,0);
\foreach \j in {3, 4}
        \draw[black] (0,{\j},0) -- (2,{\j},0);
\draw[black] (3,4,0) -- (3,5,0);
\draw[black] (3,4,0) -- (2,4,0);
\draw[black] (3,4,0) -- (3,4,1);
\draw[black] (1,2,0) -- (1,3,0);
\draw[black] (1,2,0) -- (0,2,0);
\draw[black] (1,2,0) -- (1,2,1);
\draw[black] (0,2,0) -- (0,2,1);
\draw[black] (3,5,1) -- (3,5,0);

\draw[black, fill=blue!10, opacity=0.5] (0,0,0) -- (0,0,1) -- (4,0,1) -- (4,0,0) -- cycle;
\draw[black, fill=blue!10, opacity=0.5] (0,0,0) -- (0,1,0) -- (0,1,1) -- (0,0,1) -- cycle;
\draw[black, fill=blue!10, opacity=0.5] (0,0,0) -- (0,1,0) -- (4,1,0) -- (4,0,0) -- cycle;

\draw[black, fill=gray!20, opacity=0.3] (0,1,0) -- (0,1,1) -- (0,5,1) -- (0,5,0) -- cycle;
\draw[black, fill=gray!20, opacity=0.3] (0,1,0) -- (0,1,1) -- (4,1,1) -- (4,1,0) -- cycle;
\draw[black, fill=gray!20, opacity=0.3] (0,1,0) -- (4,1,0) -- (4,5,0) -- (0,5,0) -- cycle;

\draw[black, fill=yellow!50, opacity=0.4] (4,1,0) -- (5,1,0) -- (5,0,0) -- (4,0,0) -- cycle;
\draw[black, fill=yellow!50, opacity=0.4] (4,1,0) -- (4,1,1) -- (4,0,1) -- (4,0,0) -- cycle;
\draw[black, fill=yellow!50, opacity=0.4] (4,1,0) -- (4,1,1) -- (5,1,1) -- (5,1,0) -- cycle;
\draw[black, fill=yellow!50, opacity=0.4] (5,0,1) -- (4,0,1) -- (4,0,0) -- (5,0,0) -- cycle;
\draw[black, fill=yellow!50, opacity=0.4] (5,0,1) -- (4,0,1) -- (4,1,1) -- (5,1,1) -- cycle;
\draw[black, fill=yellow!50, opacity=0.4] (5,0,1) -- (5,1,1) -- (5,1,0) -- (5,0,0) -- cycle;

\draw[black, fill=blue!10, opacity=0.5] (4,1,0) -- (5,1,0) -- (5,5,0) -- (4,5,0) -- cycle;
\draw[black, fill=blue!10, opacity=0.5] (5,5,1) -- (5,5,0) -- (5,1,0) -- (5,1,1) -- cycle;
\draw[black, fill=blue!10, opacity=0.5] (5,5,1) -- (4,5,1) -- (4,1,1) -- (5,1,1) -- cycle;
\draw[black, fill=blue!10, opacity=0.5] (5,5,1) -- (4,5,1) -- (4,5,0) -- (5,5,0) -- cycle;
\draw[black, fill=blue!10, opacity=0.5] (4,0,1) -- (4,1,1) -- (0,1,1) -- (0,0,1) -- cycle;

\draw[black, fill=gray!20, opacity=0.3] (4,5,1) -- (0,5,1) -- (0,5,0) -- (4,5,0) -- cycle;
\draw[black, fill=gray!20, opacity=0.3] (4,5,1) -- (4,1,1) -- (4,1,0) -- (4,5,0) -- cycle;
\draw[black, fill=gray!20, opacity=0.3] (4,5,1) -- (0,5,1) -- (0,1,1) -- (4,1,1) -- cycle;

\foreach \i in {1, 2}
        \draw[black] ({\i},3,1) -- ({\i},5,1);
\foreach \j in {3, 4}
        \draw[black] (0,{\j},1) -- (2,{\j},1);

\draw[double, ->, >=stealth, red] (1,4,0.5)--(1,5,0.5);
\draw[double, ->, >=stealth, red] (1,4,0.5)--(0,4,0.5);
\draw[double, ->, >=stealth, red] (1,4,0.5)--(0,5,0.5);
\filldraw[red] (1,4,0.5) circle (3pt);

\draw[double, ->, >=stealth, OliveGreen] (1,3,0.5)--(1,4,0.5);
\draw[double, ->, >=stealth, OliveGreen] (1,3,0.5)--(0,3,0.5);
\draw[double, ->, >=stealth, OliveGreen] (1,3,0.5)--(0,4,0.5);
\filldraw[OliveGreen] (1,3,0.5) circle (3pt);

\draw[double, ->, >=stealth, OliveGreen] (2,4,0.5)--(2,5,0.5);
\draw[double, ->, >=stealth, OliveGreen] (2,4,0.5)--(1,4,0.5);
\draw[double, ->, >=stealth, OliveGreen] (2,4,0.5)--(1,5,0.5);
\filldraw[OliveGreen] (2,4,0.5) circle (3pt);

\draw[double, ->, >=stealth, blue] (1,2,0.5)--(1,3,0.5);
\draw[double, ->, >=stealth, blue] (1,2,0.5)--(0,2,0.5);
\draw[double, ->, >=stealth, blue] (1,2,0.5)--(0,3,0.5);
\filldraw[blue] (1,2,0.5) circle (3pt);

\draw[double, ->, >=stealth, blue] (2,3,0.5)--(2,4,0.5);
\draw[double, ->, >=stealth, blue] (2,3,0.5)--(1,3,0.5);
\draw[double, ->, >=stealth, blue] (2,3,0.5)--(1,4,0.5);
\filldraw[blue] (2,3,0.5) circle (3pt);

\draw[double, ->, >=stealth, blue] (3,4,0.5)--(3,5,0.5);
\draw[double, ->, >=stealth, blue] (3,4,0.5)--(2,4,0.5);
\draw[double, ->, >=stealth, blue] (3,4,0.5)--(2,5,0.5);
\filldraw[blue] (3,4,0.5) circle (3pt);

\draw[black] (3,5,1) -- (3,4,1);
\draw[black] (2,4,1) -- (3,4,1);
\draw[black] (1,2,1) -- (1,3,1);
\draw[black] (1,2,1) -- (0,2,1);

\filldraw[black] (2.6,2.4,0.5) circle (1pt);
\filldraw[black] (2.8,2.2,0.5) circle (1pt);
\filldraw[black] (3.0,2.0,0.5) circle (1pt);

\draw[gray,->] (1,8,2.5) -- (2,8,2.5) node[anchor=north]{$x$};
\draw[gray,->] (1,8,2.5) -- (1,7,2.5) node[anchor=west]{$y$};
\draw[gray,->] (1,8,2.5) -- (1,8,1.5) node[anchor=west]{$z$};

\node[black] at (0.3,5.3,0) {$(0,0)$};
\end{tikzpicture}
\end{center}
\end{tabular}

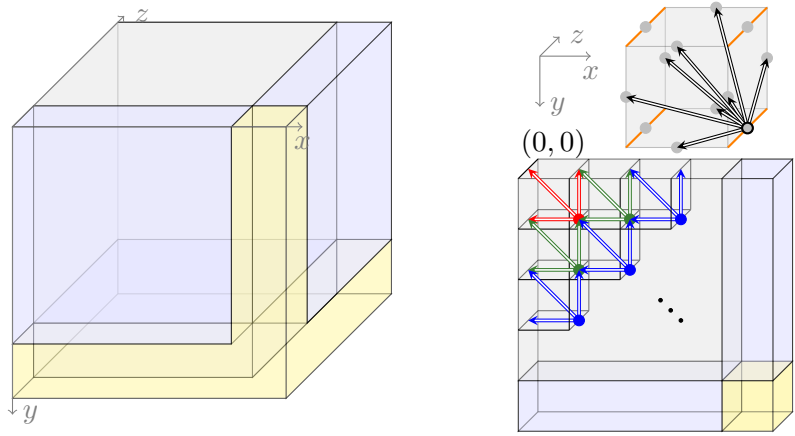
\captionof{figure}[foo]{Treating each layer of cubes as the same structure with boundaries, we initiate H gates along the edges in the z-direction of each cube and apply CNOT gates from these edges to the others in the $x$ and $y$ directions, which requires 9 steps. Then we apply CNOT gates diagonally, row by row in different colors, necessitating $6L-3$ steps. All gray layers are prepared simultaneously, followed by blue layers, leaving behind redundant cubes.} 
\label{fig:xcubesteps}
\end{table}

\FloatBarrier
\subsection{Gluing method for 3D models}\label{sec:gluing3Dmodel}
Similar to the scenario in 2D toric code case, we can simulate the ground state of 3D toric model by breaking the lattice into basic structures, simulating on and gluing them back. This results in one redundant vertex term and the excitations are quasi-particles that are able to move freely. The situation is exactly the same as 2D toric code, so we can find correcting operators to annihilate all of the excitations, which is left to readers.

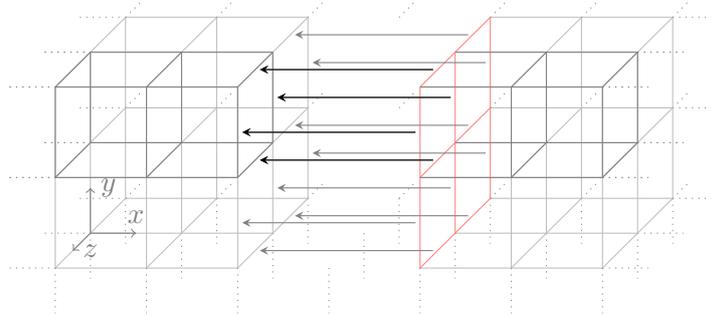
\begin{figure}[ht]
\centering
\begin{tikzpicture}[scale=0.6]
\foreach \j in {-2, 0, 2}
\foreach \k in {-2, 0, 2}
        \draw[gray!50] (-6,{\j},{\k}) -- (-2,{\j},{\k});
\foreach \j in {-2, 0, 2}
\foreach \k in {-2, 0, 2}
        \draw[dotted, gray] (-7,{\j},{\k}) -- (-6,{\j},{\k});
\foreach \i in {-6, -4, -2}
\foreach \j in {-2, 0, 2}
        \draw[dotted, gray] ({\i},{\j},-2) -- ({\i},{\j},-3);
\foreach \i in {-6, -4, -2}
\foreach \k in {-2, 0, 2}
        \draw[dotted, gray] ({\i},-2,{\k}) -- ({\i},-3,{\k});
\foreach \i in {-6, -4, -2}
\foreach \k in {-2, 0, 2}
        \draw[gray!50] ({\i},-2,{\k}) -- ({\i},2,{\k});
\foreach \i in {-6, -4, -2}
\foreach \j in {-2, 0, 2}
        \draw[gray!50] ({\i},{\j},-2) -- ({\i},{\j},2);

\foreach \j in {-2, 0, 2}
\foreach \k in {-2, 0, 2}
        \draw[gray!50] (2,{\j},{\k}) -- (6,{\j},{\k});
\foreach \j in {-2, 0, 2}
\foreach \k in {-2, 0, 2}
        \draw[dotted, gray] (6,{\j},{\k}) -- (7,{\j},{\k});
\foreach \i in {0, 2, 4, 6}
\foreach \j in {-2, 0, 2}
        \draw[dotted, gray] ({\i},{\j},-2) -- ({\i},{\j},-3);
\foreach \i in {0, 2, 4, 6}
\foreach \k in {-2, 0, 2}
        \draw[dotted, gray] ({\i},-2,{\k}) -- ({\i},-3,{\k});
\foreach \i in {4, 6}
\foreach \k in {-2, 0, 2}
        \draw[gray!50] ({\i},-2,{\k}) -- ({\i},2,{\k});
\foreach \i in {4, 6}
\foreach \j in {-2, 0, 2}
        \draw[gray!50] ({\i},{\j},-2) -- ({\i},{\j},2);

\foreach \j in {-2, 0, 2}
\foreach \k in {0, 2}
        \draw[red!50] (2,{\j},{\k}) -- (2,{\j},{\k-2});
\foreach \j in {0, 2}
\foreach \k in {-2, 0, 2}
        \draw[red!50] (2,{\j},{\k}) -- (2,{\j-2},{\k});

\foreach \j in {-2, 0, 2}
\foreach \k in {-1, 1}
        \draw[->, >=stealth, black!50] (1.9,{\j},{\k}) -- (-1.9,{\j},{\k});
\foreach \j in {-1, 1}
\foreach \k in {-2, 0, 2}
        \draw[->, >=stealth, black!50] (1.9,{\j},{\k}) -- (-1.9,{\j},{\k});

\foreach \j in {0, 2}
\foreach \k in {0, 2}
        \draw[gray] (-6,{\j},{\k}) -- (-2,{\j},{\k});
\foreach \j in {0, 2}
\foreach \k in {0, 2}
        \draw[dotted, gray] (-7,{\j},{\k}) -- (-6,{\j},{\k});
\foreach \i in {-6, -4, -2}
\foreach \k in {0, 2}
        \draw[gray] ({\i},0,{\k}) -- ({\i},2,{\k});
\foreach \i in {-6, -4, -2}
\foreach \j in {0, 2}
        \draw[gray] ({\i},{\j},0) -- ({\i},{\j},2);

\foreach \j in {0, 2}
\foreach \k in {0, 2}
        \draw[gray] (2,{\j},{\k}) -- (6,{\j},{\k});
\foreach \j in {0, 2}
\foreach \k in {0, 2}
        \draw[dotted, gray] (6,{\j},{\k}) -- (7,{\j},{\k});
\foreach \i in {4, 6}
\foreach \k in {0, 2}
        \draw[gray] ({\i},0,{\k}) -- ({\i},2,{\k});
\foreach \i in {4, 6}
\foreach \j in {0, 2}
        \draw[gray] ({\i},{\j},0) -- ({\i},{\j},2);

\foreach \j in {0, 2}
        \draw[->, >=stealth, black] (1.9,{\j},1) -- (-1.9,{\j},1);
\foreach \k in {0, 2}
        \draw[->, >=stealth, black] (1.9,1,{\k}) -- (-1.9,1,{\k});

\draw[gray,->] (-6,-2,0) -- (-5,-2,0) node[anchor=south]{$x$};
\draw[gray,->] (-6,-2,0) -- (-6,-1,0) node[anchor=west]{$y$};
\draw[gray,->] (-6,-2,0) -- (-6,-2,1) node[anchor=west]{$z$};
\end{tikzpicture}
\caption{Following the preparation of ground states on the individual lattices, we designate all qubits on one side of the gluing plane as ancilla qubits (represented by red edges). Subsequently, we apply CNOT gates in parallel from these ancilla qubits to the opposite side. This process allows us to obtain the ground state of the fused lattice after appropriately disentangling the ancilla qubits.}
\label{fig:glue3dXcube}
\end{figure}

Different methods for gluing in the X-cube model exist, and an intuitive one is shown in Figure \ref{fig:glue3dXcube}. In this method, the quantum circuits are applied to each of the individual pieces to obtain their respective ground states. Subsequently, CNOT gates are employed along the gluing plane to glue them together. It is essential to note that this process is not a simple measurement, as each edge is influenced by two cube terms. Disentanglement necessitates the implementation of measurements on all red edges and correction operators to eliminate potential excitations based on the measurement outcomes. However, the X-cube model poses greater complexity as the excitations are fractons. A systematic approach to find correcting operators is based on the following two facts:

1. There are three columns of redundant cubes as shown in Figure \ref{fig:xcubesteps}. 

2. The excitation betraying cube terms is a fracton that are not able to move freely. While a \textit{membrane operator} (see \cite{prem2017glassy} for details) creates fractons on four corners of a rectangular. 

\begin{table}[ht]
\centering
\begin{tabular}{m{6cm} m{5cm}}
\begin{center}
\begin{tikzpicture}[scale=0.65]
\draw[gray, fill=gray!20, opacity=0.1] (6,6,0) -- (6,0,0) -- (0,0,0) -- (0,6,0) -- cycle;
\draw[gray, fill=gray!20, opacity=0.1] (0,0,0) -- (0,0,6) -- (0,6,6) -- (0,6,0) -- cycle;
\draw[gray, fill=gray!20, opacity=0.1] (0,0,0) -- (0,0,6) -- (6,0,6) -- (6,0,0) -- cycle;
\draw[black, fill=blue!20, opacity=0.4] (0,5,0) -- (5,5,0) -- (5,0,0) -- (0,0,0) -- cycle;
\draw[black, fill=blue!20, opacity=0.4] (0,5,0) -- (0,5,5) -- (0,0,5) -- (0,0,0) -- cycle;
\draw[black, fill=blue!20, opacity=0.4] (5,4,5) -- (5,4,0) -- (0,4,0) -- (0,4,5) -- cycle;

\draw[black, fill=blue!60, opacity=0.6] (0,4,0) -- (0,4,0+1) -- (0,4+1,0+1) -- (0,4+1,0) -- cycle;
\draw[black, fill=blue!60, opacity=0.6] (0,4,0) -- (0,4,0+1) -- (0+1,4,0+1) -- (0+1,4,0) -- cycle;
\draw[black, fill=blue!60, opacity=0.6] (0+1,4+1,0) -- (0+1,4+1,0+1) -- (0+1,4,0+1) -- (0+1,4,0) -- cycle;
\draw[black, fill=blue!60, opacity=0.6] (0+1,4+1,0) -- (0+1,4+1,0+1) -- (0,4+1,0+1) -- (0,4+1,0) -- cycle;
\draw[black, fill=blue!60, opacity=0.6] (0+1,4+1,0) -- (0+1,4,0) -- (0,4,0) -- (0,4+1,0) -- cycle;
\draw[black, fill=blue!60, opacity=0.6] (0+1,4+1,0+1) -- (0+1,4,0+1) -- (0,4,0+1) -- (0,4+1,0+1) -- cycle;

\draw[black, fill=blue!60, opacity=0.6] (4,4,0) -- (4,4,0+1) -- (4,4+1,0+1) -- (4,4+1,0) -- cycle;
\draw[black, fill=blue!60, opacity=0.6] (4,4,0) -- (4,4,0+1) -- (4+1,4,0+1) -- (4+1,4,0) -- cycle;
\draw[black, fill=blue!60, opacity=0.6] (4+1,4+1,0) -- (4+1,4+1,0+1) -- (4+1,4,0+1) -- (4+1,4,0) -- cycle;
\draw[black, fill=blue!60, opacity=0.6] (4+1,4+1,0) -- (4+1,4+1,0+1) -- (4,4+1,0+1) -- (4,4+1,0) -- cycle;
\draw[black, fill=blue!60, opacity=0.6] (4+1,4+1,0) -- (4+1,4,0) -- (4,4,0) -- (4,4+1,0) -- cycle;
\draw[black, fill=blue!60, opacity=0.6] (4+1,4+1,0+1) -- (4+1,4,0+1) -- (4,4,0+1) -- (4,4+1,0+1) -- cycle;

\draw[black, fill=blue!60, opacity=0.6] (0,4,4) -- (0,4,4+1) -- (0,4+1,4+1) -- (0,4+1,4) -- cycle;
\draw[black, fill=blue!60, opacity=0.6] (0,4,4) -- (0,4,4+1) -- (0+1,4,4+1) -- (0+1,4,4) -- cycle;
\draw[black, fill=blue!60, opacity=0.6] (0+1,4+1,4) -- (0+1,4+1,4+1) -- (0+1,4,4+1) -- (0+1,4,4) -- cycle;
\draw[black, fill=blue!60, opacity=0.6] (0+1,4+1,4) -- (0+1,4+1,4+1) -- (0,4+1,4+1) -- (0,4+1,4) -- cycle;
\draw[black, fill=blue!60, opacity=0.6] (0+1,4+1,4) -- (0+1,4,4) -- (0,4,4) -- (0,4+1,4) -- cycle;
\draw[black, fill=blue!60, opacity=0.6] (0+1,4+1,4+1) -- (0+1,4,4+1) -- (0,4,4+1) -- (0,4+1,4+1) -- cycle;

\draw[black, fill=blue!60, opacity=0.6] (0,0,0) -- (0,0,0+1) -- (0,0+1,0+1) -- (0,0+1,0) -- cycle;
\draw[black, fill=blue!60, opacity=0.6] (0,0,0) -- (0,0,0+1) -- (0+1,0,0+1) -- (0+1,0,0) -- cycle;
\draw[black, fill=blue!60, opacity=0.6] (0+1,0+1,0) -- (0+1,0+1,0+1) -- (0+1,0,0+1) -- (0+1,0,0) -- cycle;
\draw[black, fill=blue!60, opacity=0.6] (0+1,0+1,0) -- (0+1,0+1,0+1) -- (0,0+1,0+1) -- (0,0+1,0) -- cycle;
\draw[black, fill=blue!60, opacity=0.6] (0+1,0+1,0) -- (0+1,0,0) -- (0,0,0) -- (0,0+1,0) -- cycle;
\draw[black, fill=blue!60, opacity=0.6] (0+1,0+1,0+1) -- (0+1,0,0+1) -- (0,0,0+1) -- (0,0+1,0+1) -- cycle;

\draw[black, fill=blue!60, opacity=0.6] (4,0,0) -- (4,0,0+1) -- (4,0+1,0+1) -- (4,0+1,0) -- cycle;
\draw[black, fill=blue!60, opacity=0.6] (4,0,0) -- (4,0,0+1) -- (4+1,0,0+1) -- (4+1,0,0) -- cycle;
\draw[black, fill=blue!60, opacity=0.6] (4+1,0+1,0) -- (4+1,0+1,0+1) -- (4+1,0,0+1) -- (4+1,0,0) -- cycle;
\draw[black, fill=blue!60, opacity=0.6] (4+1,0+1,0) -- (4+1,0+1,0+1) -- (4,0+1,0+1) -- (4,0+1,0) -- cycle;
\draw[black, fill=blue!60, opacity=0.6] (4+1,0+1,0) -- (4+1,0,0) -- (4,0,0) -- (4,0+1,0) -- cycle;
\draw[black, fill=blue!60, opacity=0.6] (4+1,0+1,0+1) -- (4+1,0,0+1) -- (4,0,0+1) -- (4,0+1,0+1) -- cycle;

\draw[black, fill=blue!60, opacity=0.6] (0,0,4) -- (0,0,4+1) -- (0,0+1,4+1) -- (0,0+1,4) -- cycle;
\draw[black, fill=blue!60, opacity=0.6] (0,0,4) -- (0,0,4+1) -- (0+1,0,4+1) -- (0+1,0,4) -- cycle;
\draw[black, fill=blue!60, opacity=0.6] (0+1,0+1,4) -- (0+1,0+1,4+1) -- (0+1,0,4+1) -- (0+1,0,4) -- cycle;
\draw[black, fill=blue!60, opacity=0.6] (0+1,0+1,4) -- (0+1,0+1,4+1) -- (0,0+1,4+1) -- (0,0+1,4) -- cycle;
\draw[black, fill=blue!60, opacity=0.6] (0+1,0+1,4) -- (0+1,0,4) -- (0,0,4) -- (0,0+1,4) -- cycle;
\draw[black, fill=blue!60, opacity=0.6] (0+1,0+1,4+1) -- (0+1,0,4+1) -- (0,0,4+1) -- (0,0+1,4+1) -- cycle;

\draw[black, fill=blue!20, opacity=0.2] (0,5,1) -- (5,5,1) -- (5,0,1) -- (0,0,1) -- cycle;
\draw[black, fill=blue!20, opacity=0.2] (1,5,0) -- (1,5,5) -- (1,0,5) -- (1,0,0) -- cycle;

\draw[black, fill=blue!60, opacity=0.6] (4,4,4) -- (4,4,4+1) -- (4,4+1,4+1) -- (4,4+1,4) -- cycle;
\draw[black, fill=blue!60, opacity=0.6] (4,4,4) -- (4,4,4+1) -- (4+1,4,4+1) -- (4+1,4,4) -- cycle;
\draw[black, fill=blue!60, opacity=0.6] (4+1,4+1,4) -- (4+1,4+1,4+1) -- (4+1,4,4+1) -- (4+1,4,4) -- cycle;
\draw[black, fill=blue!60, opacity=0.6] (4+1,4+1,4) -- (4+1,4+1,4+1) -- (4,4+1,4+1) -- (4,4+1,4) -- cycle;
\draw[black, fill=blue!60, opacity=0.6] (4+1,4+1,4) -- (4+1,4,4) -- (4,4,4) -- (4,4+1,4) -- cycle;
\draw[black, fill=blue!60, opacity=0.6] (4+1,4+1,4+1) -- (4+1,4,4+1) -- (4,4,4+1) -- (4,4+1,4+1) -- cycle;

\draw[black, fill=blue!20, opacity=0.4] (5,5,5) -- (5,5,0) -- (0,5,0) -- (0,5,5) -- cycle;
\draw[gray, fill=gray!20, opacity=0.3] (6,6,6) -- (0,6,6) -- (0,6,0) -- (6,6,0) -- cycle;
\draw[gray, fill=gray!20, opacity=0.3] (6,6,6) -- (6,0,6) -- (6,0,0) -- (6,6,0) -- cycle;
\draw[gray, fill=gray!20, opacity=0.3] (6,6,6) -- (6,0,6) -- (0,0,6) -- (0,6,6) -- cycle;
\draw[gray,->] (0,0,0) -- (6.3,0,0) node[anchor=north east]{$x$};
\draw[gray,->] (0,0,0) -- (0,6.3,0) node[anchor=north west]{$y$};
\draw[gray,->] (0,0,0) -- (0,0,6.3) node[anchor=south]{$z$};
\node[black!70] at (6.6,4.8,5.5) {$(i,j,k)$};
\node[black!70] at (2.6,4.8,5.5) {$(1,j,k)$};
\node[black!70] at (1,5.3,0) {$(1,j,1)$};
\node[black!70] at (4.5,5.3,0) {$(i,j,1)$};
\node[black!70] at (1.5,0,1.8) {$(1,1,1)$};
\node[black!70] at (1.5,0,5.8) {$(1,1,k)$};
\node[black!70] at (5.5,0,1.8) {$(i,1,1)$};
\end{tikzpicture}
\end{center}
&
\begin{center}
\begin{tikzpicture}[scale=0.75]
\foreach \i in {0, 1, 2, 3, 4}
\foreach \j in {0, 1, 2, 3, 4}
        \draw[black] ({\i},{\j},0) -- ({\i},{\j},1);
\foreach \i in {1, 2, 3, 4}
        \draw[black] ({\i},0,0) -- ({\i},5,0);
\foreach \j in {1, 2, 3, 4}
        \draw[black] (0,{\j},0) -- (5,{\j},0);
\draw[black, fill=blue!10, opacity=0.5] (0,0,0) -- (0,0,1) -- (0,5,1) -- (0,5,0) -- cycle;
\draw[black, fill=blue!10, opacity=0.5] (0,0,0) -- (0,0,1) -- (5,0,1) -- (5,0,0) -- cycle;
\draw[black, fill=blue!10, opacity=0.5] (0,0,0) -- (5,0,0) -- (5,5,0) -- (0,5,0) -- cycle;
\draw[black, fill=blue!10, opacity=0.5] (5,5,0) -- (5,5,1) -- (0,5,1) -- (0,5,0) -- cycle;
\draw[black, fill=blue!10, opacity=0.5] (5,5,0) -- (5,5,1) -- (5,0,1) -- (5,0,0) -- cycle;
\draw[black, fill=blue!10, opacity=0.5] (0,0,1) -- (5,0,1) -- (5,5,1) -- (0,5,1) -- cycle;
\foreach \i in {0, 1, 2, 3, 4, 5}
        \draw[black] ({\i},5,0) -- ({\i},5,1);
\foreach \j in {0, 1, 2, 3, 4, 5}
        \draw[black] (5,{\j},0) -- (5,{\j},1);
\foreach \i in {1, 2, 3, 4}
        \draw[black] ({\i},0,1) -- ({\i},5,1);
\foreach \j in {1, 2, 3, 4}
        \draw[black] (0,{\j},1) -- (5,{\j},1);
\foreach \i in {0, 4}
\foreach \j in {0, 4}
        \draw[black, fill=blue!50, opacity=0.6] ({\i},{\j},0) -- ({\i},{\j},1) -- ({\i},{\j+1},1) -- ({\i},{\j+1},0) -- cycle;
\foreach \i in {0, 4}
\foreach \j in {0, 4}
        \draw[black, fill=blue!50, opacity=0.6] ({\i},{\j},0) -- ({\i},{\j},1) -- ({\i+1},{\j},1) -- ({\i+1},{\j},0) -- cycle;
\foreach \i in {1, 5}
\foreach \j in {1, 5}
        \draw[black, fill=blue!50, opacity=0.6] ({\i},{\j},0) -- ({\i},{\j},1) -- ({\i},{\j-1},1) -- ({\i},{\j-1},0) -- cycle;
\foreach \i in {1, 5}
\foreach \j in {1, 5}
        \draw[black, fill=blue!50, opacity=0.6] ({\i},{\j},0) -- ({\i},{\j},1) -- ({\i-1},{\j},1) -- ({\i-1},{\j},0) -- cycle;
\foreach \i in {1, 5}
\foreach \j in {1, 5}
        \draw[black, fill=blue!50, opacity=0.6] ({\i},{\j},0) -- ({\i},{\j-1},0) -- ({\i-1},{\j-1},0) -- ({\i-1},{\j},0) -- cycle;
\foreach \i in {1, 5}
\foreach \j in {1, 5}
        \draw[black, fill=blue!50, opacity=0.6] ({\i},{\j},1) -- ({\i},{\j-1},1) -- ({\i-1},{\j-1},1) -- ({\i-1},{\j},1) -- cycle;
\foreach \i in {1, 2, 3, 4}
\foreach \j in {1, 2, 3, 4}
        \draw[very thick, OliveGreen] ({\i},{\j},0) -- ({\i},{\j},1);
\node[black!70] at (0.7,-0.4,1) {$(i,j,k)$};
\node[black!70] at (4.5,5.7,1) {$(i',j',k')$};
\end{tikzpicture}
\end{center}
\end{tabular}

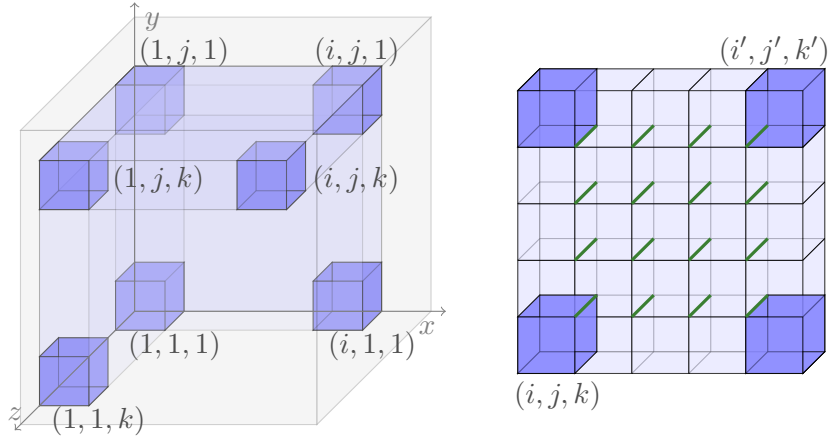
\captionof{figure}[foo]{A membrane operator consisting of $Z$ operators on green edges creates fractons at four corners; The correcting operator is a product of three membrane operators.} 
\label{fig:glue3d}
\end{table}

Illustrated in Figure \ref{fig:glue3d}, each cube in the $n^3$ cubic lattice, underlying the 3D torus topology, is assigned Cartesian coordinates $(i,j,k)$, where $1 \le i,j,k \le n$. Redundant cubes are positioned along three columns, namely $(i,1,1)$, $(1,i,1)$, and $(1,1,i)$ for $i = 1\cdots n$. A membrane operator $\mathcal{M}[(i,j,k),(i',j',k')]$, consists of $Z$ operators in the rectangle from $(i,j,k)$ to $(i',j',k')$ creates excitations at the four corners.  For instance, when addressing an excitation at $(i,j,1)$, applying $\mathcal{M}[(1,1,1),(i,j,1)]$ leads to the annihilation of the excitation and the creation of excitations at redundant cubes, which are inconsequential. When dealing with a general excitation at $(i,j,k)$, with $i,j,k \neq 1$, a multi-step procedure comes into play. Initially, $\mathcal{M}[(1,j,1),(i,j,k)]$ is applied to eliminate the excitation, generating three additional excitations at $(1,j,1)$, $(1,j,k)$, and $(i,j,1)$. Disregarding the one at the redundant cube, the other two are subsequently eliminated by $\mathcal{M}[(1,1,1),(1,j,k)]$ and $\mathcal{M}[(1,1,1),(i,j,1)]$, respectively. In essence, the product operator $\mathcal{M}[(1,1,1),(1,j,k)]$ $\mathcal{M}[(1,1,1),(i,j,1)]$ $\mathcal{M}[(1,j,1),(i,j,k)]$ is capable of annihilating general excitations.

\FloatBarrier
\section{Conclusion and outlook}
In this paper, we propose a method to prepare the ground state of a Hamiltonian consisting of local commuting projectors composed solely of Pauli $X$ and Pauli $Z$ operators.  Our approach involves finding an appropriate initial state that serves as the ground state of these projectors and applying a quantum circuit composed solely of Clifford gates to achieve the Hamiltonian's ground state.  We demonstrate the effectiveness of our method on 2D toric codes with various surface conditions, both with and without boundaries, as well as on the 3D toric model and the X-cube model.  Our method enables the preparation of arbitrary ground states for 2D and 3D toric model with a linear-depth circuit, meeting the lower bound for preparing ground states in topological phases.  It also works for any basis of the ground state in the X-cube model using a linear-depth quantum circuit.  We present these results on specific lattices, such as the 2D square lattice or 3D cubic lattice, and introduce a gluing method to facilitate ground state preparation on general 2D and 3D lattices.  This gluing method provides a trade-off between measurement usage and circuit depth and can be applied to obtain the ground state of larger lattices by assembling ground states of smaller components.

There are several future directions to proceed from this work. One natural progression involves extending our method to other 3D models of interest. Furthermore, the applicability of our approach to the non-abelian Kitaev model presents a straightforward extension, offering the potential to broaden the scope of its application.

\vspace{0.5cm}
\noindent \textbf{Acknowledgments.} 

The authors are partially supported by NSF CCF 2006667, Quantum Science Center (led by ORNL), and ARO MURI. 

\sloppy
\printbibliography

\onecolumn\newpage
\appendix
\section{2D toric code on sphere}\label{sec:2dsphere}

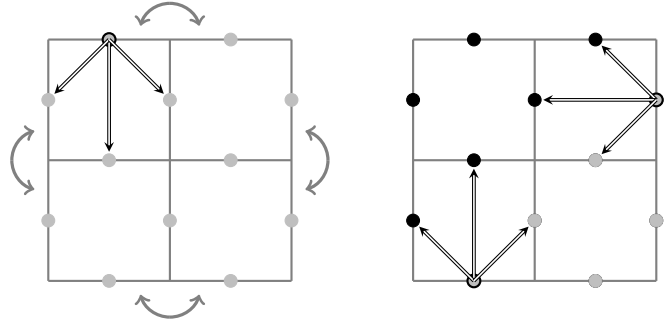
\begin{figure}[ht]
\centering
\begin{tikzpicture}[scale=0.8]
\draw[thick, gray] (-2,-2) -- (2,-2);
\draw[thick, gray] (-2,2) -- (2,2);
\draw[thick, gray] (-2,-2) -- (-2,2);
\draw[thick, gray] (2,-2) -- (2,2);
\draw[thick, gray] (-2,0) -- (2,0);
\draw[thick, gray] (0,-2) -- (0,2);
\draw[->, gray, very thick] (2.6,0) arc (0:-75:0.5);
\draw[->, gray, very thick] (2.6,0) arc (0:75:0.5);
\draw[->, gray, very thick] (-2.6,0) arc (180:255:0.5);
\draw[->, gray, very thick] (-2.6,0) arc (180:105:0.5);
\draw[->, gray, very thick] (0,2.6) arc (90:15:0.5);
\draw[->, gray, very thick] (0,2.6) arc (90:165:0.5);
\draw[->, gray, very thick] (0,-2.6) arc (-90:-15:0.5);
\draw[->, gray, very thick] (0,-2.6) arc (-90:-165:0.5);
\foreach \i in {-1, 1}
\foreach \j in {-2, 0, 2}
        \filldraw[gray!50] ({\i},{\j}) circle (3pt);
\foreach \i in {-2, 0, 2}
\foreach \j in {-1, 1}
        \filldraw[gray!50] ({\i},{\j}) circle (3pt);
\draw[thick, black] (-1,2) circle (3pt);
\draw[double, ->, >=stealth, black] (-1,2)--(-1,0.14);
\draw[double, ->, >=stealth, black] (-1,2)--(-1.9,1.1);
\draw[double, ->, >=stealth, black] (-1,2)--(-0.1,1.1);

\draw[thick, gray] (4,-2) -- (8,-2);
\draw[thick, gray] (4,2) -- (8,2);
\draw[thick, gray] (4,-2) -- (4,2);
\draw[thick, gray] (8,-2) -- (8,2);
\draw[thick, gray] (4,0) -- (8,0);
\draw[thick, gray] (6,-2) -- (6,2);
\foreach \i in {-1, 1}
\foreach \j in {-2, 0, 2}
        \filldraw[black] ({6+\i},{\j}) circle (3pt);
\foreach \i in {-2, 0, 2}
\foreach \j in {-1, 1}
        \filldraw[black] ({6+\i},{\j}) circle (3pt);
\filldraw[gray!50] (7,0) circle (3pt);
\filldraw[gray!50] (6,-1) circle (3pt);
\filldraw[gray!50] (7,-2) circle (3pt);
\filldraw[gray!50] (8,-1) circle (3pt);
\filldraw[gray!50] (8,1) circle (3pt);
\draw[thick, black] (8,1) circle (3pt);
\draw[double, ->, >=stealth, black] (8,1)--(7.1,1.9);
\draw[double, ->, >=stealth, black] (8,1)--(6.14,1);
\draw[double, ->, >=stealth, black] (8,1)--(7.1,0.1);
\filldraw[gray!50] (5,-2) circle (3pt);
\draw[thick, black] (5,-2) circle (3pt);
\draw[double, ->, >=stealth, black] (5,-2)--(5,-0.14);
\draw[double, ->, >=stealth, black] (5,-2)--(4.1,-1.1);
\draw[double, ->, >=stealth, black] (5,-2)--(5.9,-1.1);
\end{tikzpicture}
\caption{Boundary edges are identified according to the double-headed arrows.}
\label{fig:2dsphere}
\end{figure}

Similar with the example of genus 1 torus, we identify different qubit pairs to change the four plaquettes into a sphere as shown in Figure \ref{fig:2dsphere}. The bottom right plaquette is chosen to be redundant and two steps will complete the procedure. 

\section{2D toric code on genus n surface} \label{sec:2dothers}

\begin{figure}[ht]
\centering
\begin{tikzpicture}[scale=0.8]
\filldraw[gray!20] (-3,-2) rectangle (-1,2);
\draw[thick, gray] (-4,-3) rectangle (0,3);
\draw[thick, gray] (-3,-2) rectangle (-1,2);
\draw[thick, gray] (-3,-2) -- (-4,-3);
\draw[thick, gray] (-3,2) -- (-4,3);
\draw[thick, gray] (-1,2) -- (0,3);
\draw[thick, gray] (-1,-2) -- (0,-3);
\draw[thick, gray] (-3,0) -- (-4,0);
\draw[thick, dashed, gray] (-0.5,0.6) -- (-0.5,-0.6);
\draw[thick, red] (-4,0) -- (-4,3);
\draw[thick, yellow] (-4,3) -- (0,3);
\draw[thick, blue] (-4,0) -- (-4,-3);
\draw[thick, red] (-4,-3) -- (0,-3);
\filldraw[gray!50] (-4,1.5) circle (3pt);
\filldraw[gray!50] (-4,-1.5) circle (3pt);
\filldraw[gray!50] (-2,3) circle (3pt);
\filldraw[gray!50] (-2,-3) circle (3pt);
\filldraw[gray!50] (-0.5,2.5) circle (3pt);
\filldraw[gray!50] (-3.5,0) circle (3pt);
\filldraw[gray!50] (-0.5,-2.5) circle (3pt);
\filldraw[gray!50] (-3.5,-2.5) circle (3pt);
\filldraw[black] (-3,1) circle (3pt);
\filldraw[black] (-3,-1) circle (3pt);
\filldraw[black] (-2,2) circle (3pt);
\filldraw[black] (-2,-2) circle (3pt);
\filldraw[gray!50] (-3.5,2.5) circle (3pt);
\draw[thick, black] (-3.5,2.5) circle (3pt);
\draw[double, ->, >=stealth, black] (-3.5,2.5)--(-3.5,0.14);
\draw[double, ->, >=stealth, black] (-3.5,2.5)--(-3.9,1.6);
\draw[double, ->, >=stealth, black] (-3.5,2.5)--(-3.1,1.1);
\filldraw[gray!20] (2,-2) rectangle (4,2);
\draw[thick, gray] (1,-3) rectangle (5,3);
\draw[thick, gray] (2,-2) rectangle (4,2);
\draw[thick, gray] (2,-2) -- (1,-3);
\draw[thick, gray] (2,2) -- (1,3);
\draw[thick, gray] (4,2) -- (5,3);
\draw[thick, gray] (4,-2) -- (5,-3);
\draw[thick, gray] (2,0) -- (1,0);
\draw[thick, dashed, gray] (4.5,0.6) -- (4.5,-0.6);
\draw[thick, red] (1,0) -- (1,3);
\draw[thick, yellow] (1,3) -- (5,3);
\draw[thick, blue] (1,0) -- (1,-3);
\draw[thick, red] (1,-3) -- (5,-3);
\filldraw[gray!50] (1,-1.5) circle (3pt);
\filldraw[gray!50] (3,3) circle (3pt);
\filldraw[gray!50] (4.5,2.5) circle (3pt);
\filldraw[gray!50] (4.5,-2.5) circle (3pt);
\filldraw[gray!50] (1.5,-2.5) circle (3pt);
\filldraw[black] (2,-1) circle (3pt);
\filldraw[black] (3,2) circle (3pt);
\filldraw[black] (3,-2) circle (3pt);
\filldraw[black] (1.5,0) circle (3pt);
\filldraw[black] (1,1.5) circle (3pt);
\filldraw[black] (1.5,2.5) circle (3pt);
\filldraw[black] (3,-3) circle (3pt);
\filldraw[black] (2,1) circle (3pt);
\draw[thick, black] (4.5,2.5) circle (3pt);
\draw[double, ->, >=stealth, black] (4.5,2.5)--(1.64,2.5);
\draw[double, ->, >=stealth, black] (4.5,2.5)--(3.1,2.1);
\draw[double, ->, >=stealth, black] (4.5,2.5)--(3.1,2.9);
\draw[thick, dashed, gray] (5.5,0) -- (6.7,0);
\end{tikzpicture}
\caption{The shaded area represents the developed disk; Boundaries with the same color are identified to change the plaquettes into a genus n torus.}
\label{fig:2dtorusn}
\end{figure}
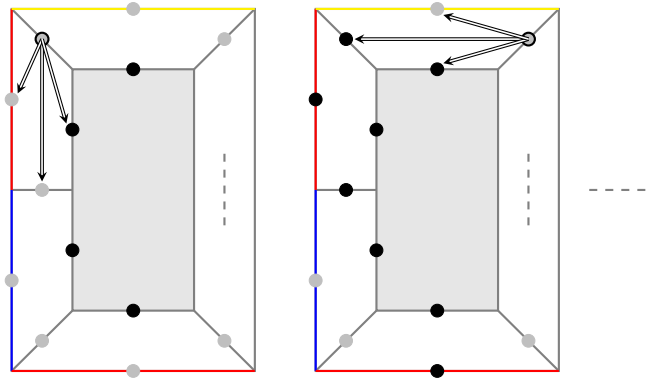

Figure \ref{fig:2dtorusn} shows a genus n surface which is a disk enclosed by a ribbon with identified edges. Beginning with $|\phi_0\rangle$, we develop a disk from inside and leave the ribbon with all identified edges undeveloped. Then we choose one edge in the ribbon to apply the method of basic structure and repeat in clockwise direction. After $2n-1$ steps for a genus n torus, we will get the ground state of the closed surface. 

\section{Local CNOT operation} \label{sec:clocalCNOT}

In the preparation of arbitrary state of 2D toric code, we use $CNOT$ to transmit the logical states vertically and horizontally. If we employ non-local CNOT gates, as illustrated on the left side of Figure \ref{fig:clocalCNOT}, it takes $\lceil log_{2}(L) \rceil$ steps. However, when CNOT gates are constrained to constant distances $d$, this procedure requires $\lceil log_{2}(d)+\frac{L}{2d} \rceil$ steps, as shown on the right side of Figure \ref{fig:clocalCNOT}. The distance $d$ is defined such that two qubits are considered to be $d$ apart if the shortest path connecting them contains $d-1$ qubits.

\begin{table}[ht]
\centering
\begin{tabular}{m{5cm} m{5cm}} 
\begin{center}
\begin{tikzpicture}[scale=0.33]
\foreach \i in {-8, -6, -4, -2, 0, 2, 4, 6, 8, 10, 12, 14, 16, 18, 20, 22, 24}
        \draw[thick, gray] (0,{\i}) -- (8,{\i});
\foreach \i in {0, 2, 4, 6, 8}
        \draw[thick, gray] ({\i},-8) -- ({\i},24);
\foreach \i in {1, 3, 5, 7}
\foreach \j in {-8, -6, -4, -2, 0, 2, 4, 6, 8, 10, 12, 14, 16, 18, 20, 22, 24}
        \filldraw[gray!50] ({\i},{\j}) circle (3pt);
\foreach \i in {0, 2, 4, 6, 8}
\foreach \j in {-7, -5, -3, -1, 1, 3, 5, 7, 9, 11, 13, 15, 17, 19, 21, 23}
        \filldraw[gray!50] ({\i},{\j}) circle (3pt);

\draw[thick, black] (4,0) -- (4,16);
\foreach \j in {-7, -5, -3, -1, 1, 3, 5, 9, 11, 13, 15, 17, 19, 21, 23}
        \filldraw[black] (4,{\j}) circle (3pt);
\filldraw[black] (4,7) circle (4.5pt);
\draw [double, ->, >=stealth, black] (4.2,7.2) to [bend right] (4.2,8.8);
\draw [double, ->, >=stealth, red] (4.2,6.8) to [bend left] (4.2,5.2);
\draw [double, ->, >=stealth, red] (4.2,9.2) to [bend right] (4.2,10.8);
\draw [double, ->, >=stealth, green] (3.8,9.2) to [bend left] (3.8,12.8);
\draw [double, ->, >=stealth, green] (3.8,11.2) to [bend left] (3.8,14.8);
\draw [double, ->, >=stealth, green] (3.8,6.8) to [bend right] (3.8,3.2);
\draw [double, ->, >=stealth, green] (3.8,4.8) to [bend right] (3.8,1.2);
\draw [double, ->, >=stealth, blue] (4.3,6.8) to [bend left] (4.3,-0.8);
\draw [double, ->, >=stealth, blue] (4.3,4.8) to [bend left] (4.3,-2.8);
\draw [double, ->, >=stealth, blue] (4.3,2.8) to [bend left] (4.3,-4.8);
\draw [double, ->, >=stealth, blue] (4.3,0.8) to [bend left] (4.3,-6.8);
\draw [double, ->, >=stealth, blue] (4.3,9.2) to [bend right] (4.3,16.8);
\draw [double, ->, >=stealth, blue] (4.3,11.2) to [bend right] (4.3,18.8);
\draw [double, ->, >=stealth, blue] (4.3,13.2) to [bend right] (4.3,20.8);
\draw [double, ->, >=stealth, blue] (4.3,15.2) to [bend right] (4.3,22.8);
\end{tikzpicture}
\end{center}
&
\begin{center}
\begin{tikzpicture}[scale=0.33]
\foreach \i in {-8, -6, -4, -2, 0, 2, 4, 6, 8, 10, 12, 14, 16, 18, 20, 22, 24}
        \draw[thick, gray] (0,{\i}) -- (8,{\i});
\foreach \i in {0, 2, 4, 6, 8}
        \draw[thick, gray] ({\i},-8) -- ({\i},24);
\foreach \i in {1, 3, 5, 7}
\foreach \j in {-8, -6, -4, -2, 0, 2, 4, 6, 8, 10, 12, 14, 16, 18, 20, 22, 24}
        \filldraw[gray!50] ({\i},{\j}) circle (3pt);
\foreach \i in {0, 2, 4, 6, 8}
\foreach \j in {-7, -5, -3, -1, 1, 3, 5, 7, 9, 11, 13, 15, 17, 19, 21, 23}
        \filldraw[gray!50] ({\i},{\j}) circle (3pt);

\draw[thick, black] (4,0) -- (4,16);
\foreach \j in {-7, -5, -3, -1, 1, 3, 5, 9, 11, 13, 15, 17, 19, 21, 23}
        \filldraw[black] (4,{\j}) circle (3pt);
\filldraw[black] (4,7) circle (4.5pt);
\draw [double, ->, >=stealth, black] (4.2,7.2) to [bend right] (4.2,8.8);
\draw [double, ->, >=stealth, red] (4.2,6.8) to [bend left] (4.2,5.2);
\draw [double, ->, >=stealth, red] (4.2,9.2) to [bend right] (4.2,10.8);
\draw [double, ->, >=stealth, green] (3.8,9.2) to [bend left] (3.8,12.8);
\draw [double, ->, >=stealth, green] (3.8,11.2) to [bend left] (3.8,14.8);
\draw [double, ->, >=stealth, green] (3.8,6.8) to [bend right] (3.8,3.2);
\draw [double, ->, >=stealth, green] (3.8,4.8) to [bend right] (3.8,1.2);
\draw [double, ->, >=stealth, blue] (4.2,2.8) to [bend left] (4.2,-0.8);
\draw [double, ->, >=stealth, blue] (4.2,0.8) to [bend left] (4.2,-2.8);
\draw [double, ->, >=stealth, orange] (3.8,-1.2) to [bend right] (3.8,-4.8);
\draw [double, ->, >=stealth, orange] (3.8,-3.2) to [bend right] (3.8,-6.8);
\draw [double, ->, >=stealth, blue] (4.2,13.2) to [bend right] (4.2,16.8);
\draw [double, ->, >=stealth, blue] (4.2,15.2) to [bend right] (4.2,18.8);
\draw [double, ->, >=stealth, orange] (3.8,17.2) to [bend left] (3.8,20.8);
\draw [double, ->, >=stealth, orange] (3.8,19.2) to [bend left] (3.8,22.8);
\end{tikzpicture}
\end{center}
\end{tabular}

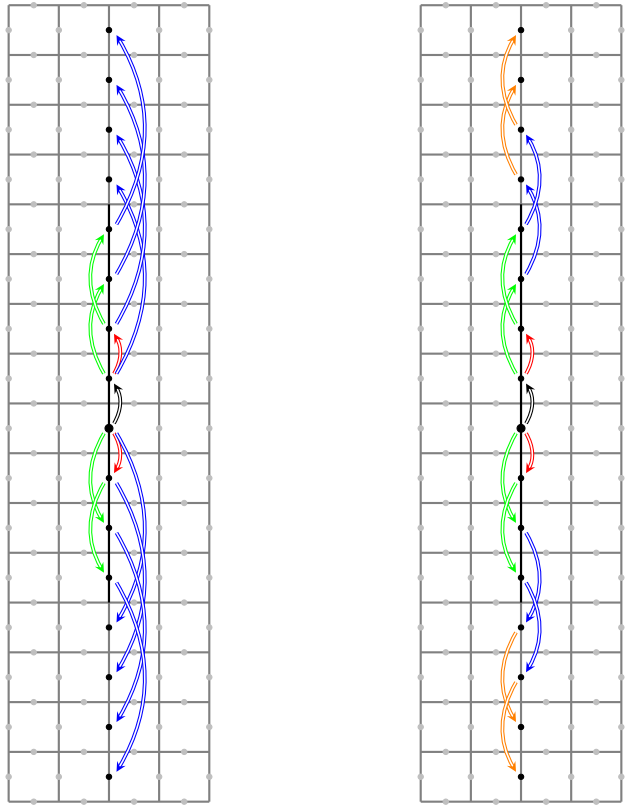
\captionof{figure}[foo]{For the case $L=16$, we illustrate the utilization of CNOT gates to vertically transmit the logical states in the sequence: black, red, green, blue, orange. On the left-hand side, there exists no constraint on the distance $d$, permitting the use of non-local CNOT gates, resulting in $log_{2}(16)=4$ steps. On the right-hand side, with the restriction of $d=2$, the process requires $log_{2}(2)+\frac{16}{4}=5$ steps.} 
\label{fig:clocalCNOT}
\end{table}

\section{3D toric model with boundary} \label{sec:3dtoricwith}

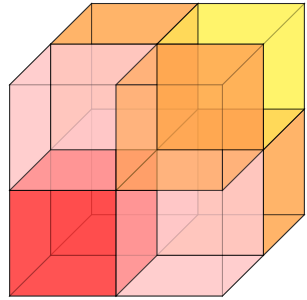
\begin{figure}[ht]
\centering
\begin{tikzpicture}[scale=0.7]
\coordinate (OO) at (0,0,0);
\coordinate (A) at (0,2,0);
\coordinate (B) at (0,2,2);
\coordinate (C) at (0,0,2);
\coordinate (D) at (2,0,0);
\coordinate (E) at (2,2,0);
\coordinate (F) at (2,2,2);
\coordinate (G) at (2,0,2);
\coordinate (H) at (4,0,0);
\coordinate (I) at (4,2,0);
\coordinate (J) at (4,2,2);
\coordinate (K) at (4,0,2);
\coordinate (L) at (0,4,2);
\coordinate (M) at (2,4,2);
\coordinate (N) at (2,4,0);
\coordinate (O) at (0,4,0);
\coordinate (P) at (0,2,-2);
\coordinate (Q) at (0,0,-2);
\coordinate (R) at (2,0,-2);
\coordinate (S) at (2,2,-2);
\coordinate (T) at (0,4,-2);
\coordinate (U) at (2,4,-2);
\coordinate (V) at (4,4,2);
\coordinate (W) at (4,4,0);
\coordinate (X) at (4,2,-2);
\coordinate (Y) at (4,0,-2);
\coordinate (Z) at (4,4,-2);
\draw[black, fill=pink!90, opacity=0.6] (P) -- (Q) -- (R) -- (S) -- cycle;
\draw[black, fill=pink!90, opacity=0.6] (P) -- (Q) -- (OO) -- (A) -- cycle;
\draw[black, fill=pink!90, opacity=0.6] (E) -- (D) -- (R) -- (S) -- cycle;
\draw[black, fill=pink!90, opacity=0.6] (A) -- (E) -- (S) -- (P) -- cycle;
\draw[black, fill=pink!90, opacity=0.6] (OO) -- (D) -- (R) -- (Q) -- cycle;
\draw[black, fill=orange!70, opacity=0.6] (P) -- (S) -- (U) -- (T) -- cycle;
\draw[black, fill=orange!70, opacity=0.6] (O) -- (N) -- (U) -- (T) -- cycle;
\draw[black, fill=orange!70, opacity=0.6] (E) -- (N) -- (U) -- (S) -- cycle;
\draw[black, fill=orange!70, opacity=0.6] (A) -- (O) -- (T) -- (P) -- cycle;
\draw[black, fill=orange!70, opacity=0.6] (R) -- (S) -- (X) -- (Y) -- cycle;
\draw[black, fill=orange!70, opacity=0.6] (S) -- (X) -- (I) -- (E) -- cycle;
\draw[black, fill=orange!70, opacity=0.6] (H) -- (I) -- (X) -- (Y) -- cycle;
\draw[black, fill=orange!70, opacity=0.6] (R) -- (Y) -- (H) -- (D) -- cycle;
\draw[black, fill=yellow!80, opacity=0.6] (Z) -- (W) -- (I) -- (X) -- cycle;
\draw[black, fill=yellow!80, opacity=0.6] (Z) -- (W) -- (N) -- (U) -- cycle;
\draw[black, fill=yellow!80, opacity=0.6] (Z) -- (U) -- (S) -- (X) -- cycle;
\draw[black, fill=red!80, opacity=0.6] (OO) -- (C) -- (G) -- (D) -- cycle;
\draw[black, fill=red!80, opacity=0.6] (OO) -- (A) -- (E) -- (D) -- cycle;
\draw[black, fill=red!80, opacity=0.6] (OO) -- (A) -- (B) -- (C) -- cycle;
\draw[black, fill=red!80, opacity=0.6] (D) -- (E) -- (F) -- (G) -- cycle;
\draw[black, fill=red!80, opacity=0.6] (C) -- (B) -- (F) -- (G) -- cycle;
\draw[black, fill=red!80, opacity=0.6] (A) -- (B) -- (F) -- (E) -- cycle;
\draw[black, fill=pink!90, opacity=0.6] (H) -- (I) -- (J) -- (K) -- cycle;
\draw[black, fill=pink!90, opacity=0.6] (G) -- (F) -- (J) -- (K) -- cycle;
\draw[black, fill=pink!90, opacity=0.6] (F) -- (E) -- (I) -- (J) -- cycle;
\draw[black, fill=pink!90, opacity=0.6] (E) -- (D) -- (H) -- (I) -- cycle;
\draw[black, fill=pink!90, opacity=0.6] (D) -- (G) -- (K) -- (H) -- cycle;
\draw[black, fill=pink!90, opacity=0.6] (O) -- (N) -- (M) -- (L) -- cycle;
\draw[black, fill=pink!90, opacity=0.6] (B) -- (F) -- (M) -- (L) -- cycle;
\draw[black, fill=pink!90, opacity=0.6] (F) -- (E) -- (N) -- (M) -- cycle;
\draw[black, fill=pink!90, opacity=0.6] (A) -- (E) -- (N) -- (O) -- cycle;
\draw[black, fill=pink!90, opacity=0.6] (A) -- (B) -- (L) -- (O) -- cycle;
\draw[black, fill=orange!70, opacity=0.6] (M) -- (N) -- (W) -- (V) -- cycle;
\draw[black, fill=orange!70, opacity=0.6] (M) -- (V) -- (J) -- (F) -- cycle;
\draw[black, fill=orange!70, opacity=0.6] (J) -- (I) -- (W) -- (V) -- cycle;
\draw[black, fill=orange!70, opacity=0.6] (E) -- (I) -- (W) -- (N) -- cycle;
\end{tikzpicture}
\caption{The beginning cube is colored red. The pink, orange and yellow cube represent the cubes connected with one, two or three faces developed.}
\label{fig:3dmodel}
\end{figure}

The generation from 2D toric code to 3D toric model with boundary is complicated but direct. We can continue to use a plaquette as the basic structure but consider four different types of cubes. Let us take the eight cubes in Figure \ref{fig:3dmodel} as an example. We begin with the red cube and develop it into pink cubes. Orange cubes are the next and the yellow cube completes the model. In the following, we will divide the method into four steps, each step describes one type of cubes.

To develop the qubits in the beginning red cube, we need to develop five rather than six faces as the cube is a closed surface with one redundant face. As shown in Figure \ref{fig:3dtoruswith1}, we develop a face first and choose the four qubits on the opposite face to repeat the basic structure. After that, considering the pink cube shares a face with developed cube, we only need to develop four more faces as the second cube is also a closed surface. We choose the four qubits on the face opposite to the developed cube to repeat the basic structure.

\begin{table}[ht]
\centering
\begin{tabular}{m{2.4cm} m{2.4cm} m{6cm}} 
\begin{center}
\begin{tikzpicture}[scale=0.8]
\coordinate (OO) at (0,0,0);
\coordinate (A) at (0,2,0);
\coordinate (B) at (0,2,2);
\coordinate (C) at (0,0,2);
\coordinate (D) at (2,0,0);
\coordinate (E) at (2,2,0);
\coordinate (F) at (2,2,2);
\coordinate (G) at (2,0,2);
\draw[thick, gray] (OO) -- (C) -- (G) -- (D) -- cycle;
\draw[thick, gray] (OO) -- (A) -- (E) -- (D) -- cycle;
\draw[thick, gray] (OO) -- (A) -- (B) -- (C) -- cycle;
\draw[thick, gray] (D) -- (E) -- (F) -- (G) -- cycle;
\draw[thick, gray] (C) -- (B) -- (F) -- (G) -- cycle;
\draw[thick, gray] (A) -- (B) -- (F) -- (E) -- cycle;
\filldraw[gray!50] (0,1,0) circle (3pt); 
\filldraw[gray!50] (1,0,0) circle (3pt);
\filldraw[gray!50] (2,1,0) circle (3pt);
\filldraw[gray!50] (1,2,0) circle (3pt);
\filldraw[gray!50] (0,1,2) circle (3pt);
\filldraw[gray!50] (1,0,2) circle (3pt);
\filldraw[gray!50] (2,1,2) circle (3pt);
\filldraw[gray!50] (1,2,2) circle (3pt);
\filldraw[gray!50] (0,0,1) circle (3pt);
\filldraw[gray!50] (2,0,1) circle (3pt);
\filldraw[gray!50] (0,2,1) circle (3pt);
\filldraw[gray!50] (2,2,1) circle (3pt);
\draw[thick, black] (2,2,1) circle (3pt); 
\draw[double, ->, >=stealth, black] (2,2,1)--(2,0.14,1); 
\draw[double, ->, >=stealth, black] (2,2,1)--(2,1.1,0.1);
\draw[double, ->, >=stealth, black] (2,2,1)--(2,1.1,1.9);
\end{tikzpicture}
\end{center}
&
\begin{center}
\begin{tikzpicture}[scale=0.8]
\coordinate (OO) at (0,0,0);
\coordinate (A) at (0,2,0);
\coordinate (B) at (0,2,2);
\coordinate (C) at (0,0,2);
\coordinate (D) at (2,0,0);
\coordinate (E) at (2,2,0);
\coordinate (F) at (2,2,2);
\coordinate (G) at (2,0,2);
\draw[thick, gray] (OO) -- (C) -- (G) -- (D) -- cycle;
\draw[thick, gray] (OO) -- (A) -- (E) -- (D) -- cycle;
\draw[thick, gray] (OO) -- (A) -- (B) -- (C) -- cycle;
\draw[thick, gray] (D) -- (E) -- (F) -- (G) -- cycle;
\draw[thick, gray] (C) -- (B) -- (F) -- (G) -- cycle;
\draw[thick, gray] (A) -- (B) -- (F) -- (E) -- cycle;
\filldraw[gray!50] (0,1,0) circle (3pt); 
\filldraw[gray!50] (1,0,0) circle (3pt);
\filldraw[black] (2,1,0) circle (3pt);
\filldraw[gray!50] (1,2,0) circle (3pt);
\filldraw[gray!50] (0,1,2) circle (3pt);
\filldraw[gray!50] (1,0,2) circle (3pt);
\filldraw[black] (2,1,2) circle (3pt);
\filldraw[gray!50] (1,2,2) circle (3pt);
\filldraw[gray!50] (0,0,1) circle (3pt);
\filldraw[black] (2,0,1) circle (3pt);
\filldraw[gray!50] (0,2,1) circle (3pt);
\filldraw[black] (2,2,1) circle (3pt);
\draw[thick, black] (0,1,0) circle (3pt); 
\draw[thick, black] (0,1,2) circle (3pt);
\draw[thick, black] (0,0,1) circle (3pt);
\draw[thick, black] (0,2,1) circle (3pt);
\draw[double, ->, >=stealth, black] (0,2,1)--(1.86,2,1); 
\draw[double, ->, >=stealth, black] (0,2,1)--(0.9,2,0.1);
\draw[double, ->, >=stealth, black] (0,2,1)--(0.9,2,1.9);
\draw[double, ->, >=stealth, black] (0,0,1)--(1.86,0,1); 
\draw[double, ->, >=stealth, black] (0,0,1)--(0.9,0,0.1);
\draw[double, ->, >=stealth, black] (0,0,1)--(0.9,0,1.9);
\draw[double, ->, >=stealth, black] (0,1,0)--(0.9,0.1,0); 
\draw[double, ->, >=stealth, black] (0,1,0)--(0.9,1.91,0);
\draw[double, ->, >=stealth, black] (0,1,0)--(1.86,1,0);
\draw[double, ->, >=stealth, black] (0,1,2)--(0.9,0.1,2); 
\draw[double, ->, >=stealth, black] (0,1,2)--(0.9,1.91,2);
\draw[double, ->, >=stealth, black] (0,1,2)--(1.86,1,2);
\end{tikzpicture}
\end{center}
&
\begin{center}
\begin{tikzpicture}[scale=0.8]
\coordinate (OO) at (0,0,0);
\coordinate (A) at (0,2,0);
\coordinate (B) at (0,2,2);
\coordinate (C) at (0,0,2);
\coordinate (D) at (2,0,0);
\coordinate (E) at (2,2,0);
\coordinate (F) at (2,2,2);
\coordinate (G) at (2,0,2);
\coordinate (H) at (4,0,0);
\coordinate (I) at (4,2,0);
\coordinate (J) at (4,2,2);
\coordinate (K) at (4,0,2);
\draw[thick, gray] (OO) -- (C) -- (G) -- (D) -- cycle;
\draw[thick, gray] (OO) -- (A) -- (E) -- (D) -- cycle;
\draw[thick, gray] (OO) -- (A) -- (B) -- (C) -- cycle;
\draw[thick, gray] (D) -- (E) -- (F) -- (G) -- cycle;
\draw[thick, gray] (C) -- (B) -- (F) -- (G) -- cycle;
\draw[thick, gray] (A) -- (B) -- (F) -- (E) -- cycle;
\draw[thick, gray] (H) -- (I) -- (J) -- (K) -- cycle;
\draw[thick, gray] (G) -- (F) -- (J) -- (K) -- cycle;
\draw[thick, gray] (F) -- (E) -- (I) -- (J) -- cycle;
\draw[thick, gray] (E) -- (D) -- (H) -- (I) -- cycle;
\draw[thick, gray] (D) -- (G) -- (K) -- (H) -- cycle;
\filldraw[black] (0,1,0) circle (3pt);
\filldraw[black] (1,0,0) circle (3pt);
\filldraw[black] (2,1,0) circle (3pt);
\filldraw[black] (1,2,0) circle (3pt);
\filldraw[black] (0,1,2) circle (3pt);
\filldraw[black] (1,0,2) circle (3pt);
\filldraw[black] (2,1,2) circle (3pt);
\filldraw[black] (1,2,2) circle (3pt);
\filldraw[black] (0,0,1) circle (3pt);
\filldraw[black] (2,0,1) circle (3pt);
\filldraw[black] (0,2,1) circle (3pt);
\filldraw[black] (2,2,1) circle (3pt);
\filldraw[gray!50] (3,0,0) circle (3pt);
\filldraw[gray!50] (3,2,0) circle (3pt);
\filldraw[gray!50] (3,0,2) circle (3pt);
\filldraw[gray!50] (3,2,2) circle (3pt);
\filldraw[gray!50] (4,1,0) circle (3pt);
\filldraw[gray!50] (4,1,2) circle (3pt);
\filldraw[gray!50] (4,0,1) circle (3pt);
\filldraw[gray!50] (4,2,1) circle (3pt);
\draw[thick, black] (4,1,0) circle (3pt); 
\draw[thick, black] (4,1,2) circle (3pt);
\draw[thick, black] (4,0,1) circle (3pt);
\draw[thick, black] (4,2,1) circle (3pt);
\draw[double, ->, >=stealth, black] (4,2,1)--(2.14,2,1); 
\draw[double, ->, >=stealth, black] (4,2,1)--(3.1,2,0.1);
\draw[double, ->, >=stealth, black] (4,2,1)--(3.1,2,1.9);
\draw[double, ->, >=stealth, black] (4,0,1)--(2.14,0,1); 
\draw[double, ->, >=stealth, black] (4,0,1)--(3.1,0,0.1);
\draw[double, ->, >=stealth, black] (4,0,1)--(3.1,0,1.9);
\draw[double, ->, >=stealth, black] (4,1,0)--(3.1,0.1,0); 
\draw[double, ->, >=stealth, black] (4,1,0)--(3.1,1.91,0);
\draw[double, ->, >=stealth, black] (4,1,0)--(2.14,1,0);
\draw[double, ->, >=stealth, black] (4,1,2)--(3.1,0.1,2); 
\draw[double, ->, >=stealth, black] (4,1,2)--(3.1,1.91,2);
\draw[double, ->, >=stealth, black] (4,1,2)--(2.14,1,2);
\end{tikzpicture}
\end{center}
\end{tabular}

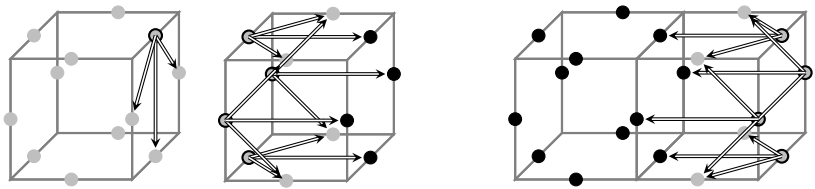
\captionof{figure}[foo]{The left two cubes describe the first step to develop the red cube. The right cubes describes the second step to develop the pink cube.} 
\label{fig:3dtoruswith1}
\end{table}

Similarly, we need to develop three faces for the orange cubes and two faces for the yellow cube as shown in Figure \ref{fig:3dtoruswith2}. The four steps complete the procedure to simulate the ground state of toric model on the eight cubes lattices. And we are able to develop any size cubes with boundary using the method described above.

\begin{table}[ht]
\centering
\begin{tabular}{m{5cm} m{6cm}} 
\begin{center}
\begin{tikzpicture}[scale=0.8]
\foreach \i in {0, 2, 4}
\foreach \j in {0, 2, 4}
        \draw[thick, gray] ({\i},{\j},2) -- ({\i},{\j},0);
\foreach \i in {0, 2, 4}
\foreach \k in {0, 2}
        \draw[thick, gray] ({\i},0,{\k}) -- ({\i},4,{\k});
\foreach \j in {0, 2, 4}
\foreach \k in {0, 2}
        \draw[thick, gray] (0,{\j},{\k}) -- (4,{\j},{\k});
\foreach \i in {0, 2, 4}
\foreach \j in {1, 3}
\foreach \k in {0, 2}
        \filldraw[black] ({\i},{\j},{\k}) circle (3pt);
\foreach \i in {0, 2, 4}
\foreach \j in {0, 2, 4}
        \filldraw[black] ({\i},{\j},1) circle (3pt);
\foreach \i in {1, 3}
\foreach \j in {0, 2, 4}
\foreach \k in {0, 2}
        \filldraw[black] ({\i},{\j},{\k}) circle (3pt);
\filldraw[gray!50] (3,4,0) circle (3pt);
\filldraw[gray!50] (3,4,2) circle (3pt);
\filldraw[gray!50] (4,4,1) circle (3pt);
\filldraw[gray!50] (4,3,0) circle (3pt);
\filldraw[gray!50] (4,3,2) circle (3pt);
\draw[thick, black] (3,4,2) circle (3pt); 
\draw[thick, black] (3,4,0) circle (3pt);
\draw[thick, black] (4,3,0) circle (3pt);
\draw[double, ->, >=stealth, black] (3,4,2)--(2.1,3.1,2); 
\draw[double, ->, >=stealth, black] (3,4,2)--(3.9,3.1,2);
\draw[double, ->, >=stealth, black] (3,4,2)--(3,2.14,2);
\draw[double, ->, >=stealth, black] (3,4,0)--(2.1,4,0.9);
\draw[double, ->, >=stealth, black] (3,4,0)--(3.9,4,0.9);
\draw[double, ->, >=stealth, black] (3,4,0)--(3,4,1.86);
\draw[double, ->, >=stealth, black] (4,3,0)--(4,3.9,0.9);
\draw[double, ->, >=stealth, black] (4,3,0)--(4,2.1,0.9);
\draw[double, ->, >=stealth, black] (4,3,0)--(4,3,1.86);
\end{tikzpicture}
\end{center}
&
\begin{center}
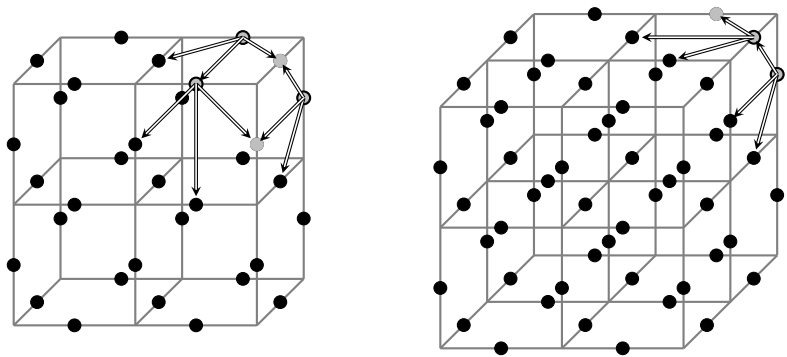

\begin{tikzpicture}[scale=0.8]
\foreach \i in {0, 2, 4}
\foreach \j in {0, 2, 4}
        \draw[thick, gray] ({\i},{\j},2) -- ({\i},{\j},-2);
\foreach \i in {0, 2, 4}
\foreach \k in {-2, 0, 2}
        \draw[thick, gray] ({\i},0,{\k}) -- ({\i},4,{\k});
\foreach \j in {0, 2, 4}
\foreach \k in {-2, 0, 2}
        \draw[thick, gray] (0,{\j},{\k}) -- (4,{\j},{\k});
\foreach \i in {0, 2, 4}
\foreach \j in {1, 3}
\foreach \k in {-2, 0, 2}
        \filldraw[black] ({\i},{\j},{\k}) circle (3pt);
\foreach \i in {0, 2, 4}
\foreach \j in {0, 2, 4}
\foreach \k in {-1, 1}
        \filldraw[black] ({\i},{\j},{\k}) circle (3pt);
\foreach \i in {1, 3}
\foreach \j in {0, 2, 4}
\foreach \k in {-2, 0, 2}
        \filldraw[black] ({\i},{\j},{\k}) circle (3pt);
\filldraw[gray!50] (3,4,-2) circle (3pt);
\filldraw[gray!50] (4,4,-1) circle (3pt);
\filldraw[gray!50] (4,3,-2) circle (3pt);
\draw[thick, black] (4,4,-1) circle (3pt); 
\draw[thick, black] (4,3,-2) circle (3pt);
\draw[double, ->, >=stealth, black] (4,4,-1)--(3.1,4,-0.1); 
\draw[double, ->, >=stealth, black] (4,4,-1)--(3.1,4,-1.9);
\draw[double, ->, >=stealth, black] (4,4,-1)--(2.16,4,-1);
\draw[double, ->, >=stealth, black] (4,3,-2)--(4,2.1,-1.1);
\draw[double, ->, >=stealth, black] (4,3,-2)--(4,3.9,-1.1);
\draw[double, ->, >=stealth, black] (4,3,-2)--(4,3,-0.14);
\end{tikzpicture}
\end{center}
\end{tabular}
\captionof{figure}[foo]{The left figure describes the step of orange cubes, and we need to develop the face in front first. The right figure describes the final step to develop the yellow cube, and we need to develop the face above first.} 
\label{fig:3dtoruswith2}
\end{table}

\section{3D toric model without boundary} \label{sec:3dtoricwithout}
In Figure \ref{fig:3dtoruswithout}, the opposite faces are identified together to represent the 3D torus. We begin with $|\phi_0\rangle$ and choose four free qubits in the lower layer to take the procedure in basic structure. After this step and identification of opposite faces, we get the lattice with the middle untouched. Finally, choose three more free qubits to repeat the basic structure and leave a vertex redundant.

\begin{table}[ht]
\centering
\begin{tabular}{m{5.5cm} m{5.5cm}}
\begin{center}
\begin{tikzpicture}[scale=0.7]
\foreach \i in {0, 2, 4}
\foreach \j in {0, 2, 4}
        \draw[thick, gray] ({\i},{\j},2) -- ({\i},{\j},-2);
\foreach \i in {0, 2, 4}
\foreach \k in {-2, 0, 2}
        \draw[thick, gray] ({\i},0,{\k}) -- ({\i},4,{\k});
\foreach \j in {0, 2, 4}
\foreach \k in {-2, 0, 2}
        \draw[thick, gray] (0,{\j},{\k}) -- (4,{\j},{\k});
\foreach \i in {0, 2, 4}
\foreach \j in {1, 3}
\foreach \k in {-2, 0, 2}
        \filldraw[gray!50] ({\i},{\j},{\k}) circle (3pt);
\foreach \i in {0, 2, 4}
\foreach \j in {0, 2, 4}
\foreach \k in {-1, 1}
        \filldraw[gray!50] ({\i},{\j},{\k}) circle (3pt);
\foreach \i in {1, 3}
\foreach \j in {0, 2, 4}
\foreach \k in {-2, 0, 2}
        \filldraw[gray!50] ({\i},{\j},{\k}) circle (3pt);
\foreach \i in {0, 2}
\foreach \k in {0, 2}
        \draw[thick, black] ({\i},1,{\k}) circle (3pt); 
\foreach \i in {0, 2}
\foreach \k in {0, 2}
        \draw[double, ->, >=stealth, black] ({\i},1,{\k})--({\i},-1,{\k}); 
\foreach \i in {0, 2}
\foreach \k in {0, 2}
        \draw[double, ->, >=stealth, black] ({\i},1,{\k})--({\i},0,{\k+1});
\foreach \i in {0, 2}
\foreach \k in {0, 2}
        \draw[double, ->, >=stealth, black] ({\i},1,{\k})--({\i+1},0,{\k});
\foreach \i in {0, 2}
\foreach \k in {0, 2}
        \draw[double, ->, >=stealth, black] ({\i},1,{\k})--({\i},0,{\k-1});
\foreach \i in {0, 2}
\foreach \k in {0, 2}
        \draw[double, ->, >=stealth, black] ({\i},1,{\k})--({\i-1},0,{\k});
\end{tikzpicture}
\end{center}
&
\begin{center}
\begin{tikzpicture}[scale=0.7]
\foreach \i in {0, 2, 4}
\foreach \j in {0, 2, 4}
        \draw[thick, gray] ({\i},{\j},2) -- ({\i},{\j},-2);
\foreach \i in {0, 2, 4}
\foreach \k in {-2, 0, 2}
        \draw[thick, gray] ({\i},0,{\k}) -- ({\i},4,{\k});
\foreach \j in {0, 2, 4}
\foreach \k in {-2, 0, 2}
        \draw[thick, gray] (0,{\j},{\k}) -- (4,{\j},{\k});
\foreach \i in {0, 2, 4}
\foreach \j in {1, 3}
\foreach \k in {-2, 0, 2}
        \filldraw[black] ({\i},{\j},{\k}) circle (3pt);
\foreach \i in {0, 2, 4}
\foreach \j in {0, 4}
\foreach \k in {-1, 1}
        \filldraw[black] ({\i},{\j},{\k}) circle (3pt);
\foreach \i in {1, 3}
\foreach \j in {0, 4}
\foreach \k in {-2, 0, 2}
        \filldraw[black] ({\i},{\j},{\k}) circle (3pt);
\foreach \i in {0, 2, 4}
\foreach \k in {-1, 1}
        \filldraw[gray!50] ({\i},2,{\k}) circle (3pt);
\foreach \i in {1, 3}
\foreach \k in {-2, 0, 2}
        \filldraw[gray!50] ({\i},2,{\k}) circle (3pt);
\draw[thick, black] (3,2,-2) circle (3pt); 
\draw[double, ->, >=stealth, black] (3,2,-2)--(1,2,-2); 
\draw[double, ->, >=stealth, black] (3,2,-2)--(2,1,-2);
\draw[double, ->, >=stealth, black] (3,2,-2)--(2,3,-2);
\draw[double, ->, >=stealth, black] (3,2,-2)--(2,2,-1);
\draw[double, ->, >=stealth, black] (3,2,-2)--(2,2,-3);
\foreach \i in {0, 2}
        \draw[thick, black] ({\i},2,1) circle (3pt); 
\foreach \i in {0, 2}
        \draw[double, ->, >=stealth, black] ({\i},2,1)--({\i},2,-1); 
\foreach \i in {0, 2}
        \draw[double, ->, >=stealth, black] ({\i},2,1)--({\i},1,0);
\foreach \i in {0, 2}
        \draw[double, ->, >=stealth, black] ({\i},2,1)--({\i},3,0);
\foreach \i in {0, 2}
        \draw[double, ->, >=stealth, black] ({\i},2,1)--({\i+1},2,0);
\foreach \i in {0, 2}
        \draw[double, ->, >=stealth, black] ({\i},2,1)--({\i-1},2,0);
\end{tikzpicture}
\end{center}
\end{tabular}

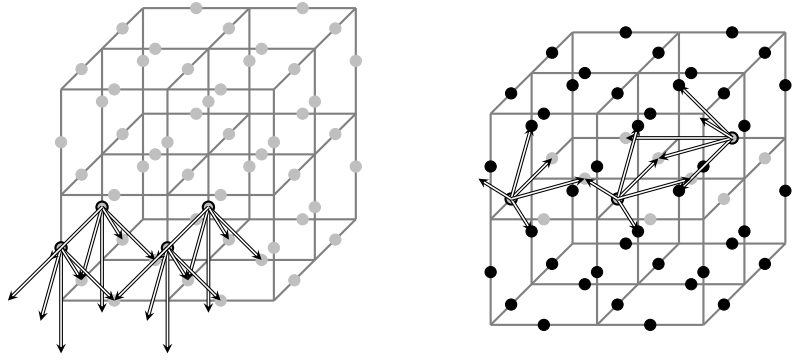
\captionof{figure}[foo]{A qubit $|0\rangle$ is placed at each gray dot at the beginning. The color changes to black when a quantum gate is applied on the qubit.} 
\label{fig:3dtoruswithout}
\end{table}

\section{X-cube model simple example} \label{sec:Xcubesimple}

To illustrate the method, we take the eight cubes case as a simple example shown in Figure \ref{fig:xcubesimple}. Considering the redundant cubes in yellow, we only need to develop four cubes left. The initial state is $|\phi_0\rangle$, and we begin with the cube at the right front higher corner to apply the basic structure. After this step and identifying opposite faces, we get the result on the right-hand side of Figure \ref{fig:xcubesimple}. Then we choose three more free qubits from each cube connecting with the developed cube to repeat the procedure of basic structure and the ground state is completed.

\begin{table}[ht]
\centering
\begin{tabular}{m{5cm} m{5cm}} 
\begin{center}
\begin{tikzpicture}[scale=0.8]
\coordinate (OO) at (0,0,0);
\coordinate (A) at (0,2,0);
\coordinate (B) at (0,2,2);
\coordinate (C) at (0,0,2);
\coordinate (D) at (2,0,0);
\coordinate (E) at (2,2,0);
\coordinate (F) at (2,2,2);
\coordinate (G) at (2,0,2);
\coordinate (H) at (4,0,0);
\coordinate (I) at (4,2,0);
\coordinate (J) at (4,2,2);
\coordinate (K) at (4,0,2);
\coordinate (L) at (0,4,2);
\coordinate (M) at (2,4,2);
\coordinate (N) at (2,4,0);
\coordinate (O) at (0,4,0);
\coordinate (P) at (0,2,-2);
\coordinate (Q) at (0,0,-2);
\coordinate (R) at (2,0,-2);
\coordinate (S) at (2,2,-2);
\coordinate (T) at (0,4,-2);
\coordinate (U) at (2,4,-2);
\coordinate (V) at (4,4,2);
\coordinate (W) at (4,4,0);
\coordinate (X) at (4,2,-2);
\coordinate (Y) at (4,0,-2);
\coordinate (Z) at (4,4,-2);
\draw[black, fill=yellow!60, opacity=0.4] (OO) -- (A) -- (P) -- (Q) -- cycle; 
\draw[black, fill=yellow!60, opacity=0.4] (OO) -- (D) -- (R) -- (Q) -- cycle;
\draw[black, fill=yellow!60, opacity=0.4] (S) -- (P) -- (Q) -- (R) -- cycle;
\draw[black, fill=yellow!60, opacity=0.4] (O) -- (N) -- (U) -- (T) -- cycle; 
\draw[black, fill=yellow!60, opacity=0.4] (O) -- (N) -- (E) -- (A) -- cycle;
\draw[black, fill=yellow!60, opacity=0.4] (O) -- (T) -- (P) -- (A) -- cycle;
\draw[black, fill=yellow!60, opacity=0.4] (O) -- (T) -- (U) -- (N) -- cycle;
\draw[black, fill=yellow!60, opacity=0.4] (S) -- (P) -- (T) -- (U) -- cycle;
\draw[black, fill=yellow!60, opacity=0.4] (S) -- (P) -- (A) -- (E) -- cycle;
\draw[black, fill=yellow!60, opacity=0.4] (H) -- (D) -- (R) -- (Y) -- cycle; 
\draw[black, fill=yellow!60, opacity=0.4] (H) -- (D) -- (E) -- (I) -- cycle;
\draw[black, fill=yellow!60, opacity=0.4] (I) -- (X) -- (Y) -- (H) -- cycle;
\draw[black, fill=yellow!60, opacity=0.4] (I) -- (X) -- (S) -- (E) -- cycle;
\draw[black, fill=yellow!60, opacity=0.4] (S) -- (R) -- (D) -- (E) -- cycle;
\draw[black, fill=yellow!60, opacity=0.4] (S) -- (R) -- (Y) -- (X) -- cycle;
\draw[black, fill=yellow!60, opacity=0.4] (OO) -- (C) -- (B) -- (A) -- cycle; 
\draw[black, fill=yellow!60, opacity=0.4] (OO) -- (C) -- (G) -- (D) -- cycle;
\draw[black, fill=yellow!60, opacity=0.4] (F) -- (G) -- (D) -- (E) -- cycle;
\draw[black, fill=yellow!60, opacity=0.4] (F) -- (G) -- (C) -- (B) -- cycle;
\draw[black, fill=yellow!60, opacity=0.4] (A) -- (E) -- (F) -- (B) -- cycle;
\draw[black, fill=yellow!60, opacity=0.4] (A) -- (E) -- (D) -- (OO) -- cycle;
\draw[black, fill=gray!20, opacity=0.3] (S) -- (U) -- (Z) -- (X) -- cycle; 
\draw[black, fill=gray!20, opacity=0.3] (S) -- (U) -- (N) -- (E) -- cycle;
\draw[black, fill=gray!20, opacity=0.3] (Z) -- (W) -- (N) -- (U) -- cycle;
\draw[black, fill=gray!20, opacity=0.3] (Z) -- (W) -- (I) -- (X) -- cycle;
\draw[black, fill=gray!20, opacity=0.3] (S) -- (E) -- (I) -- (X) -- cycle;
\draw[black, fill=gray!20, opacity=0.3] (H) -- (I) -- (J) -- (K) -- cycle; 
\draw[black, fill=gray!20, opacity=0.3] (H) -- (I) -- (E) -- (D) -- cycle;
\draw[black, fill=gray!20, opacity=0.3] (G) -- (K) -- (H) -- (D) -- cycle;
\draw[black, fill=gray!20, opacity=0.3] (G) -- (K) -- (J) -- (F) -- cycle;
\draw[black, fill=gray!20, opacity=0.3] (G) -- (D) -- (E) -- (F) -- cycle;
\draw[black, fill=gray!20, opacity=0.3] (O) -- (A) -- (B) -- (L) -- cycle; 
\draw[black, fill=gray!20, opacity=0.3] (O) -- (A) -- (E) -- (N) -- cycle;
\draw[black, fill=gray!20, opacity=0.3] (M) -- (L) -- (O) -- (N) -- cycle;
\draw[black, fill=gray!20, opacity=0.3] (M) -- (L) -- (B) -- (F) -- cycle;
\draw[black, fill=gray!20, opacity=0.3] (A) -- (B) -- (F) -- (E) -- cycle;
\draw[black, fill=gray!20, opacity=0.3] (F) -- (M) -- (V) -- (J) -- cycle; 
\draw[black, fill=gray!20, opacity=0.3] (F) -- (M) -- (N) -- (E) -- cycle;
\draw[black, fill=gray!20, opacity=0.3] (E) -- (I) -- (J) -- (F) -- cycle;
\draw[black, fill=gray!20, opacity=0.3] (E) -- (I) -- (W) -- (N) -- cycle;
\draw[black, fill=gray!20, opacity=0.3] (W) -- (V) -- (M) -- (N) -- cycle;
\draw[black, fill=gray!20, opacity=0.3] (W) -- (V) -- (J) -- (I) -- cycle;

\filldraw[gray!50] (2,3,0) circle (3pt);
\filldraw[gray!50] (3,2,0) circle (3pt);
\filldraw[gray!50] (4,3,0) circle (3pt);
\filldraw[gray!50] (2,2,1) circle (3pt);
\filldraw[gray!50] (4,4,1) circle (3pt);
\filldraw[gray!50] (3,4,0) circle (3pt);
\filldraw[gray!50] (2,4,1) circle (3pt);
\filldraw[gray!50] (4,2,1) circle (3pt);
\filldraw[gray!50] (2,3,2) circle (3pt);
\filldraw[gray!50] (3,4,2) circle (3pt);
\filldraw[gray!50] (3,2,2) circle (3pt);
\draw[double, ->, >=stealth, black] (4,3,2)--(2,3,0);
\draw[double, ->, >=stealth, black] (4,3,2)--(3,2,0);
\draw[double, ->, >=stealth, black] (4,3,2)--(4,3,0);
\draw[double, ->, >=stealth, black] (4,3,2)--(2,2,1);
\draw[double, ->, >=stealth, black] (4,3,2)--(4,4,1);
\draw[double, ->, >=stealth, black] (4,3,2)--(3,4,0);
\draw[double, ->, >=stealth, black] (4,3,2)--(2,4,1);
\draw[double, ->, >=stealth, black] (4,3,2)--(4,2,1);
\draw[double, ->, >=stealth, black] (4,3,2)--(2,3,2);
\draw[double, ->, >=stealth, black] (4,3,2)--(3,4,2);
\draw[double, ->, >=stealth, black] (4,3,2)--(3,2,2);
\filldraw[gray!50] (4,3,2) circle (3pt);
\draw[thick, black] (4,3,2) circle (3pt);
\end{tikzpicture}
\end{center}
&
\begin{center}
\begin{tikzpicture}[scale=0.8]
\foreach \i in {0, 2, 4}
\foreach \j in {0, 2, 4}
        \draw[thick, gray] ({\i},{\j},2) -- ({\i},{\j},-2);
\foreach \i in {0, 2, 4}
\foreach \k in {-2, 0, 2}
        \draw[thick, gray] ({\i},0,{\k}) -- ({\i},4,{\k});
\foreach \j in {0, 2, 4}
\foreach \k in {-2, 0, 2}
        \draw[thick, gray] (0,{\j},{\k}) -- (4,{\j},{\k});
\foreach \i in {0, 2, 4}
\foreach \j in {1, 3}
\foreach \k in {-2, 0, 2}
        \filldraw[gray!50] ({\i},{\j},{\k}) circle (3pt);
\foreach \i in {0, 2, 4}
\foreach \j in {0, 2, 4}
\foreach \k in {-1, 1}
        \filldraw[gray!50] ({\i},{\j},{\k}) circle (3pt);
\foreach \i in {1, 3}
\foreach \j in {0, 2, 4}
\foreach \k in {-2, 0, 2}
        \filldraw[gray!50] ({\i},{\j},{\k}) circle (3pt);
\filldraw[black] (2,3,0) circle (3pt);
\filldraw[black] (2,3,2) circle (3pt);
\filldraw[black] (2,2,1) circle (3pt);
\filldraw[black] (2,4,1) circle (3pt);
\filldraw[black] (3,2,0) circle (3pt);
\filldraw[black] (3,4,0) circle (3pt);
\filldraw[black] (3,2,2) circle (3pt);
\filldraw[black] (3,4,2) circle (3pt);
\filldraw[black] (4,3,2) circle (3pt);
\filldraw[black] (4,3,0) circle (3pt);
\filldraw[black] (4,2,1) circle (3pt);
\filldraw[black] (4,4,1) circle (3pt);
\filldraw[black] (0,3,2) circle (3pt);
\filldraw[black] (0,3,0) circle (3pt);
\filldraw[black] (0,2,1) circle (3pt);
\filldraw[black] (0,4,1) circle (3pt);
\filldraw[black] (2,3,-2) circle (3pt);
\filldraw[black] (3,2,-2) circle (3pt);
\filldraw[black] (3,4,-2) circle (3pt);
\filldraw[black] (4,3,-2) circle (3pt);
\filldraw[black] (2,0,1) circle (3pt);
\filldraw[black] (3,0,0) circle (3pt);
\filldraw[black] (3,0,2) circle (3pt);
\filldraw[black] (4,0,1) circle (3pt);
\filldraw[black] (0,0,1) circle (3pt);
\filldraw[black] (0,3,-2) circle (3pt);
\filldraw[black] (3,0,-2) circle (3pt);
\draw[thick, black] (1,4,2) circle (3pt);
\draw[thick, black] (4,1,2) circle (3pt);
\draw[thick, black] (4,4,-1) circle (3pt);
\end{tikzpicture}
\end{center}
\end{tabular}

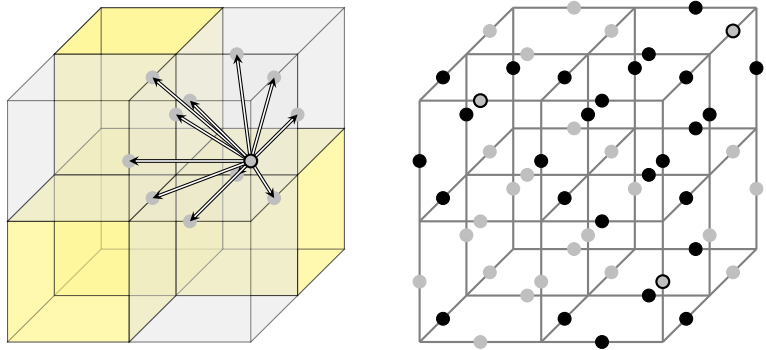
\captionof{figure}[foo]{The left figure is an example of X-cube model with opposite faces identified. The right figure shows the result after the first step and the free qubits for next step are circled.} 
\label{fig:xcubesimple}
\end{table}

\end{document}